\documentclass{svjour3}                     
\smartqed  
\usepackage{graphicx}
\usepackage{epstopdf}
\usepackage{lscape} 
\usepackage{txfonts}
\newcommand{\edt}[1]{#1}
\usepackage{natbib} 
\bibpunct{(}{)}{;}{a}{}{,} 

\newcommand{\aj}{AJ}
\newcommand{\aap}{A\&A}
\newcommand{\aapr}{A\&A~Rev.}
\newcommand{\aaps}{A\&AS}
\newcommand{\apj}{ApJ}
\newcommand{\apjl}{ApJL}
\newcommand{\apjs}{ApJS}
\newcommand{\apss}{ApSS}
\newcommand{\araa}{ARA\&A}
\newcommand{\grl}{Geophys.~Res.~Lett.}
\newcommand{\jgr}{J. Gephys. Res.}
\newcommand{\memsai}{Mem.~Soc.~Astron.~Italiana}
\newcommand{\mnras}{MNRAS}
\newcommand{\nat}{Nature}
\newcommand{\pasj}{PASJ}
\newcommand{\solphys}{Sol. Phys.}
\newcommand{\ssr}{Space~Sci.~Rev.}          
\newcommand{\rmxaa}{Rev. Mex. de Astron. y Astrof.} 
\begin{document}

\title{Solar science with the Atacama Large Millimeter/submillimeter Array -- {A new} view of our Sun} 

\titlerunning{Solar science with ALMA}        

\author{S.~Wedemeyer  
\and T.~Bastian  
\and R.~Braj{\v s}a  
\and H.~Hudson  
\and G.~Fleishman  
\and M.~Loukitcheva  
\and B.~Fleck  
\and E.~P.~Kontar  
\and B.~De Pontieu  
\and P.~Yagoubov  
\and S.~K.~Tiwari  
\and R.~Soler  
\and J.~H.~Black  
\and P.~Antolin  
\and E.~Scullion  
\and S.~Gun\'ar   
\and N.~Labrosse  
\and H.-G.~Ludwig  
\and A.~O.~Benz  
\and S.~M.~White   
\and P.~Hauschildt  
\and J.~G.~Doyle  
\and V.~M.~Nakariakov  
\and T.~Ayres   
\and P.~Heinzel  
\and M.~Karlicky  
\and T.~Van~Doorsselaere  
\and D.~Gary  
\and C.~E.~Alissandrakis  
\and A.~Nindos   
\and S.~K.~Solanki  
\and L.~Rouppe~van~der~Voort  
\and M.~Shimojo  
\and Y.~Kato 
\and T.~Zaqarashvili   
\and E.~Perez 
\and {C.~L.~Selhorst} 
\and M.~Barta 
}

\authorrunning{Wedemeyer et al.} 

\institute{
Sven Wedemeyer 
\at Institute of Theoretical Astrophysics, University of Oslo, Postboks 1029 Blindern, N-0315 Oslo, Norway
    \email{svenwe@astro.uio.no}
\at European ARC, Czech node, Astronomical Institute ASCR, Ondrejov, Czech Republic\\
Tel. 	:	+47-228 56 520\\
Fax 	:	+47-228 56 505\\
E-mail 	:	\email{sven.wedemeyer@astro.uio.no}
\and T.~Bastian 
\at National Radio Astronomy Observatory (NRAO), 520 Edgemont Road, Charlottesville, VA 22903, USA
\and R.~Braj{\v s}a 
\at Hvar Observatory, Faculty of Geodesy, University of Zagreb, Croatia
\at European ARC, Czech node, Astronomical Institute ASCR, Ondrejov, Czech Republic
\and H.~Hudson 
\at Space Sciences Laboratory, 7 Gauss Way, University of California, Berkeley, CA 94720-7450, USA 
\at SUPA, School of Physics \& Astronomy, University of Glasgow,   Glasgow,  G12 8QQ, UK
\and G.~Fleishman 
\at Center For Solar-Terrestrial Research, Physics Department, New Jersey Institute of Technology, 323 MLK blvd, Newark, NJ 07102, USA
\and M.~Loukitcheva 
\at Astronomical Institute, Saint-Petersburg University, Universitetskii pr. 28, 198504 Saint-Petersburg, Russia 
\at Max-Planck-Institut f\"ur Sonnensystemforschung, Justus-von-Liebig-Weg 3, 37077 G\"ottingen, Germany
\and B.~Fleck 
\at ESA Science Operations Department, c/o NASA Goddard Space Flight Center, Greenbelt, MD 20771, USA
\and E.~P.~Kontar 
\at SUPA, School of Physics \& Astronomy, University of Glasgow,   Glasgow,  G12 8QQ, UK
\and B.~De Pontieu 
\at Lockheed Martin Solar \& Astrophysics Laboratory, 3251 Hanover Street, Org. A021S, B. 252, Palo Alto, CA 94304, USA
\at Institute of Theoretical Astrophysics, University of Oslo, Postboks 1029 Blindern, N-0315 Oslo, Norway
\and P.~Yagoubov 
\at European Organisation for Astronomical Research in the Southern Hemisphere (ESO), Karl-Schwarzschild-Strasse 2, D-85748 Garching bei M\"unchen,  Germany
\and S.~K.~Tiwari 
\at NASA Marshall Space Flight Center, ZP 13, Huntsville, AL 35805, USA
\and R.~Soler 
\at Departament de Fisica, Universitat de les Illes Balears, E-07122 Palma de Mallorca, Spain
\and J.~H.~Black 
\at Chalmers University of Technology, Dept. of Earth and Space Sciences, Onsala Space Observatory, SE-43992 Onsala, Sweden
\and P.~Antolin 
\at National Astronomical Observatory of Japan, 2-21-1 Osawa, Mitaka, Tokyo 181-8588, Japan
\and E.~Scullion 
\at Trinity College Dublin, College Green, Dublin 2, Ireland
\and S.~Gun\'ar 
\at School of Mathematics and Statistics, University of St Andrews, North Haugh, St Andrews, KY16 9SS, UK
\at Astronomical Institute, Academy of Sciences, Fric\u{o}va 298, 251 65 Ond\u{r}ejov, Czech Republic (on leave)
\and N.~Labrosse 
\at SUPA, School of Physics \& Astronomy, University of Glasgow, UK
\and H.-G.~Ludwig 
\at ZAH, Landessternwarte K{\"o}nigstuhl 12, D-69117 Heidelberg, Germany  
\and A.~O.~Benz 
\at FHNW, Institute for 4D Technologies, Windisch, Switzerland
\and S.~M.~White 
\at Space Vehicles Directorate, AFRL, 3550 Aberdeen Avenue SE, Bldg 427, Kirtland AFB, NM 87117-5776, USA
\and P.~Hauschildt
\at Hamburger Sternwarte, Gojenbergsweg 112, 21029 Hamburg, Germany  
\and J.~G.~Doyle 
\at Armagh Observatory, College Hill BT61 9DG, N. Ireland
\and V.~M.~Nakariakov 
\at Centre for Fusion, Space and Astrophysics, Department of Physics, University of Warwick, Coventry CV4 7AL, UK
\and T.~Ayres 
\at Center for Astrophysics and Space Astronomy, University of Colorado, Boulder, CO 80309, USA
\and P.~Heinzel 
\at Astronomical Institute, Academy of Sciences, Fric\u{o}va 298, 251 65 Ond\u{r}ejov, Czech Republic
\and M.~Karlicky 
\at Astronomical Institute, Academy of Sciences, Fric\u{o}va 298, 251 65 Ond\u{r}ejov, Czech Republic
\at European ARC, Czech node, Astronomical Institute ASCR, Ondrejov, Czech Republic
\and T.~Van~Doorsselaere 
\at Centre for mathematical Plasma Astrophysics, Mathematics Department, KU~Leuven, Celestijnenlaan 200B bus 2400, B-3000 Leuven, Belgium
\and D.~Gary
\at Center For Solar-Terrestrial Research, Physics Department, New Jersey Institute of Technology, 323 MLK blvd, Newark, NJ 07102, USA
\and C.~E.~Alissandrakis
\at Department of Physics, University of Ioannina, GR-45110 Ioannina, Greece.
\and A.~Nindos 
\at Department of Physics, University of Ioannina, GR-45110 Ioannina, Greece.
\and S.~K.~Solanki
\at Max-Planck-Institut for Sonnensystemforschung, Justus-von-Liebig-Weg 3, 37077 G\"ottingen, Germany
\at School of Space Research, Kyung Hee University, Yongin, Gyeonggi, Republic of Korea
\and L.~Rouppe~van~der~Voort 
\at Institute of Theoretical Astrophysics, University of Oslo, Postboks 1029 Blindern, N-0315 Oslo, Norway 
\and M.~Shimojo
\at National Astronomical Observatory of Japan, 2-21-1 Osawa, Mitaka, Tokyo 181-8588, Japan
\and Y.~Kato
\at Institute of Theoretical Astrophysics, University of Oslo, Postboks 1029 Blindern, N-0315 Oslo, Norway
\and T.~Zaqarashvili 
\at Institute of Physics, University of Graz, UniversitŠtsplatz 5, 8010 Graz, Austria
\at Abastumani Astrophysical Observatory at Ilia State University, University St. 2, Tbilisi, Georgia
\and E.~Perez
\at SUPA, School of Physics \& Astronomy, University of Glasgow, UK
\and {C.~L.~Selhorst} 
\at  University of Vale do Para\'{i}ba (UNIVAP), Av. Shishima Hifumi, 2911 - Urbanova - CEP 12244-000,  S\~{a}o Jos\'{e} dos Campos, S\~{a}o Paulo, Brazil
\and M.~Barta
\at European ARC, Czech node, Astronomical Institute ASCR, Ondrejov, Czech Republic
\at Astronomical Institute, Academy of Sciences, Fric\u{o}va 298, 251 65 Ond\u{r}ejov, Czech Republic
}

\date{Received: April 24, Accepted: December 10, 2015}

\maketitle

\begin{abstract}
The Atacama Large Millimeter/submillimeter Array (ALMA) is a new powerful tool for observing the Sun at high spatial, temporal, and spectral resolution. 
These capabilities can address a broad range of fundamental scientific questions in solar physics. 
The radiation observed by ALMA originates mostly from the chromosphere -- a complex and dynamic region  between the photosphere and corona, which plays a crucial role in the transport of energy and matter and, ultimately, the heating of the outer layers of the solar atmosphere.
Based on first solar test observations, strategies for regular solar campaigns are currently being developed. 
State-of-the-art numerical simulations of the solar atmosphere and modeling of instrumental effects can help constrain and optimize future observing modes for ALMA. 
Here we present a short technical description of ALMA and an overview of past efforts and future possibilities for solar observations at submillimeter and millimeter wavelengths.  
In addition, selected numerical simulations and observations at other wavelengths demonstrate ALMA's scientific potential for studying the Sun for a large range of science cases.   
\end{abstract}

\section{Introduction}
\begin{figure}
\centering
\resizebox{\textwidth}{!}
{\includegraphics[]{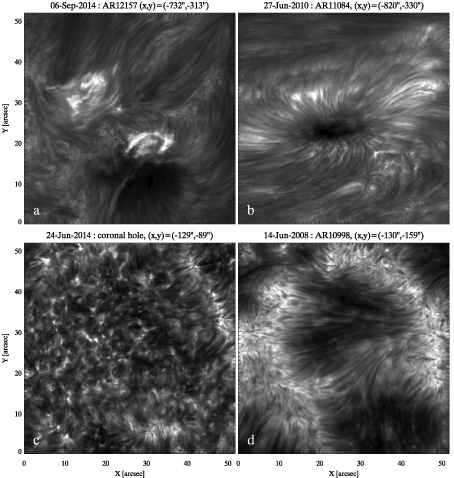}}
\caption{Examples of chromospheric images taken in the core of the Ca\,II\,infrared triplet line at $\lambda = 854.2$\,nm with the CRISP instrument at the Swedish 1-m Solar Telescope (SST). 
(a)~An active region with an on-going M1.1~class flare.  
(b)~The chromosphere above a sunspot close to the limb.  
(c)~A quiet region inside a coronal hole close to disk center.  
(d)~A decaying active region with a pronounced magnetic network cell close to disk center.
}
\label{fig:solarimages}
\end{figure}

Despite decades of intensive research, the chromosphere of our Sun -- an important atmospheric layer between the photosphere and the corona -- is still elusive owing to the complicated nature of the partially ionized chromospheric plasma and its intricate interaction with radiation and the arising technical challenges for observing the emergent radiation (see Fig.~\ref{fig:solarimages}). 
While advances in instrumentation during recent years have led to improved chromospheric diagnostics, sparking    
renewed attention for this important interface layer, the difficulties with interpreting the few currently accessible diagnostics with complicated formation mechanisms have hampered progress. 
In contrast, radiation at millimeter wavelengths has been known to have great diagnostic potential but could not be exploited to full extent due to technical limitations so far, in particular, because 
technically achievable telescope apertures had been small for the relatively long 
wavelengths. 
Consequently, single dish observations did not have sufficient temporal or spatial resolution, and small interferometric arrays could not measure the full range of spatial scales in the chromosphere.

The Atacama Large Millimeter/submillimeter Array (ALMA), which recently commenced operation, 
is now an enormous leap forward in terms of spatial resolution at millimeter and submillimeter wavelengths, offering 
an unprecedented  view of the solar chromosphere and its interface with the upper photosphere and low transition region. 
For extended sources like the Sun, ALMA will achieve a spatial resolution that is close to what is currently possible at visible and infrared wavelengths. 
In addition, ALMA observes with high temporal resolution and can change between different receivers and thus wavelength bands, which are subdivided in a large number of spectral channels, both polarised and unpolarised. 
The major advantage of ALMA compared to other diagnostics is that it serves as a nearly linear thermometer of the solar plasma because the observed brightness temperature is closely related to the gas temperature of the continuum forming layer.  
It thus probes the local plasma conditions that are, at least in principle, much easier to interpret. 
Hence, ALMA will be strongly complementary to other ground-based and space-borne observatories. 
Coordinated observations with other observatories will therefore provide new science opportunities to improve our understanding of the dynamics and energetics of this highly complex interface between the solar photosphere and the corona.

\begin{figure}
\centering
\resizebox{10cm}{!}
{\includegraphics[]{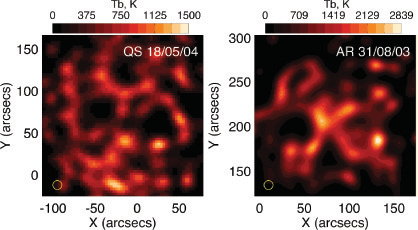}}
\caption{Reconstructed interferometric images of a Quiet Sun region (left) and active region (right) from observations with the Berkeley-Illinois-Maryland Array (BIMA).  
The yellow circles in the lower left corner of each panel represent the restored beam with a diameter 10''. 
Adopted from \citet{2009A&A...497..273L,2014A&A...561A.133L}. 
}
\label{fig:bimaimages}
\end{figure}
\begin{figure}[h!]
\centering
\resizebox{\textwidth}{!}
{\includegraphics[]{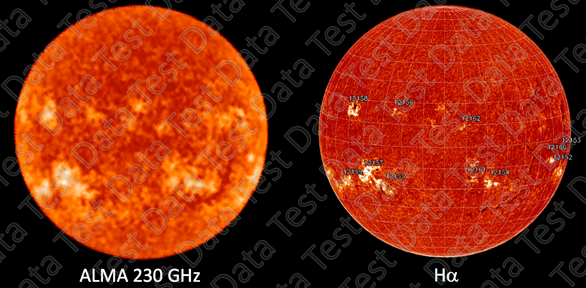}}
\caption{Full-disk maps of the Sun taken on September 7th, 2014.
\textit{Left:}~Brightness temperature map obtained by scanning the solar disk with a single ALMA total power antenna (PM01) at a frequency of 230\,GHz \mbox{($\lambda = 1.3$\,mm)} in a double-circle pattern, which was completed in roughly 10\,min. 
\textit{Right:}~Corresponding H$\alpha$ map. 
Courtesy of  \citet{Phillips2015}. 
Please note that this commissioning data is shown here for illustration purposes only in accordance with ALMA data policy. 
}
\label{fig:almafulldisk}
\end{figure}

ALMA's unprecedented capabilities will allow addressing a wide range of topics and  major questions in contemporary solar physics \citep{bastian02,2011SoPh..268..165K,2012IAUSS...6E.205B,ssalmon_espm15}, including the dynamics, thermal structure and energy transport in the ``quiet'' solar chromosphere, active regions and sunspots, spicules, prominences and filaments, and flares.
An example of the challenging complex structure of the solar chromosphere is shown in Fig.~\ref{fig:solarimages}. 
These high-resolution images were obtained by the CRISP (CRisp Imaging SpectroPolarimeter) instrument at the Swedish 1-m Solar Telescope (SST\footnote{Please note that the abbreviation SST is used for both the Swedish 1-m Solar Telescope and the Solar Submillimeter Telescope.}) in the core of the Ca\,II\,infrared triplet line at a wavelength of 854.2\,nm. 
This spectral line is among the currently most used chromospheric diagnostics. 
The images exhibit structures on spatial scales down to the resolution limit of the telescope ($\sim $0.''2 at 854.2\,nm but down to $\sim $0.''1 for, e.g., Ca~II~K). 
In the millimeter range, however, the spatial resolution of observations has been significantly lower in the pre-ALMA era 
{as can be seen from the otherwise elaborate examples in Fig.~\ref{fig:bimaimages}, which have been obtained with the Berkeley-Illinois-Maryland Array (BIMA) at a wavelength of $\lambda = 3.5$\,mm \citep{2009A&A...497..273L,2014A&A...561A.133L}.} 
The magnetic network is clearly visible in the image of a quiet Sun region (left panel) but no features smaller than the restored beam size of 10'' can be discerned. 
Consequently, the intricate small-scale chromospheric structure inside the network regions, which can be clearly seen in the Ca\,II~image in Fig.~\ref{fig:solarimages}c, remains unresolved in the BIMA maps.  
This is true for all other instruments operating at millimeter wavelengths, so that much of the rich chromospheric fine-structure could not be observed at these long wavelengths so far. 
The situation now changes completely with ALMA, which opens up a new domain of spatial resolution at millimeter wavelengths.

In preparation of regular solar observations, which are expected to start in Cycle~4 in 2016 after the initial solar observing modes have been demonstrated (see Fig.~\ref{fig:almafulldisk}), this review article provides a summary of ALMA's capabilities and outlines potential science cases. 
An introduction to the origins of the radiation at millimeter and submillimeter wavelengths in the solar atmosphere is given in Sect.~\ref{sec:mmsunalma}, followed by a description of the technical parameters and capabilities of ALMA and its diagnostic potential for solar observations in Sect.~\ref{sec:mmtechobservations}. 
Section~\ref{sec:sciencecases} provides a broader and more detailed overview of potential science cases. 
Conclusions and an outlook are given in Sect.~\ref{sec:concoutlook}.  
This article was produced as a joint activity of the Solar Simulations for the Atacama Large Millimeter Observatory Network (SSALMON\footnote{The network is open to everybody who has a professional interest in contributing to potential ALMA solar science, which include or require simulations. More information and registration at http://ssalmon.uio.no.}, http://ssalmon.uio.no), which is an international network aiming at anticipating key solar science with ALMA through simulation studies \citep{ssalmon_espm15}.

\section{The formation of millimeter radiation in the Sun} 
\label{sec:mmradiation}
\label{sec:mmsunalma}

In the absence of strong magnetic fields, e.g. in quiet Sun regions, the opacity at millimeter wavelengths is mainly due to {electron-ion free-free absorption\footnote{{This process is also referred to as ``inverse thermal bremsstrahlung'' in the sense that it is the absorption process reverse to photon emission by bremsstrahlung  \citep[e.g.,][]{2015EPJD...69..132F}.\label{note:inversebrems}}} 
and -- to lesser extent -- H$^-$~free-free absorption 
\citep[cf., e.g.,][and references therein]{2006A&A...456..697W,2011SoPh..273..309S}. 
The resulting thermal} radiation and their diagnostic potential are described in Sect.~\ref{sec:thermalradiation}. 
Radio recombination lines and molecular lines at (sub-)mm wavelengths are discussed in Sects.~\ref{sec:spectroscopy} and \ref{sec:carbonmonoxide}, respectively. 
As detailed in Sect.~\ref{sec:gyroemission}, the gyroresonance process does not contribute significantly to the thermal emission at millimeter wavelengths, while the gyrosynchrotron mechanism can produce non-thermal emission in connection with magnetic fields and high-energy particles. 
Additional emission mechanisms, which are discussed  in connection with a yet unexplained high-frequency (i.e., sub-THz) flare emission component \citep[see~Sect.~\ref{sec:flares} and][]{2010ApJ...709L.127F,2013A&ARv..21...58K} include synchrotron emission from relativistic positrons and electrons, diffusive radiation, and possibly Cherenkov emission but require further investigation.

\subsection{{Thermal continuum radiation at millimeter and submillimeter wavelengths}} 
\label{sec:thermalradiation} 

{The two major opacity sources relevant for thermal continuum radiation at millimeter and submillimeter wavelengths in the solar atmosphere are due to free-free absorption of the following two kinds: 
(1)~Electron-ion free-free absorption$^3$ 
($\mathrm{H}^+  +  \mathrm{e}^-   + \gamma \longrightarrow  \mathrm{H}^+  +  \mathrm{e}^{-\,*}$, i.e. 
interaction of a free electron with radiation in the Coulomb field of a charged ion). 
(2)~H$^-$  free-free absorption 
($\mathrm{H} + \mathrm{e}^- + \gamma  \longrightarrow \mathrm{H}  +  \mathrm{e}^{-\,*}$),  
occurring} in the field of a neutral hydrogen atom which is polarized by the Coulomb field of the passing electron.
At millimeter wavelengths, {free-free absorption} can be treated either by quantum as well as -- by virtue of the correspondence principle -- classical theory \citep[see][]{1961RvMP...33..525O}. 
Due to its importance a lot of work was dedicated to the treatment of this process. 
Different approaches and approximations led to a multitude of prescriptions of the absorption coefficient whose ranges of validity and accuracy are not always obvious to the end-user \citep[for a discussion, see][]{1961ApJ...134.1010O}. 
However, the possibility to treat the problem fairly rigorously leaves these ambiguities on a level which is typically immaterial for application in solar physics. 
We point the reader to the widely used quantum mechanical calculations of the Gaunt factor by \citet{1961ApJS....6..167K}.

Here, we follow \citet{1985ARA&A..23..169D} and write the {free-free} absorption coefficient due to electron-ion free-free absorption$^3$ 
in the \edt{absence of magnetic field\footnote{\edt{See Sect.~\ref{sec:magneticfield} for the magnetized case.}}} semi-classically as 
\begin{equation} 
\chi_\mathrm{ions,ff} \approx \frac{1}{3c} \left(\frac{2}{\pi}\right)^\frac{1}{2} 
\frac{\nu^2_\mathrm{p}}{\nu^2} \frac{4\pi e^4}{\sqrt{m_\mathrm{e}} (k_\mathrm{B}T)^{3/2}}
\sum_i Z^2_i n_i \, \frac{\pi}{\sqrt{3}} g_\mathrm{ions,ff}(T,\nu) 
\label{e:ionsff}
\end{equation} 
where $g_\mathrm{ions,ff}$ is the thermally averaged Gaunt 
factor
\footnote{Note, that Eq.~18 of \citet{1985ARA&A..23..169D} is likely in error,
  the numerical value of 18.2 of Dulk's Eq.~20 cannot be reproduced using
  Eq.~18 as is. It is likely that the Gaunt factor should read as given in
  Eq.~\ref{e:gauntionsff} here which brings Dulk's Gaunt factor much closer to
  results of \citet{1961ApJS....6..167K} and \citet{1970A&A.....9..318O}. The
  numerical evaluation given later is corrected accordingly.}
\begin{equation} 
g_\mathrm{ions,ff}(T,\nu) = \frac{\sqrt{3}}{\pi} 
\ln\left(\frac{(2 k_\mathrm{B}T)^{3/2}}{2 \pi\Gamma
  \nu\sqrt{m_\mathrm{e}}e^2}\right)
\label{e:gauntionsff}
\end{equation} 
valid for temperatures $T< 2\times10^5\,\mathrm{K}$.
$c$ is the speed of light, $\Gamma\approx 1.781$ is Euler's constant, $k_\mathrm{B}$ is the Boltzmann constant, $\nu$ the photon frequency, $T$ the (kinetic) electron temperature, $n_\mathrm{e}$ the electron density, $n_i$ the density of ion $i$ of charge $Z_i$, $e$ the electron charge ($4.8\ldots\times 10^{-10}\,\mathrm{esu}$), $m_\mathrm{e}$ the electron mass, and 
\begin{equation}
\nu_\mathrm{p}=\sqrt{\frac{e^2 n_\mathrm{e}}{\pi  m_\mathrm{e} }} 
\label{e:plasmafreq}
\end{equation}
the plasma frequency. 
Numerically, Eq.~\ref{e:ionsff} evaluates to (in cgs units)
\begin{equation}
\chi_\mathrm{ions,ff} \approx 9.78\times 10^{-3} \frac{n_\mathrm{e}}{\nu^2 T^{3/2}} \sum_i Z^2_i n_i 
\times (17.9 + \ln T^{3/2} - \ln\nu)  
\qquad [\mathrm{cm}^{-1}].
\label{e:ionsffnumerical}
\end{equation}
which leads to the following simplified expression \citep{1965sra..book.....K}
\begin{equation}
\chi = \xi(T,e) \frac{n_e^2}{n \nu^2 T^{3/2}}
\end{equation}
where $n_e=n_i$ is assumed, $n$ is the index of refraction and $\xi(T,e)$ is a slowly varying function of electron temperature and density, which evaluates to $\simeq 0.12$ in the chromosphere and $\simeq 0.2$ in the corona.

\begin{figure}
\centering
\resizebox{9cm}{!}{\includegraphics{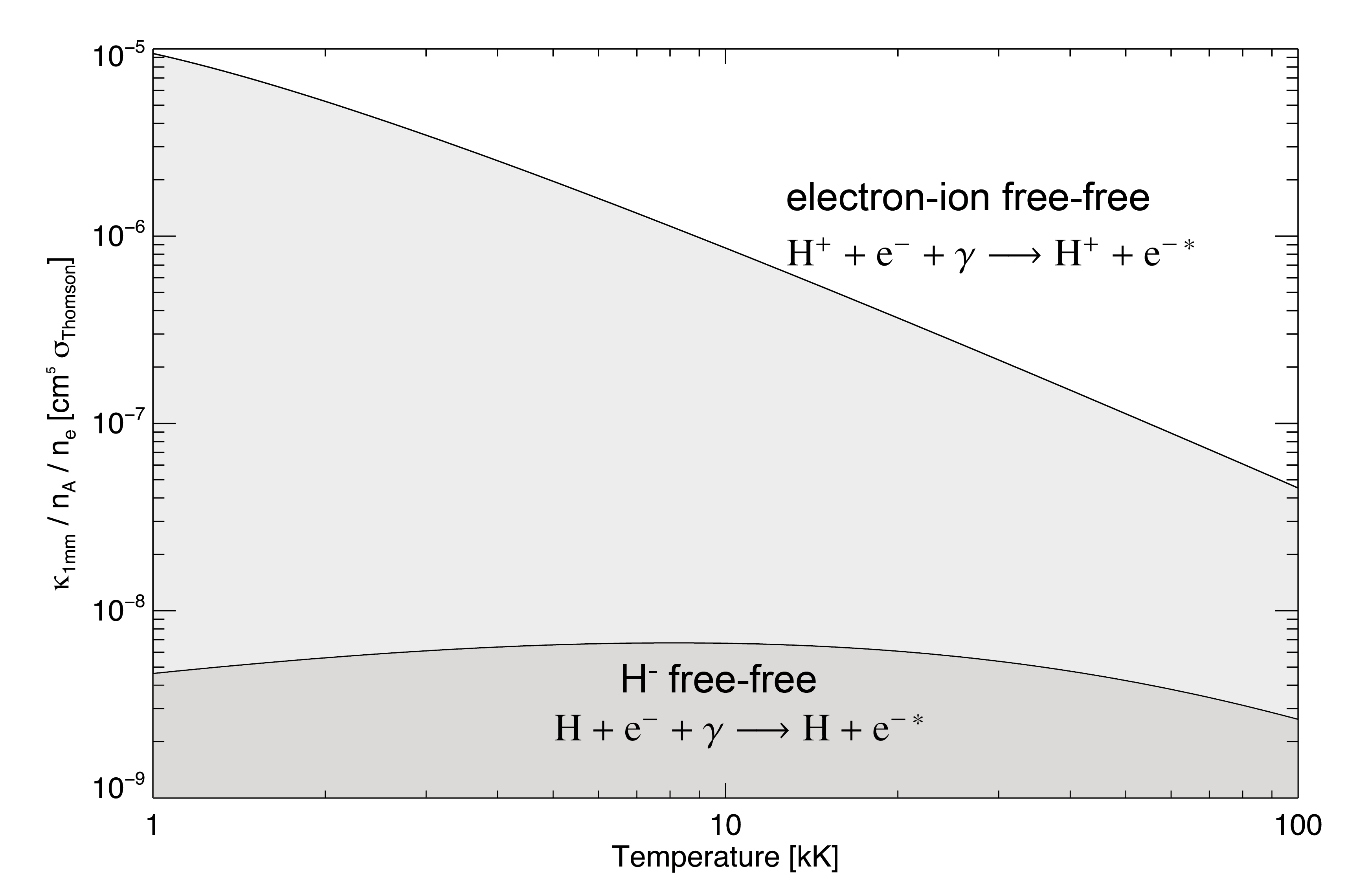}}
\caption{Opacity at 1\,mm wavelength due to {electron-ion free-free absorption} and H$^-$
  free-free absorption as a function of temperature in units of the Thomson
  cross section for electron scattering ($\sigma_\mathrm{Thomson} = 6.65\,\times\,10^{-29}\,\mathrm{m}^2$). 
  The opacity was normalised to the electron density $n_\mathrm{e}$ and to the effective density of absorbing ions or atoms, i.e. to $Z_i^2 n_i$ in the case of {electron-ion} bremsstrahlung and to the 
    number density of neutral H-atoms in the case of H$^-$. 
  Please note that the H$^-$ opacity is affected by the low hydrogen ionisation and thus 
  low election density at low temperatures.}    
\label{fig:mmabsorption}
\end{figure}

Figure~\ref{fig:mmabsorption} shows the resulting opacity for a reference wavelength of 1\,mm together with the free-free opacity due to H$^-$ according to \citet{1974ApJ...187..179S}. 
In a plasma mainly composed of hydrogen, electron-ion free-free absorption$^3$ 
is the main opacity source in the considered temperature range as long as the ionisation fraction of hydrogen is not very small. 
Chromospheric electron densities are in the range $n_\mathrm{e}\approx 10^9\ldots 10^{11} \,\mathrm{cm}^{-3}$, implying densities of ions and neutral hydrogen of similar order. As a consequence, Thomson scattering -- as opposed to the solar corona -- is an unimportant opacity source in the chromosphere.

According to Eq.~\ref{e:plasmafreq} the plasma frequency increases with electron density. 
In a region of rather high electron density of $10^{11}\,\mathrm{cm}^{-3}$ the plasma frequency corresponds to a wavelength of $\approx 10\,\mathrm{cm}$. 
For a wave of a rather long (by ALMA standards) wavelength of 1\,cm this results in an $0.5\,\%$ reduction of the index of refraction from one. 
This implies that to good precision light rays at ALMA wavelengths travel on straight lines. 
Moreover, the effect on the absorption coefficients is minor since it scales inversely proportional to the index of refraction.

Since the opacity is due to particle interactions, the assumption of LTE conditions holds and source function of the radiation continuum is Planckian .
At millimeter and radio wavelengths $\lambda$ (and for chromospheric temperatures $T$), the Rayleigh-Jeans limit of the Planck function can be used in good approximation (since    $hc  / \lambda \ll k_\mathrm{B}\,T$) so that the spectral radiance, i.e. the energy emitted per unit area of emitting surface, per unit time, per unit frequency and per unit solid angle, due to thermal emission is given by
\begin{equation} 
    B_\lambda\,(T)  =  2 \ h \ c^2 \ \left({e^\frac{h\,c}{\lambda\,k_\mathrm{B}\,T} \ - \ 1}\right)^{-1} \ \lambda^{-5} \ \approx \ 2 \ c \ k_\mathrm{B} \ T \ \lambda^{-4}
    \label{eq:blambda}
\end{equation} 
or
\begin{equation} 
\label{eq:bnutherm}
    B_\nu\,(T)  =  \frac{2 \ h}{c^2}  \left({e^\frac{h\nu}{k_\mathrm{B}T} - 1}\right)^{-1} \  \nu^3 \ 
    \approx 
    \frac{2 \ k_\mathrm{B}}{c^2} \  T \ \nu^2 .
\end{equation} 
The emergent specific intensity  $I_\nu$ is then derived from integration of the source function (here the Planck function $B_\lambda\,(T(\tau))$ or  $B_\nu\,(T(\tau))$)  over the optical depth $\tau$ along the line of sight:  
\begin{equation} 
  \label{eq:inuintegral}
  I_\nu  =  \int_0^{\tau_\mathrm{max}\,(\nu)}   B_\nu\,(T\,(\tilde{\tau})) \ e^{-\tilde{\tau}} \ d\tilde{\tau} 
  , 
\end{equation} 
where $\tau_\mathrm{max}\,(\nu)$ is the maximum of the optical depth range at the considered frequency (or wavelength) where significant contributions to the source function occur. 
In the optically thick case $\tau_\mathrm{max} = \infty$.

In the radio and millimeter range, it is 
\edt{more common to use the brightness temperature $T_\mathrm{b}$ instead of 
the continuum intensity $I_\nu$ or $I_\lambda$, which are related in the following way (as derived from Eqs.~\ref{eq:blambda}-\ref{eq:inuintegral}):}  
\begin{equation} 
T_\mathrm{b} 
=    
\frac{c^2}{2 \ k_\mathrm{B}} \ \nu^{-2} \ I_\nu 
= 
\frac{\lambda^4}{2 \ c \ k_\mathrm{B}} \ I_\lambda .
\label{eq:tbnu}
\end{equation} 
A common unit for the flux density at these wavelengths is Jansky with \linebreak$1~\mathrm{Jy} = 10^{-26}\ \mathrm{W m}^{-2}\ \mathrm{Hz}^{-1} = 10^{-23}\ \mathrm{erg\ s}^{-1} \mathrm{cm}^{-2} \mathrm{Hz}^{-1}$.
In solar physics, however, it is more common to use solar flux units (1\,sfu$=10^4$\,Jy). 
%
\edt{Typical values for the Quiet Sun flux density integrated over the solar disk are on the order of 
$\sim 1.2\,10^6$\,sfu at 0.3\,mm (1\,THz),  
$\sim 1.2\,10^5$\,sfu at 1.0\,mm (300\,GHz), 
$\sim 1.5\,10^4$\,sfu at 3.0\,mm (100\,GHz), and 
$\sim 2\,10^3$\,sfu at 9.0\,mm (33\,GHz), respectively, 
based on observed brightness temperatures compiled by \citet{2004A&A...419..747L}.}
The flux from active regions (the ``S-component'') can be comparable, whereas flares are much brighter. 
The extreme case shown in Fig.~\ref{fig:flare_subTHz} even reaches a peak value of \edt{$\sim 5~10^4$\,sfu at 3\,mm (100\,GHz)}.

Numerical modelling of the solar atmosphere and the emergent radiation has proved useful for predicting which properties of the radiation can be exploited as plasma diagnostic, how to set up the instrument, and how to interpret the resulting data. 
For instance, radiative transfer calculations have been carried out based on model atmospheres for different regions, including the quiet Sun, coronal holes, filaments and active regions \citep{2007SoPh..245..167B,2009A&A...493..613B,2009CEAB...33...79R}.
The resulting synthetic brightness temperatures for a wavelength of 8\,mm are in good agreement with corresponding observational results. 
A more extensive study for a broader wavelength range from 0.3\,mm to 10.0\,mm is currently in progress 
\citep{Brajsa2015}.

\begin{figure}
\centering
\resizebox{9cm}{!}{\includegraphics[]{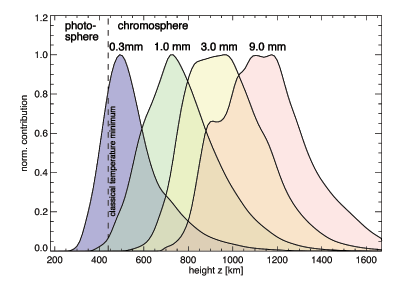}}
\caption{
Contribution functions of the continuum intensity at four selected wavelengths between 0.3\,mm and 9.0\,mm at solar disk-center based on a numerical model snapshot of the solar atmosphere under quiet Sun conditions. 
For this model (see model~E described in Sect.~\ref{sec:quietsunmodel}), time-dependent hydrogen ionisation and with it non-equilibrium electron densities have been taken into account. 
The contribution functions have been averaged horizontally and normalised to their maximum.  
The dashed vertical line labelled ``classical temperature minimum'' marks the boundary between the photosphere and the model chromosphere. 
Data adapted from \citet{2007A&A...471..977W}. 
}
\label{fig:contf}
\end{figure}

The linear dependence on the gas temperature means that ALMA effectively acts as a linear thermometer of the solar chromosphere and provides measurements of localised plasma conditions, which are in principle much easier to interpret than other chromospheric diagnostics in the ultraviolet, visible or infrared
spectral range.
However, the opacity depends on the number of free electrons and protons and thus on the ionisation degree of hydrogen.
The millimeter continuum radiation is therefore affected by deviations from the ionisation equilibrium of hydrogen due to finite recombination timescales  in the chromosphere.
Departures from Maxwellian electron velocity distributions may also have detectable effects
\citep[see, e.g.,][]{2014ApJ...781...77F}, including not only flare-related kappa distributions \citep[e.g.,][]{2015ApJ...799..129O}, but also equilibrium distributions \citep[e.g.,][]{2014ApJ...796..142B}.

Furthermore, the opacity increases with wavelength so that the atmosphere becomes optically thick at increasing height.
In other words, the effective formation height of the millimeter continuum radiation increases with height from the temperature minimum at the shortest wavelengths to the high chromosphere (and transition region) at the longest wavelengths accessible with ALMA \citep{2007A&A...471..977W}.
This effect is clearly visible in the results of the brightness temperature synthesis based on  3D numerical simulations of quiet Sun regions, which are described in more detail in Fig.~\ref{sec:quietsunmodel}.

The contribution functions of the continuum intensity at ALMA wavelengths is shown in Fig.~\ref{fig:contf}
for a selected model, for which important non-equilibrium effects have been taken into account (see model~E described in Sect.~\ref{sec:quietsunmodel} for more details). 
These contribution functions \citep[published originally by][]{2007A&A...471..977W} describe how much of the emergent radiation is contributed over height (here for a view directly from the top, thus corresponding to solar disk centre). 
On average, the contribution maximum occurs at a height of 490\,km for $\lambda = 0.3$\,mm, 730\,km for $\lambda = 1.0$\,mm, 960\,km for $\lambda = 3.0$\,mm, and  1170\,km for $\lambda = 9.0$\,mm 
{in their quiet Sun model, respectively.  
For comparison, \citet{2015A&A...575A..15L} find average formation heights of 
700\,km for $\lambda = 0.4$\,mm, 1000\,km for $\lambda = 1$\,mm, 1550\,km for $\lambda = 3$\,mm, 1700\,km for $\lambda = 4.5$\,mm and 2000\,km for $\lambda = 9$\,mm, respectively, based on a model of an enhanced network region (see Sect.~\ref{sec:quietsunmodel}).
} 
The height of the maximum contribution ($z_\mathrm{m}$) varies 
{with spatial position and view angle} 
due to the substantial thermal inhomogeneities in the chromosphere. 
{In the quiet Sun model by \citet{2007A&A...471..977W}, the standard deviation $z_\mathrm{m} (x,y)$ ranges 
}
from $\sim 90$\,km for $\lambda = 0.3$\,mm up to $\sim 160$\,km for $\lambda = 9$\,mm, which is $\Delta z_\mathrm{m} (\lambda) / z_\mathrm{m,avg} (\lambda)  \sim 15$\,\% of the average height of the contribution maximum for all wavelengths. 
Alternatively, the FWHM of the average contribution functions are 220\,km, 360\,km, 440\,km, and 480\,km, respectively.

It is important to note that the formation height can thus vary substantially in space {but} also in time \citep{2004A&A...419..747L,2007A&A...471..977W,2015A&A...575A..15L} on time scales of a few 10\,s when the variation is due to propagating shock waves.                 
On the other hand, accounting for the time-dependent non-equilibrium behaviour of hydrogen ionisation leads to less extended formation height ranges compared to the local thermodynamic equilibrium (LTE) conditions (see Sect.~\ref{sec:quietsunmodel} for more details on small-scale fluctuations).
In summary, the continuum intensity from quiet Sun regions at ALMA wavelengths originates from relatively narrow layers, whose effective heights increase with wavelength, although with a significant overlap in the contribution functions. 
This property allows ALMA to observe different chromospheric layers by changing the wavelength.

\edt{In the general magnetized case,} the brightness temperature spectrum noticeably responds to the magnetic field along the line of sight in the continuum forming layer due to the dependence of the free-free opacity on the local magnetic field strength \edt{(see Sect.~\ref{sec:magneticfield} for details).} 
This phenomenon offers a straightforward way to measure magnetic fields based on the observed degree of polarisation in the spatially resolved spectrum.

\subsection{Recombination lines}
\label{sec:spectroscopy} 
%
A transition between Rydberg states of an atomic ion yields a line at radio frequencies
\begin{equation} 
\nu \approx 2\ c\ R_M\ Z^2\ \frac{\Delta n}{n^3}
\end{equation}
for an ion of effective charge $Z$ ($Z=1$ for H\,I, 2 for He\,II, etc.) and neutral atomic mass $M$ in atomic mass units (amu), where $c$ is the speed of light, $n$ is the principal quantum number of the lower state of transition $(n+\Delta n) \to n$, and $R_M$ is related to the Rydberg constant $R_{\infty} = 109737.315685$\,cm$^{-1}$ by
\begin{equation} 
R_M = \frac{R_{\infty}}{1 + (1822.88848\ M - Z)^{-1}}  \quad.
\end{equation} 
Rydberg states of atoms are populated by recombination processes in plasmas and such recombination lines have been widely observed in photoionized nebulae and interstellar matter. 
\citet{dupree1968} proposed that such lines might be seen in the solar corona where dielectronic recombination naturally overpopulates excited states of abundant ions like Fe\,XV and O\,VI. 
Searches for coronal lines at radio frequencies have remained fruitless \citep{berger-simon1972}, the main reason being that Rydberg transitions with $n\geq 100$ suffer severe collisional line broadening that merges them with the continuum even at coronal densities \citep{1974SvA....17..634D,greve1977,hoang-binh1982,hoang-binh1987}.

Meanwhile, limb-brightened, Zeeman-split, mid-infrared emission lines at wavelengths near 12\,$\mu$m  \citep{murcray1981,brault-noyes1983} were identified as Rydberg transitions $n=7-6$ of Mg\,I and Al\,I arising in the low chromosphere \citep{chang-noyes1983} and subsequently modelled in some detail \citep{carlsson1992,1995A&A...293..225B}.  
\citet{boreiko-clark1986} detected  emission lines of H\,I ($n=13,14,15$) in the far-infrared, \mbox{$\tilde\nu=\nu/c$ $\sim 25 - 90$ cm$^{-1}$}, with a balloon-borne Fourier-transform spectrometer (FTS). 
Finally, sub-mm-wave measurements with a FTS at the James Clerk Maxwell Telescope (JCMT) demonstrated that recombination lines of H\,I are detectable near the solar limb at frequencies accessible to ALMA: $n=20-19$ at 888.047\,GHz \citep{clark2000a} and $n=22-21$ at 662.404\,GHz \citep{clark2000b}. 
The 21$\alpha$ line appeared at a level of approximately 11\,\%\ of the continuum with a width ten times broader than thermal Doppler broadening expected at the temperature of the lower chromosphere. These pioneering sub-millimeter  observations had limited angular resolution ($19''$ at 662\,GHz). 
Observations with ALMA will provide much better sensitivity, spectral and spatial resolution, and fidelity of line profiles.

The brightnesses of spectral lines and continuum will provide direct probes of the temperature structure in the chromosphere and transition zone while the line profiles can be used to trace microturbulence, flows, and density. 
Higher in the chromosphere, lines of He\,I might become detectable. 
With increasing angular distance from the limb, it might even be possible to measure the long-sought coronal lines of highly ionized species.  
In this connection it is worth noting that coronal O\,VI $87\alpha$ at 353.5874\,GHz could be observed in the same tuning as chromospheric H\,I $26\alpha$ at 353.6228\,GHz, and many other such combinations are possible.

Observations of Rydberg transitions will provide exquisite tests of models of the upper photosphere, chromosphere, and corona. 
Detailed comparisons with models will also demand improved understanding of interesting physics, especially regarding collisional line-broadening, which remains controversial for low-density plasmas \citep[e.g.,][]{peach2014,hey2014}.

There are other types of atomic transitions that occur at sub-mm wavelengths. 
Consider, for example, the forbidden fine-structure transition in [C\,III] \mbox{1s$^2$2s2p $^3$P$^o_1$ - $^3$P$^o_0$} at 710.2\,GHz frequency. 
The upper state of this transition is also the upper state of the C\,III\,190.9\,nm line which appears in the chromospheric ultraviolet spectrum of the Sun and solar-type stars. 
The fractional abundance of C$^{+2}$ maximizes at a temperature $T=10^{4.85}$\,K. 
In the solar atmospheric model of \citet{1993ApJ...406..319F}, this occurs at a height $h\,\approx\,2210$\,km above the visible photosphere, where the electron density is approximately $9\times 10^9$\,cm$^{-3}$. 
The excitation of the five lowest levels of C$^{+2}$ is calculated for these conditions, with use of the collision strengths and transition probabilities of \citet{fernandez2014}. 
Such a simple model that reproduces the observed intensity of the C\,III\,97.7\,nm line \citep{Warren2001} implies that the 710\,GHz line should appear weakly in absorption against the sub-mm-wave continuum that arises deeper in the atmosphere, with an estimated flux of $-2.3$\,mJy when averaged over a 5\,arcsec diameter on the solar disk. 
Finally, the positronium analog of the 21-cm line lies at 203\,GHz and may appear under flare conditions \citep[e.g.,][]{2009ApJ...707..457E} \edt{(see the end of Sect.~\ref{sec:majorflares}).}

\begin{figure}
\centering
\resizebox{\textwidth}{!}
{\includegraphics[]{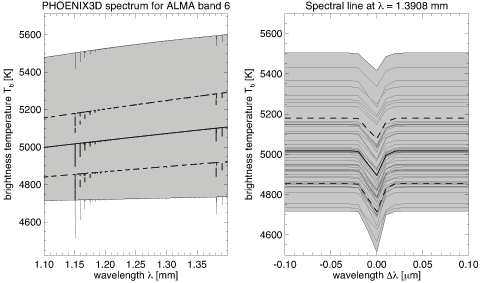}}
\caption{Synthetic spectrum for the wavelength range covered by ALMA band~6 (left) and a close-up for a spectral line at a wavelength of $\lambda = 1.3908$\,mm (right). 
The spectrum has been calculated in full  3D with the radiative transfer code PHOENIX/3D \citep{2010A&A...509A..36H}
in LTE but includes scattering. 
For this preliminary result, 33$\,\times\,$33 columns from the  3D magnetohydrodynamic model by \citep{2012Natur.486..505W} has been used as input.  
The grey area marks the range in brightness temperature covered by all considered positions in the model, whereas the thick black line represents the horizontally averaged spectrum. 
The thick dashed lines mark the average plus / minus the standard deviation. 
The right panel contains a few exemplary line profiles, too. 
}
\label{fig:phoenixspec}
\end{figure}

\subsection{Molecular lines of carbon monoxide}
\label{sec:carbonmonoxide}
As {further detailed} in Sect.~\ref{sec:quietsunthermal}, carbon monoxide molecules have been found to exist in cool pockets in the upper photosphere and low chromosphere, both of which are dynamically changing due to the interaction of propagating shock waves and magnetic fields.  
CO might be directly detectable with ALMA because rotational transitions of the CO molecule lie throughout the millimeter and sub-millimeter parts of the spectrum. 
At 3500\,K, the CO $J=6\to 5$ and $7\to 6$ lines at 691\,GHz and 806\,GHz, respectively, might have brightness temperatures of the order of 20\,K, to be consistent with the observed equivalent widths and line widths of vibration-rotation lines analyzed by \citet{2006ApJS..165..618A}. 
The line widths are expected to be large due to pressure broadening. 
Thus calibration and baseline subtraction will be a challenge.
However, we know today that the chromosphere exhibits substantial temperature fluctuations on short time-scales and on small spatial scales, resulting in corresponding fluctuations of the CO number densities as well as the continuum optical depths. 
It is difficult to estimate to which extent CO rotational lines will be detectable with ALMA in the presence of such fluctuations. 
Numerical simulations involving CO \citep[e.g.,][]{asensio03,2005A&A...438.1043W,2006ASPC..354..301W,2007A&A...462L..31W} will help answer this question. 
Such detailed predictions of CO line profiles would be key to prepare and interpret ALMA observations. 
Observations of molecular lines of CO are interesting as regarding the thermal structure of the solar chromosphere,  but also for the determination of elemental abundances and isotopic ratios \citep[see, e.g.,][and references therein]{2013ApJ...765...46A}.

A first attempt at predicting detailed ALMA spectra is presented in Fig.~\ref{fig:phoenixspec}. 
The  3D magnetohydrodynamic model by \citet[][see Sect.~\ref{sec:quietsunmodel}, incl. Table~\ref{tab:quietsunmodels}
and Fig.~\ref{fig:Tbquietsun}]{2012Natur.486..505W} has been used as input for the radiative transfer code PHOENIX/3D \citep{2010A&A...509A..36H,2013A&A...550A.104B}. 
The resulting radiative flux  for each of the selected 33$\,\times\,$33 model columns for 30,000 wavelength points was converted to brightness temperatures for Fig.~\ref{fig:phoenixspec}. 
Already this early result shows that the continuum level is expected to vary significantly whereas the included spectral lines are very narrow. 
The line depths are on average up to $\sim120$\,K and in extreme cases up to 250\,K. 
The calculations have so far been carried out under the assumption of LTE for the occupation numbers and the molecular CO number densities, whereas background continuum scattering is included.

The LTE assumption for the occupation numbers is more likely to be valid, owing to the fast collision rates (with pervasive atomic hydrogen) that rapidly thermalize the pure rotational levels, especially given the long radiative decay times.  
The instantaneous chemical equilibrium assumption for the CO number densities is more questionable, owing to the increasing difficulty of making the molecules at progressively lower densities at higher altitudes above the photosphere; and the likely frequent interruptions of the formation process due to passing hot waves and shocks.  
Future detailed non-LTE calculations \citep{2014A&A...566A..89H} may thus change the details of the spectral line profiles. 
Nevertheless, the calculations presented here demonstrate that a powerful instrument like ALMA with its high spatial and spectral resolution is indeed required to make full use of the diagnostically interesting lines at mm wavelengths.

It also is possible that the CO pure rotational transitions could go into emission, if they arise in warmer material than is present deeper where the sub-millimeter continuum becomes optically thick.  
It is unlikely, however, that the sub-mm CO lines will appear in emission off the limb, like their thermal-IR cousins, because the sub-millimeter continuum forms much higher than the 5~$\mu$m counterpart, and the slant-angle effect likely will cause the sub-mm radiation to overwhelm the potential CO rotational transitions.  
In other words, the effective altitude where the sub-millimeter continuum would become transparent at the limb would be higher than the low-chromosphere region where the molecules apparently can survive.  
Nevertheless, such phenomena should be looked for, because CO has revealed many \edt{surprises} in the past.

\subsection{Gyroresonance and gyrosynchrotron radiation} 
\label{sec:gyroemission} 

Gyroresonance emission \citep{1962ApJ...136..975K,1962AZh....39....5Z,1962SvA.....6....3Z}, which is responsible for the microwave emission \edt{above} sunspots, is produced by 
thermal electrons in the presence of magnetic fields but is rather unlikely to be observed
in the millimeter range. 
Corresponding  intensity contributions would be measurable only at the lowest gyroharmonics, $s=2-5$, which are below 30\,GHz and thus outside ALMA's frequency range even for the strongest magnetic fields occurring in the solar atmosphere. 
For instance, a magnetic field strength of 
{3570}\,G would be required for a significant ({third} harmonic) emission contribution at a wavelength of 10\,mm.

In contrast, non-thermal gyrosynchrotron (GS) emission from energetic electrons should be detectable because the dominant gyroharmonics rapidly increase with electron energy. 
For example, for a relativistic electron with the Lorenz factor $\gamma$, the dominant emission frequency is by a factor of $\gamma^2$ larger than the gyrofrequency, which can fall everywhere in the ALMA spectral domain. 
Such energetic particles are present in the solar atmosphere primarily during flares. 
Based on experience from the known parameter space (resolution and sensitivity) we therefore expect ALMA to detect non-thermal gyrosynchrotron emission due to high-energy electrons during times of flares or other major activity. 
However, with the new parameter space opened up by ALMA, we may also find non-thermal signatures on small spatial scales and short time scales in phenomena associated with intense magnetic flux concentrations in network or plage regions or due to reconnection events in the quiet Sun,  micro- or ``nanoflares'' (see Sect.~\ref{sec:flares}).

In general, the gyrosynchrotron absorption and emission coefficients are complicated functions of the energetic electron distribution, the magnetic field and the physical parameters of the ambient plasma \citep{1969ApJ...158..753R}.
The standard spectrum of the gyrosynchrotron radiation has a bell shape, on which some harmonic structure can be seen at the low-frequency range \citep{1968Ap&SS...2..171M,1969ApJ...158..753R}. 
The spectral peak frequency {$\nu_\mathrm{peak}$} can be specified by either the condition $\tau\sim1$, demarcating the optically thick and thin regimes, or by the effect of high plasma density (Razin effect, 
{$\nu_\mathrm{peak} \approx 20\ n_e/B$).}
Although neither low- nor high-frequency slope is a perfect power-law, the shape of this spectrum is often characterized by the low- and high-frequency power-law spectral indices, $\alpha_{lf}$ and $\alpha_{hf}$. 
In the optically thick regime $\alpha_{lf}=2-3$ in a uniform source, while in the conditions of the dominating Razin effect, the spectrum can be much steeper, $\alpha_{lf}> 3$. 
The source non-uniformity can make the observed spectrum much shallower, however. 
The high-frequency spectral index $\alpha_{hf}$ is primarily determined by the fast electron energy spectrum and their pitch-angle anisotropy
{\citep{1968Ap&SS...2..171M,1969ApJ...158..753R,2003ApJ...587..823F,2011ApJ...742...87K}.}
There is no simple and accurate analytical formula reliably linking the slope of the particle distribution in the energy (and/or angular) domain and the radio spectral index $\alpha_{lf}$; the available analytical
approximations are too crude to be quantitatively applied. 
The exact numerical expressions \citep{1968Ap&SS...2..171M,1969ApJ...158..753R} are very cumbersome and computationally slow to be routinely used. 
Fortunately, fast {gyrosynchrotron} codes have recently been developed \citep{2010ApJ...721.1127F}, which, for instance, as integral part of the 3D modeling tool GX Simulator \citep{2015ApJ...799..236N} facilitate the quick computing of the gyrosynchrotron emission from realistic 3D volumes (see Sect.~\ref{sec:armodelling} for more details).

\subsection{{Polarisation and magnetic field measurements}} 
\label{sec:magneticfield}

{The chromospheric magnetic field is fundamentally important but poorly known because the currently available diagnostics for measuring the chromospheric magnetic field have shortcomings. 
There are a few magnetic-sensitive lines, e.g. Ca\,II\,854.2\,nm and  He\,I\,1083\,nm, that can be used for chromospheric magnetometry but these lines are commonly characterized by a weak polarisation signal, wide formation height range and require a careful treatment of non-LTE\footnote{Non-LTE, also abbreviated NLTE, refers to  departures from local thermodynamic equilibrium (LTE).} effects \citep[e.g.,][]{2010MmSAI..81..716D}. 
Most recent work on the weak chromospheric magnetic field has focused on the analysis of the Ca\,II\,IR triplet lines (849.8\,nm, 854.2\,nm and 866.2\,nm), e.g., employing the Hanle effect \citep[e.g.,][]{2014arXiv1412.5386C}, and on the He\,I\,1083\,nm line \citep[e.g.,][]{2003Natur.425..692S,2012ApJ...759...16M,2013ApJ...768..111S}. 
Nevertheless, measuring the magnetic field is fundamental for advancing our understanding of the chromospheric plasma because it is strongly coupled to the magnetic structures. 
This intricate connection is vividly seen in detailed 3D MHD simulations \citep[e.g.,][]{2011A&A...531A.154G,2006ASPC..354..345S} and has been widely confirmed by observations \citep[e.g. in H$\alpha$ images,][]{2007ApJ...655..624D}. 
The role of the chromospheric magnetic field in dynamics,  wave propagation, and (local) heating
can hardly be overestimated. 
}

{A method for} measuring chromospheric magnetic fields by evaluating thermal brems\-strahlung was developed by \citet{1980SoPh...67...29B}  and by \citet{2000A&AS..144..169G} and tested on the RATAN-600 and Nobeyama Radio Heliograph (NoRH) data at short cm wavelengths. 
The proposed technique relies on the fact that, in the general case, the longitudinal component $B_\mathrm{l}$ of the magnetic field can be derived from the observed degree of circular polarisation $\mathcal{P}$ and the brightness temperature spectrum $T_\mathrm{b} (\nu)$.

{The propagating of electromagnetic radiation in the magnetized solar medium is well-described by the magneto-ionic theory, the observable modes being the ordinary mode (o-mode) and the extraordinary mode (x-mode).} 
In a magnetized plasma the absorption coefficient and thus the opacity $\kappa$ for free-free emission is higher in the extraordinary mode than in the ordinary ($\kappa_\mathrm{x} > \kappa_\mathrm{o}$). 
Due to the resulting dependence of the opacity on the magnetic field the natural modes become optically thick in different atmospheric layers so that $z\,(\tau_\mathrm{x}\,(\nu)=1) >   z\,(\tau_\mathrm{o}\,(\nu)=1)$. 
Thus, at a given frequency we observe the x- and o-mode emission from slightly different heights with deeper (cooler) layers seen in the o-mode and higher (hotter) layers seen in the x-mode in the atmosphere with $T$ increasing outwards.

Depending on the sign of the longitudinal magnetic field component, a mode (i.e., the x-mode) can then correspond to either left circular polarisation (LCP) or right circular polarisation~(RCP). 
In the case of a magnetized atmosphere with a temperature gradient, the temperature difference between these layers produces then an observable net circular {polarisation.} 
The stronger the magnetic field is, the larger is the difference of the formation heights of the x- and o-modes, resulting in a proportionally stronger net polarisation. 
See Fig.~\ref{fig:simactiveregion} and the corresponding text in Sect.~\ref{sec:armodelling} for simulated examples, illustrating how the degree of net circular polarisation is determined from brightness temperature maps for left and right circular polarisation.

\citet{1980SoPh...67...29B} and \citet{2000A&AS..144..169G} demonstrated that the slope of the (observed) free-free emission brightness temperature  spectrum ($T_\mathrm{b}(\nu)$)
\begin{equation}
\zeta \ = \ \frac{d \log(T_\mathrm{b}\,(\nu))}{d \log(\nu)} 
\end{equation}
carries information about the gas temperature gradient in the radiation forming layers in the atmosphere and controls the degree of the circular polarisation $\mathcal{P}$, such that 
{
\begin{equation} 
\mathcal{P} \ = \ \zeta \ \frac{\nu_B}{\nu} \ \cos{\theta}\quad,
\label{eq:polarization} 
\end{equation} 
}
where $\nu$ is the observing frequency, {$\nu_B$} is the (electron) gyrofrequency 
{
\begin{equation} 
\nu_B = \frac{1}{2\,\pi} \frac{e B}{m_e\,c} =  2.8 \times 10^6 \mathrm{Hz} \times \vec{B}  \quad,
\label{eq:gyrofreq}
\end{equation} 
}
and $\theta$ is the angle between the line of sight and the magnetic field vector. 
In the simple case of an optically thin homogeneous atmospheric slab with a constant gas temperature the quadratic dependence of the observed brightness temperature $T_\mathrm{b}$ on the observing frequency~$\nu$ (cf.~Eq.~\ref{eq:inuintegral})) results in a local slope $\zeta = 2$. 
The circular polarisation, $\mathcal{P}$, is then simply \citep[cf.][]{1965sra..book.....K} 
{
\begin{equation}
\mathcal{P} \ \simeq \ 2 \ \frac{\nu_B}{\nu} \ \cos{\theta}\quad.
\end{equation}
}
This expression is valid for quasi-longitudinal propagation, which prevails unless $\theta$ is close to 90~degrees. 
Putting numerical values we get:
\begin{equation}
\mathcal{P}\mbox{~(\%)~}\simeq 1.85 \times 10^{-3} \lambda~\mbox{(mm)}~B\mbox{~(G)~} \cos\theta
\end{equation}
which gives $\mathcal{P} \simeq2 $\% at 1\,mm for a field of 1\,kG as it occurs above sunspots \citep[see, e.g.,][and references therein]{2003A&ARv..11..153S}. 
In the optically thin case the longitudinal component of the magnetic field $B_l$ can thus be found from observations at a single wavelength, but spectral observations are still desirable to confirm the free-free nature of the emission and its thickness. 
Detailed modeling for the quiet Sun (see Sect.~\ref{sec:qsmag}) predicts for the optically thick chromosphere  a circular polarisation of 0.05\% at 1\,mm and thus gives an estimate of the required accuracy for polarisation measurements, 
{although the polarization can reach a few \% above sunspots (see Sect.~\ref{sec:armodelling}).}

The expression in Eq.~(\ref{eq:polarization}) offers an elegant means of measuring the longitudinal component of the magnetic field using the final expression {(together with Eq.~(\ref{eq:gyrofreq}))}
{
\begin{equation} 
B_l = |\vec{B}| \cos \theta  = 
\frac{\mathcal{P}\ \nu}{\zeta} \   (2.8\,\times\,10^6)^{-1}\ \mathrm{Hz}^{-1} \ (G) \quad. 
\end{equation}
}
The restored magnetic field refers to the height where the emission at a given frequency is formed
{so that the magnetic field at different heights can be measured by changing the frequency
(see Sect.~\ref{sec:almamagneticfield} for details for the case of ALMA).}

It should be noted that the approach outlined above implicitly assumes a constant magnetic field within the layer between {$\tau_\mathrm{x}=1$} and $\tau_\mathrm{o}=1$ and thus neglects the complex magnetic field structure on small spatial scales as it is known from contemporary high-resolution observations, although, depending on the local magnetic field strength,  the height difference between  {$\tau_\mathrm{x}=1$} and $\tau_\mathrm{o}=1$ can be small enough to allow for such an assumption.  
{Please refer to Sect.~\ref{sec:almamagneticfield} for a description of ALMA's potential for measuring the chromospheric magnetic field based on the method introduced here.}

\section{Observations of the Sun at millimeter and submillimeter wavelengths} 
\label{sec:mmtechobservations}

\subsection{{Solar radio observations in the pre-ALMA era}} 
\label{sec:prealma}

{In spite of the important progress during the last forty years, the mm and sub-mm range has remained by large unexplored, except for single dish observations  \citep[e.g.][and references therein]{2011SoPh..273..339T} and some observations with BIMA \citep{2006A&A...456..713L,2006A&A...456..697W}. 
An important step forward after the first interferometric observations of \citet{1959AnAp...22....1K} was the use of non-solar aperture synthesis instruments for solar observations. 
The first high resolution radio imaging observations of the Sun were obtained with the Westerbork Synthesis Radio Telescope (WSRT) about forty years ago, first of active regions at 6 and 20\,cm \citep{1975Natur.257..465K,1977A&A....61...79C} and bursts \citep{1978ApJ...222..342A} and subsequently of the quiet Sun \citep{1979ApJ...234.1122K}. 
Detailed modeling of the sunspot associated emission \citep{1980A&A....82...30A} gave important information both on the emission mechanism and the structure of the atmosphere, in particular the magnetic field structure in the transition region and low corona, while modeling of the microwave emission from flaring loops \citep{1984A&A...139..507A} made possible the association of their observable characteristics with the properties of the trapped energetic electrons. 
Quiet Sun observations showed a marked similarity to the chromospheric network as it was known from observations in other wavelength ranges. 
Ever since, solar observations with the Very Large Array (VLA) have provided increased spatial and temporal resolution and extended the frequency range up to 15\,GHz 
\citep[2\,cm, see, e.g.,][]{1998ARA&A..36..131B,2007SSRv..133...73L,2011SoPh..273..309S,2004ASSL..314.....G}.  
In parallel with the occasional very high spatial resolution observations with the WSRT and the VLA, solar two-dimensional images with moderate resolution were taken daily with 
the Nobeyama Radioheliograph \citep{1994IEEEP..82..705N} at 17 and 34\,GHz (1.76 and 0.88\,cm), 
the Metsahovi Radio Observatory \citep{1992SoPh..137...67V} at 37\,GHz (8\,mm), 
the Siberian Solar Radio Telescope \citep[SSRT,][]{2003SoPh..216..239G} at 5.74\,GHz (5.2\,cm), 
 the \mbox{Nan\c cay} Radioheliograph \citep{1997LNP...483..192K} at frequencies between 150 to 450\,MHz (67 to 200\,cm), 
 together with the one-dimensional, high frequency resolution data of RATAN-600 \citep{2011AstBu..66..190B,2011AstBu..66..205B} from 3 to 18.2\,GHz (1.65 to 10\,cm). 
The Owens Valley Solar Observatory (OVSO) has also provided valuable data.
}

\subsection{The Atacama Large Millimeter/submillimeter Array}
\label{sec:almatech}

ALMA is an international partnership between Europe, North America, and East Asia in cooperation with the Republic of Chile \edt{and minor partners} to build and operate a millimetre/submillimeter interferometer on the Chajnantor plateau in the Chilean Andes at an altitude of 5000\,m.
The 66~antennas of this aperture synthesis telescope are arranged in \edt{two major groups:}  The \textit{12-m Array} and the \textit{Atacama Compact Array} (ACA). 
The 12-m array consists of 50~movable antennas with 12\,m~diameter. 
The ACA, also known as ``Morita Array'', combines 12 antennas with 7\,m diameter for interferometry \edt{(the ACA 7-m Array)} and 4 antennas with 12\,m diameter for single dish observations surrounding them \edt{(also referred to as the ACA Total Power (TP) Array).}  
The ACA antennas are installed in a very compact configuration, while the more extended 12-m Array can be rearranged to form compact or more widely spread configurations with baselines, i.e. distances between the individual antennas, of up to 16\,km.  
\edt{The ACA~TP~Array provides information equivalent}  to an interferometer with baselines from 0\,m up to 12\,m. 
\edt{The ACA~7-m~Array} samples baselines from 8.5\,m to 33\,m, bridging the baseline sampling gap between the 12-m Array and the \edt{ACA~TP~Array}.

The maximum spatial resolution is determined by the observing frequency and the longest baseline of the 12-m~Array,  and can reach a few milliarcseconds.  
Images with high fidelity at the maximum resolution, however, can only be expected if the brightness distribution is dominated by bright compact emission.
For complex images of chromospheric structures, where the detected emission fills the whole beam, the effective spatial resolution may be limited to a few 100~milliarcsec at a wavelength of 1\,mm. 
This will be quantified more accurately on the basis of further observations (see also Sect.~\ref{sec:almadiagspat}).

The \edt{ACA~TP array} is an important part of the overall array because without it ALMA would suffer from a problem common to all synthesis instruments: 
The lack of short baselines and consequent short {spatial} frequencies would make extended sources invisible. 
The maximum recoverable size, as defined in the ALMA technical handbook\footnote{http://almascience.eso.org/proposing/call-for-proposals/technical-handbook\label{note:almahandbook}}, is 
\begin{equation}
\vartheta_{max} = \frac{0.6 \lambda}{L_{min}} \mbox{rad} = \frac{37100}{L_{min}\ \nu} \mathrm{arcsec}
\end{equation}
where $L_{min}$ is the minimum baseline in meters and $\nu$ the observing frequency in GHz.  
For $L_{min}=8.4$\,m (7-m array, NS), this gives $\vartheta_{max}\approx44$\,arcsec\ at 100\,GHz. 
This simply means that, without the \edt{ACA~TP array}, the quiet sun background and also large sunspots
would be invisible although still fine structures in the chromospheric network and in solar flares would be detectable. 
See Sect.~\ref{sec:almainterfer} for a discussion of interferometric observations of the Sun and the role of the array components.

The instantaneous field of view (FOV) is conventionally given by the full-width-half-maximum (FWHM) of the main lobe of a single antenna and depends on the wavelength $\lambda$, antenna diameter $D$ and its illumination function. 
The actual ALMA 12-m antennas have a measured primary beam FWHM, which scales with wavelength approximately as 
\begin{equation} 
\theta \approx 1.13 \times \frac{\lambda}{D} \approx 19'' \times \frac{\lambda}{1\,\mathrm{mm}} \quad  \mathrm{for\ }D=12\,\mathrm{m.} 
\end{equation} 
Therefore ALMA can observe only a small fraction of the Sun at a given time. 
However, the total FOV can be increased by rapid mosaicing (see Fig.~\ref{fig:promvis})  or  
{by ``on-the-fly'' interferometric scanning, i.e. by continuously changing the pointing}.

Every antenna will eventually be equipped with receivers covering ten frequency bands in the range from 35\,GHz to 950\,GHz, {corresponding to wavelengths} from 8.6\,mm to 0.3\,mm. 
ALMA can observe with one receiver at a time but the hardware is designed to switch between up to three pre-selected receivers in about 1.5\,s. 
The exact achievable band-to-band cadence is determined by various overheads and has yet to be determined exactly.

\label{sec:fdmmode} 
Each of the ALMA bands has two polarisation channels with an instantaneous total bandwidth of 8\,GHz per polarisation. 
These are down-converted into 2\,GHz wide basebands, digitized and sent to a Correlator. 
The Correlator for the 12-m array can operate in two main modes, TDM (Time Division Mode) for low resolutions wideband continuum observations, and FDM (Frequency Division Mode) for higher spectral resolutions. 
In FDM mode, each 2\,GHz baseband is subdivided by digital filtering by up to 32~sub-bands, each with 62.5\,MHz bandwidth (usable bandwidth 58.6\,MHz to avoid edge and aliasing effects). 
These sub-bands can be individually tuned across the baseband and allow to control the analysed bandwidth and thus the frequency resolution. 
The maximum number of frequency channels in FDM dual polarisation mode is 3840. 
Thus, if all sub-bands are stitched together, the full baseband will be covered with spectral channel spacing of 488\,kHz. 
On the other side, if all Correlator resources are used to analyse one sub-band for ultimate spectral resolution, the corresponding channel spacing is 15.3\,kHz. 
The absolute minimum channel spacing achievable is 3.8\,kHz with the 29.3\,MHz bandwidth and in single polarisation mode.
The resulting spectral resolution is 7.6\,kHz (larger than the channel spacing), which corresponds to a velocity resolution of 0.02\,km\,s$^{-1}$ at $\lambda = 2.73$\,mm ($\nu = 110$\,GHz). 
Averaging of spectral channels is available in {multiples} of 2, to reduce data rate in FDM mode.
The TDM mode provides up to 128~channels per baseband ($\sim 120$~usable channels due to truncation of the edge channels in offline data processing), covers the full baseband bandwidth and provides channel spacing of 15.6\,MHz in dual polarisation mode. 
This mode has an advantage of lower data rate and faster dump time, so it is used for all observations where high spectral resolution is not required.
For more information, see the ALMA Technical Handbook$^7$.

{While ALMA is a general purpose telescope, provisions have been made to support solar observing. 
First, the surface of the antennas has been roughened to ensure that it scatters much of the flux at visible/IR wavelengths so that the receivers are not damaged\footnote{\label{note:almawebsite}Please refer to the ALMA websites for more technical information (http://www.almaobservatory.org/).}. 
Furthermore,  the detectors of ALMA are very sensitive, while the solar radio flux is very strong. 
}
At the same time most of the instrumental calibration steps, like pointing, focus corrections, and phase calibrations have to be done using much weaker radio sources, e.g. quasars. 
{When pointing at the Sun, it is necessary to reduce the high flux at millimeter wavelengths in order to avoid saturation of the receivers.}  
This is done by either inserting solar filters in the optical path, or by electronic detuning of the detectors.
Test observations have shown that observations of the quiet Sun and moderate solar flares can proceed without the use of the solar attenuators, via the detuning alternative method {(although usage of the filters remains an option)}. 
{This may also be possible for observing major flares given that the stronger flux in big events comes from a bigger area; thus, even the most energetic flares will still look 'moderate' within the restricted ALMA FOV. 
The exact procedure has yet to be tested on stronger flares.
}
The receiver sensitivity is degraded in the detuned mode, with the noise temperature typically in the range 100\,-\,1000\,K, depending on the required attenuation.
However, on the solar disk, the sensitivity will be still limited mostly by temporal and spatial fluctuations (e.g., p-mode oscillations and propagating waves)  of solar origin -- ``signal'' rather than ``noise'' in some sense. 
The test data from late 2014 have convincingly confirmed this conclusion.

\subsection{ALMA's diagnostic potential for the Sun} 
\label{sec:almadiag}

The technical specifications summarised in Sect.~\ref{sec:almatech} promise impressive diagnostic opportunities for solar observations with ALMA. 
{In the following, we discuss the advantages and challenges of observing the Sun with an interferometer and highlight key capabilities. 
}

\subsubsection{{High-resolution interferometric imaging at millimeter wavelengths}}
\label{sec:almainterfer}

The Sun is a very difficult object to observe with an interferometer, and it can be argued that the solar chromosphere is the most difficult target of all. 
The reason is that the Sun fills any field of view with emission on a wide range of spatial scales, from the large--scale background down to the smallest scales of magnetic field structures, but  interferometers are designed to sample a fixed range of spatial scales corresponding to the range of baseline lengths in the array\footnote{The angular scale on the sky sampled by an interferometer formed from a pair of antennas is proportional to the 
wavelength divided by the physical separation of the antennas projected onto the plane
perpendicular to the line of sight to the source.}. 
As an example, a single observation with the Very Large Array (VLA) covers about a factor of 20 in spatial scales. 
Synthesis of a large aperture by this technique requires reasonably dense sampling of spatial scales for successful calibration and robust imaging, so one cannot simply impose an arbitrary range of spatial
scales and achieve a successful observation \citep{1999ASPC..180.....T}.
Arrays such as the VLA and ALMA can get around this limitation of a single configuration by observing the same object with multiple configurations, each covering complementary ranges of spatial scales: thus the VLA can sample a factor of order 750 in spatial scales by observing the same object in all 4 available configurations. 
However, since configuration changes for the VLA and ALMA typically take at least a week, this technique for expanding the range of spatial scales only works for objects that do not vary on corresponding timescales, and thus is completely useless for the Sun.

We are therefore limited to the range of spatial scales present in a single ALMA configu\-ration$^7$. 
When only the 12\,m antennas are used, the minimum spacing in the interferometer array is around 14\,m and this limits the largest spatial scale sampled in all configurations. 
Most of the ALMA configurations currently in use have a range of about 20 in spatial scale, but this can be enhanced by including the array of 7\,m antennas for which the minimum spacing is around 8\,m. 
As mentioned earlier, for bright compact sources such as solar flares where emission is dominated by sources with a limited range of spatial scales, this range of spatial scales is quite adequate for excellent imaging
{(see Sect.~\ref{sec:majorflares} regarding imaging of flares).} 
This is also true of VLA observations of coronal sources, which are typically at least several times brighter than the background solar emission.
However, the solar chromosphere at millimeter wavelengths is composed of emission on all spatial scales, with contrasts of up to 1000\,K brightness temperature that must be imaged against a continuously changing background of order 4000\,-\,7000\,K. 
Thus the spatial structure to be imaged by the interferometer is much weaker than the background. 
Since it has components on all spatial scales from subarcsecond out to the field of view of the dishes, imaging is challenging.

The issues involved in imaging the solar chromosphere at millimeter wavelengths were discussed by \citet{2006A&A...456..697W} in the context of the BIMA observations shown in Fig.~\ref{fig:bimaimages}. 
The critical issue is sampling the largest spatial scales of the emission: if this is not sampled adequately then images can be dominated by spurious small-scale features. 
\citet{2006A&A...456..697W} addressed this problem by using the most compact configuration available at BIMA and by deconvolving using {the maximum entropy method (MEM)} to take advantage of the information on large spatial scales present in the primary beam response of the dishes \citep{1979ASSL...76...61E}.
However, since short baselines are essential to sample the large-scale structure and interferometers usually sample a limited range of spatial scales, one usually has to sacrifice spatial resolution to achieve successful imaging.

ALMA possesses three critical features that should allow successful imaging of the chromosphere with much better spatial resolution than previously. 
(i)~The larger number of dishes (up to 62 for interferometry) provides much more information in each integration. 
(ii)~The second advantage is the availability of the ACA array of 7\,m dishes: as long as it is in a compact configuration it can provide the shorter spacings lacking from the wider 12-m array that provides high spatial resolution. 
Just how wide a range of spatial scales can be achieved in this fashion remains to be demonstrated: it may be that we will not be able to successfully image the Sun at the highest spatial resolution of ALMA because the amount of information needed to reconstruct a filled field of view (up to 10$^5$ resolution elements) is not adequately sampled.  
The fact that the field of view of the 7\,m dishes is almost twice as large as the field of view of the 12\,m dishes may also cause problems in deconvolution. 
However, preliminary experiments using CARMA observations with both 6\,m and 10\,m dishes indicate that successful imaging is still possible (White, Loukitcheva \& Solanki, in preparation). 
(iii)~The other feature of ALMA that will be a major factor is the availability of the 12\,m TP dishes. 
These dishes have been designed with the ability for scanning the sky at very fast rates, and have already been shown to be capable of imaging the full solar disk in 8\,min (Hills \& Phillips, private communication). 
This means that the smaller targeted interferometer regions can be scanned, and the zero spacing flux recovered, on timescales shorter than 1~min, fast enough to track coherent 3-minute oscillations in the chromosphere. 
The images from the TP measurements in combination with the interferometer data, with the 7-m array providing intermediate spatial scales, is expected to improve the imaging ability of ALMA dramatically. 
In addition the single-dish data will provide absolute measurements of the total solar brightness temperature, including the background disk level that is filtered out by the interferometer, and thus provide interferometer maps with the correct temperature scaling.
{Fast-scanning of the whole solar disk (or later a small FOV) with the TP antennas will most likely be offered as an integral part of interferometric observations, aiding the image reconstruction and providing context information.} 

\label{sec:almadiagspat}

{As briefly discussed before,} 
the effective spatial resolution might be much lower when reconstructing an image for an extended area source like the Sun from a limited number of baselines
{compared to what is theoretically possible with very long baselines for separating two point sources.}  
The effective spatial resolution of a reconstructed image is determined by the coverage in the spatial Fourier space (the {u-v}-space) and the corresponding synthesised point spread function of the combined array.
The spatial resolution depends then on the array configuration, the observed wavelength and the position on the sky (due to the elevation angle of the antennas and the resulting foreshortening of the effective baseline). 
Current estimates based on numerical simulations of ALMA's imaging capabilities \citep[e.g.][]{ssalmon_alma_imaging15} imply a spatial resolution of 0''.3$\,\times\,\lambda/1$\,mm to 0''.4$\,\times\,\lambda/1$\,mm although it has to be quantified more accurately on the basis of actual  observations. 
Reconstructed images at the shortest wavelengths, i.e. 1\,mm and below, will have an effective spatial resolution close to the best possible with today's optical solar telescopes. 
Due to the increase of continuum formation height with wavelength (see Fig.~\ref{fig:contf}), ALMA samples the low chromosphere with higher spatial resolution than the upper chromosphere. 
However, it is important to emphasise that each observation still contains information on even smaller scales, produced by the longest baselines. 
There will certainly be scientific applications for which the visibilities, i.e. the measured signal on a baseline, can be used directly without the need of image reconstruction, thus utilising information on much smaller spatial scales.   
In principle, this is true for all studies that are carried out in spatial Fourier space. 
A potential application could be the determination of the spectral characteristics and levels of MHD turbulence in the solar chromosphere, which is an important question in solar physics. 
Nevertheless, the anticipated spatial resolution of reconstructed ALMA maps will be substantially higher than everything that has been possible at millimeter wavelengths before (cf.~Fig.~\ref{fig:bimaimages}).

\subsubsection{High cadence}

In interferometry, the integration time is determined by two factors: the sensitivity and the coverage of the 
{u-v} 
plane. 
Sensitivity is not a problem because the Sun is a very bright source at sub-millimeter and millimeter wavelengths and because ALMA is a very sensitive instrument. 
For interferometric observations of simple structures with strong contrast, such as bursts, snapshot mode (i.e., instantaneous measurements) and thus ultra-high cadence imaging should be possible with ALMA.
The situation is more complicated if the observational target has a complicated structure, which fills the primary beam. 
As mentioned above, in snapshot mode, ALMA samples the {u-v} plane at discrete points. 
If the sampling is uniform, in steps of $\Delta b$ wavelengths, the beam pattern is repeated at intervals of $1/\Delta b$ radians. 
Even if the sampling of the {u-v} plane is not uniform, as is the case with ALMA, confusion will still arise if the field of view has a complex structure as, e.g., in the quiet sun. 
The  {u-v} coverage required for resolving the structure puts some constraints on the achievable cadence, usually leading to a compromise between spatial and temporal resolution, which has to be adjusted according to the requirements of individual scientific applications. 
It is possible to take advantage of Earth's rotation for increasing the number of sampled points in the {u-v} plane (then in small arcs), 
{which then requires longer integration times. 
Integrating for several minutes or even up to 40\,min, as it has been done during the fifth Commissioning and Science Verification campaign in December 2014, is reasonable for studies of slowly evolving features but is prohibitive for observing fast evolving phenomena. 
For the latter, a time resolution of less than 1\,min is required.
}

A cadence well below 1\,s should be achievable for snapshot imaging, which enables the study of high-frequency waves in the solar chromosphere. 
These waves are interesting regarding their role in the energy transport and heating (see Sects.
\ref{sec:heatingmagreg} and \ref{sec:oscillwaves}) but also provide a means of probing the atmospheric structure (``chromospheric seismology''). 
For instance, eclipse observations were previously used to detect waves with periods of 6\,s in low coronal loops \citep{williams2002}. 
Such coronal loops will be observable also with ALMA, but much more often and for longer time sequences than  possible during solar eclipses.
It is also expected that the miniature loops found by \citet{2013A&A...556A.104P,2014Sci...346C.315P} have high frequency oscillations, because of the smaller length scales (but wave speeds comparable to regular-sized active region loops). 
ALMA will provide the temporal and spatial resolution needed to observe such oscillations. 
Moreover, the high cadence ALMA observations will be useful in observing oscillations observed in flare light curves, the so called quasi-periodic pulsations (QPP, for a review see e.g. \citealt{2009SSRv..149..119N}). 
Please refer to Sect.~\ref{sec:qpp} for quasi-periodic pulsations  and Sect.~\ref{sec:oscillwaves} for more details on chromospheric waves.

Apart from high-frequency waves, the relevant temporal scales for phenomena in the solar chromosphere are {typically} on the order of 10\,s  {and in some cases even longer}. 
Multi-band (and thus multi-layer) data cubes could be obtained when switching  
{receiver bands quickly with two to three of them being ready for instant use.} 
It is not entirely clear yet what effective band-to-band cadence can be achieved and if it would be fast enough for resolving the shortest timescales.

The achievable cadence for mosaicing depends on many factors such as the required FOV and thus the number of pointings, the required integration time per pointing, and the required spectral resolution, which determines the correlator mode and thus its minimum dump time (with the FDM mode being generally slower, see Sect.~\ref{sec:fdmmode}),  and other overheads like the  antenna slew rate. 
{On-the-fly\footnote{{In contrast to a sequence of discrete pointings, the antennas keep moving during an ``on-the-fly'' observation, thus changing the pointing continuously and scanning the target along predefined trajectories.}} interferometry} (OTFI) is probably faster for low time per pointing but has yet to be commissioned.
{The simultaneous use of sub-arrays acting in different frequency bands is also possible in
principle, but not in the initial solar observations.} 
The exact setup of mosaicing observations and the resulting cadence can and should be adjusted to the requirements of the considered scientific aim, finding an optimum compromise between FOV, temporal and spectral resolution.

\subsubsection{Spectral capabilities} 
\label{sec:spectralcap}

The high achievable spectral resolution and the large number of simultaneous frequency channels enables ALMA to record precise {spectra}. 
{Recombination lines (i.e. Rydberg transitions, see Sect.~\ref{sec:spectroscopy}) and molecular pure rotational transitions (Sect.~\ref{sec:carbonmonoxide}) could thus be utilized as powerful complementary diagnostic tools with ALMA.  
}
Among the ionic species with lines in the wavelength range accessible to ALMA are, for instance, H\,I, C\,III, O\,IV, O\,V, O\,VI, Ne\,VII, Si\,XI , Fe\,XV. 
While the continuum radiation is mostly formed in the chromosphere, some of the recombination lines originate in the corona, thus adding 
{coronal} 
diagnostics to ALMA's tool kit.  
Some of these recombination lines probably are severely broadened, which has made detection with previous instruments difficult. 
The high spectral resolution and high sensitivity of ALMA could finally lead to their detection. 
Carbon monoxide is probably the most promising molecule with its pure rotational spectrum in the ALMA wavelength range, especially because it provides a {local} kinematic and thermal probe for the dynamic solar chromosphere.  
Observations of CO  also will benefit from the high spatial and temporal resolution because the CO number densities are affected by {hot, propagating} shock waves on small spatial scales.

\subsubsection{{Polarization capabilities and chromospheric magnetic field measurements}}
\label{sec:almamagneticfield}

ALMA's capability of measuring the polarisation state at multiple wavelengths will provide a powerful diagnostic tool for the magnetic field over a substantial range of chromospheric heights.  
For a magnetized atmosphere with a temperature gradient, like it is the case for the Sun, the temperature difference between two layers produce an observable net circular polarisation (see Sect.~\ref{sec:magneticfield}), which can be measured by ALMA.  
As mentioned in Sect.~\ref{sec:magneticfield}, the height at which the magnetic field is determined varies with frequency. 
Strictly speaking, the difference in formation heights between the o- and x-modes should be taken into account
but models show that the difference is small (see Sect.~\ref{sec:qsmag}). 
The contribution function of the continuum intensity displayed in Fig.~\ref{fig:contf} can thus also be understood as the effective height where the magnetic field is probed. 
In the case of ALMA, it gives access to magnetic field information from the temperature minimum layer up to the transition region over the ALMA spectral range.

The technical requirement for the accuracy of the polarized flux$^{7,8}$ 
 is 0.1\% of the total intensity, after calibration. With the current on-axis calibration scheme, it is feasible to achieve this accuracy for a compact source at the beam centre. Polarisation imaging accuracy degrades off-axis, up to 0.3\% for angular source sizes of FWHM/3 of the primary beam. 
To achieve 0.1\% accuracy for extended sources, the off-axis instrumental polarization calibration has to be performed. 
As a result, high-precision multi-frequency observations of the polarized emission with ALMA should provide a wealth of information for the restoration of the longitudinal component of the magnetic field over a wide range of heights in the chromosphere. 
More detailed studies based on state-of-the-art numerical models and high-resolution observations (including ALMA) are currently carried out in order to investigate the limits and reliability of the approach.  
Based on these results, magnetic field measurements in active regions (see more details in Sect.~\ref{sec:armodelling}) will clearly be possible while such measurements for quiet Sun regions (more details in Sect.~\ref{sec:qsmag}) are at least very likely to be feasible given ALMA's high sensitivity.
Polarization measurements will also be extremely valuable during flares, which will put strong constraints on the pitch-angle anisotropy of accelerated electrons and will be able to distinguish electron and positron contributions (see Sect.~\ref{sec:flares}).

\subsubsection{ALMA observations in co-ordination with other observatories} 
\label{sec:coordobs}

\begin{figure}
\centering
\resizebox{\textwidth}{!}
{\includegraphics[]{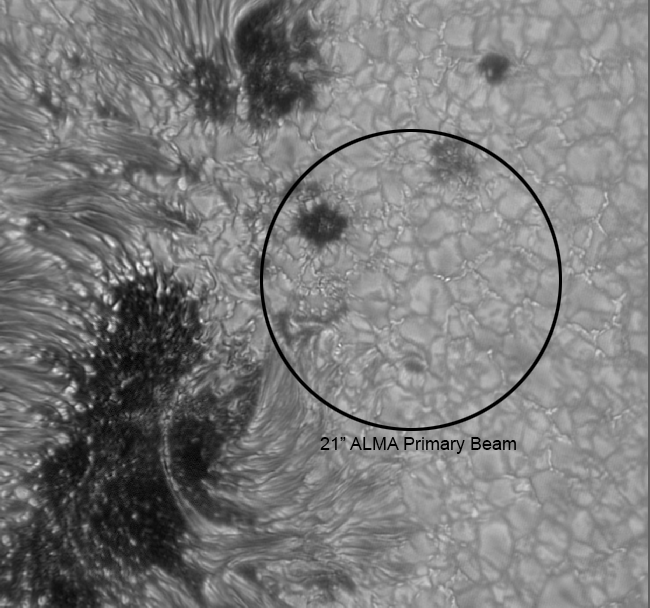}}
\caption{TiO band (705.7\,nm) continuum image from NST of a sunspot and nearby granulation on 2014 Jul 31 (courtesy of Vasyl Yurchyshyn, BBSO).  
The image has been resampled with a pixel size of 0.07\,arcsec/pixel, whereas the original image from the camera had 0.''034~pixels. 
The angular resolution is close to the diffraction limit of the telescope, which is  0.''09.
{The circle illustrates ALMA's primary beam at wavelengths around 1\,mm.}  
}
\label{fig:bbsosunspot}
\end{figure}

\begin{figure}
\centering
\resizebox{\textwidth}{!}{\includegraphics[]{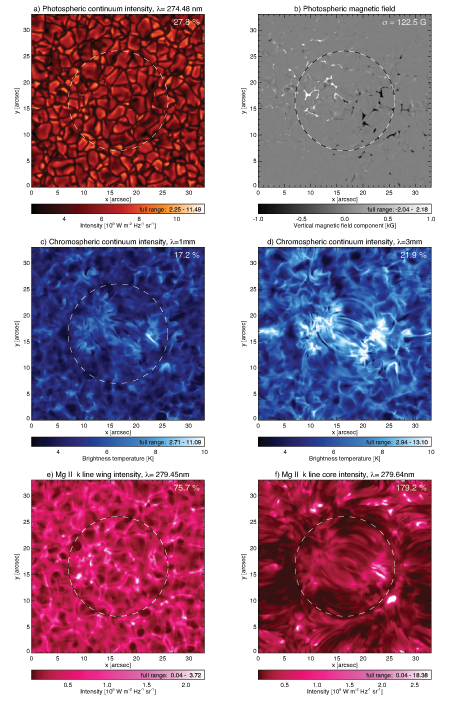}}
\caption{
Synthetic maps based on a  3D numerical simulation of an enhanced network region with magnetic loops connecting patches of opposite polarity:  
(a)~UV~continuum close of the Mg\,II\,k line showing the granulation at the bottom of the photosphere.
(b)~Vertical magnetic field component $B_z$ at the bottom of the photosphere. 
(c)~Brightness temperature at 1\,mm. 
(d)~Brightness temperature at 3\,mm. 
(e)~Intensity in the blue line wing of the Mg\,II\,k line. 
(f)~Intensity in the  line core of the Mg\,II\,k line. 
The contrast is written in the upper right corner of each panel with exception of panel~b for which the standard deviation is specified. 
The circle marks the primary beam of ALMA at $\lambda = 1$\,mm but is not shown for the 3\,mm map because, in that case, the primary beam is three times in diameter and is thus larger than the depicted region. 
The numerical model  \citep{carlssonbifrostmodel} was computed with the Bifrost code \citep[][see Sect.~\ref{sec:quietsunmodel} for more details]{2011A&A...531A.154G}.  
The intensity images for panels~a,e and f are produced with the radiative transfer code MULTI3D by \citet{2013ApJ...772...89L}, while the millimeter maps in panels c-d were calculated by \citet{2015A&A...575A..15L}. 
The magnetic field component $B_z$ shown in panel~b is taken directly from the numerical model.
}
\label{fig:bifrostmmvsmg}
\end{figure}

To make the most of ALMA's unprecedented spatial resolution, and also to place its intensity and polarisation measurements into a broader context, observations at other wavelengths are essential.  
In particular, several ground-based optical observatories with 1+-meter-class apertures, such as Big Bear Solar Observatory's 1.6-m New Solar Telescope \citep[NST,][]{2012SPIE.8444E..03G} and the European GREGOR 1.5-m telescope \citep{2012AN....333..796S}, are producing diffraction-limited observations in the visible and near-infrared.  
In addition, other ground-based observatories can provide important context data, such as the Swedish 1-m Solar Telescope, in particular known for its chromospheric observations with the CRISP\footnote{The Swedish 1-m Solar Telescope will soon be equipped with the CHROMIS instrument which will have a narrow-band Ca\,II\,K filter suitable for chromospheric observations.} instrument, Themis, VTT and Chrotel on the Canary Islands, Kanzelh{\"o}he Solar Observatory in Austria, Hvar Observatory in Croatia, and also the Daniel K. Inouye Solar Telescope \citep[DKIST, formerly the Advanced Technology Solar Telescope, ATST,][currently under construction, possibly operational by 2019/2020]{2011ASPC..437..319K} and the future European Solar Telescope \citep[EST, ][]{2010AN....331..615C}. 
For instance, the DL-NIRSP (Diffraction limited Near Infra-Red Spectropolarimeter) instrument on DKIST will provide valuable chromospheric (and possibly coronal) magnetic field measurements with a spatial resolution of $\sim 60$\,km in the NIR and down to $\sim 25$\,km for H$\alpha$ (in the most ideal case). 
The New Solar Telescope produces images, which have an angular resolution close to the diffraction limit of the telescope, i.e. 0.''09 at 705.7\,nm (in the TiO~band, see Fig.~\ref{fig:bbsosunspot}) and 0.''16 at \mbox{1560\,nm}.
With the ability to measure magnetic fields on this sub-arcsecond scale with the NIRIS instrument \citep{2012ASPC..463..291C} and others, these observations of both the deeper (at \mbox{$1,560\,$nm}) and higher layers (e.g., in H$\alpha$  or He\,I\,1083\,nm) will provide an ideal complement to the ALMA high-resolution interferometric observations of the quiet Sun and active regions.  
At Big Bear Solar Observatory, another complementary instrument to ALMA is Cyra, a cryogenic Near Infrared (NIR) echelle spectrograph \citep{2013SoPh..287..315G} capable of spatially-scanned, high spectral resolution slit imaging out to \mbox{$5000\,$nm}.  
This includes the CO molecular lines at \mbox{$4,667\,$nm}, which can directly support ALMA's exploration of the coolest parts of the solar atmosphere (see Sects.~\ref{sec:quietsun} and \ref{sec:spectroscopy}).

At GREGOR, the GRIS (GRegor Infrared Spectropraph) instrument \citep{2012AN....333..872C} is providing measurements close to the diffraction limit of the velocity and magnetic field in the upper chromosphere in the He\,I\,1083\,nm line, as well as in the photosphere at 1560\,nm, thus nicely straddling the layers. 
The ability to measure the full magnetic vector by GRIS will complement the longitudinal field measurements made by ALMA and will greatly facilitate the interpretation of the ALMA magnetic field data. 
Conversely, the very high cadence achievable with ALMA with greatly extend the capabilities of sampling the dynamics of the chromospheric magnetic field. Also, the Gregor Fabry Perot Interferometer \citep[GFPI,][]{2012ASPC..463..423P}  will provide high-resolution images and magnetic field data in various photospheric and chromospheric spectral lines.

Simultaneous campaigns with other ground-based observatories would be limited by their geographic locations  and resulting time differences with respect to the ALMA site, although some scientific applications may not require exact simultaneous observations. 
In contrast, space-borne telescopes can provide simultaneous data around the clock (except for eclipse periods) and offer complementary diagnostics, which are not accessible from the ground (e.g., EUV). 
The most relevant space missions in this respect are the Solar Dynamics Observatory \citep[SDO,][]{2012SoPh..275...17L}, the Interface Region Imaging Spectrograph \citep[IRIS,][]{2014SoPh..289.2733D},  
Hinode \citep{2007SoPh..243....3K}, the Reuven Ramaty High Energy Solar Spectroscopic Imager \citep[RHESSI,][]{2002SoPh..210....3L}, and the Solar and Heliospheric Observatory \citep[SoHO,][]{1995SSRv...72...81D,1995SoPh..162....1D,1995SoPh..162..233H,1995SoPh..162..313K}. 
Most instruments have a limited field of view just like ALMA itself and thus require detailed planning prior to the observation. 
SDO, on the other hand, observes the whole Sun at all times and offers amongst many things vector magnetograms of the photosphere recorded with the Helioseismic and Magnetic Imager  \citep[HMI,][]{2012SoPh..275..207S}, as well as filtergrams for diagnostics isolating many layers in the solar atmosphere \citep[e.g., with the Atmospheric Image Assembly, AIA,][]{2012SoPh..275...17L}. 
Furthermore, for some applications, coordination with other radio telescopes or arrays could be desirable (see Sect.~\ref{sec:armagnetometryothertelescopes} for a short list of relevant radio facilities). 
It should also be mentioned that the  Spectrometer Telescope for Imaging X-rays \citep[STIX,][]{2012SPIE.8443E..3LB} on board Solar Orbiter will be the only high-resolution X-ray instrument beyond the lifetime of RHESSI.

{For instance, the combination of ALMA and IRIS promises a set of 
complementary plasma diagnostics with great scientific potential. 
Since its launch in June~2013 as a NASA Small Explorer, IRIS has already provided fascinating new insights into the dynamics of the chromosphere and possible new heating mechanisms \citep[e.g.,][]{2014Sci...346D.315D,Tian2014,Testa2014,2014Sci...346C.315P,Hansteen2014}. 
It has also become clear, however, that the interpretation of the spectra recorded in the optically thick Mg\,II\,h and k~lines is very difficult and will require sophisticated new analysis tools in combination with state-of-the-art forward modeling. 
}
An example for a complementary ALMA-IRIS data set is illustrated based on a numerical model snapshot in Fig.~\ref{fig:bifrostmmvsmg}. 
It combines (synthetic) ALMA maps at different wavelengths, thus probing different layers of the chromosphere, with important context data (here the intensity and magnetic field in the low photosphere) and complementary other diagnostics, namely (synthetic) IRIS maps in the Mg\,II\,k line at a wavelength of 279.6\,nm.
Beside the differences in formation height, the Mg\,II\,k line core intensity is a highly non-linear measure of the atmospheric conditions.
Both the ALMA 3\,mm map and the IRIS Mg\,II\,k line core map show an imprint of the magnetic loops in the numerical model but the contrast is $\delta T_\mathrm{b} / <T_\mathrm{b}> = 21.9$\,\% for the 3\,mm map and  $\delta I / <I> = 179.2$\,\%  for the Mg\,II\,k line core map. 
Much of the contrast in the Mg~map is due to very localised brightenings, which can outshine the surrounding region by a factor $\sim 10$. 
This extreme effects make a physical interpretation difficult, especially when compared to the linear behaviour of the ALMA maps, but nevertheless probes different aspects of the chromospheric plasma conditions and thus contributes to comprehensive diagnostic data set. 
While ALMA alone will without any doubt produce excellent data, its scientific output can be significantly enhanced in combination with other observatories.

\section{Potential solar science cases for ALMA} 
\label{sec:sciencecases}

This section outlines some potential science cases for ALMA. 
In this respect, numerical simulations of the solar chromosphere can play an important role for the planning, optimization  and interpretation of observations with ALMA. 
Synthetic brightness temperature maps  calculated from numerical models can be used to simulate what ALMA might observe. 
Different instrumental configurations can be tested and adjusted to the scientific requirements, finding the optimal set-up for different scientific applications. 
In addition, coordinated observations of other chromospheric diagnostics, say at optical, infrared (IR) or ultraviolet (UV) wavelengths, can address open questions whose solution benefits from a broad-based approach to which ALMA could contribute significantly.

\subsection{Central questions in solar physics to be addressed with ALMA} 
\label{sec:bigquest}
ALMA will be able to advance our knowledge of the chromosphere in all its 
different flavours ranging from quiet Sun regions to flares. 
In particular, substantial progress can be expected regarding the following Grand-Challenge topics in contemporary solar physics. 
%
\label{sec:bigquestchroheat}
{\paragraph{Coronal and chromospheric heating} is a long-standing problem in modern astrophysics \citep[see, e.g.,][]{Ulmschneider1991}.}
{Despite much progress during the past decades,} 
several fundamental issues concerning the dynamics of the solar chromosphere are still unresolved. 
{For instance, the verdict as to what mechanisms contribute most to heating the network and the internetwork  chromosphere is still open. 
Equally, the source of the so-called ``basal flux'' \citep{Schrijver1987} is still debated.
}
Is it due to acoustic waves or processes connected to the Sun's magnetic field or both \citep[see, e.g.,][and references therein]{Schroeder2012,Cuntz2007,Judge2003,Judge1998,Buchholz1998,Schrijver1995}? 
Or in other words, are those heating processes of the alternating current (AC) or the direct current (DC) type? 
Magnetohydrodynamic waves {\citep[e.g.,][]{2007Sci...318.1574D}} can in general contribute directly or indirectly to heating the atmospheric plasma, i.e. through perturbation of the magnetic field, resulting in damping of the waves and the associated release of magnetic energy. 
{Do the problems require a fully kinetic model approach?  
And is there a magnetically driven basal flux connected to the presence of an ubiquitous, local, turbulent dynamo \citep{2007A&A...465L..43V,2010A&A...513A...1D,2011ApJ...737...52L}?  
These are important questions} not only in solar physics, but also in stellar physics because there are many low-activity dwarf stars and giants with emission line fluxes close to the basal flux limit, and in fact define it \citep[e.g.][]{Perez2011,1994A&A...285..233D}.

{ALMA has the potential to substantially contribute to answering these fundamental long-standing questions.
By probing the  3D thermal structure and dynamics of the solar chromosphere, observations with ALMA give} 
important clues on the sources and sinks of heating in the chromosphere 
{and with it essential} clues on the heating of the corona since the energy required for heating the corona must pass through the chromosphere. 
Atmospheric heating is addressed in more detail in the context of solar flares in Sect.~\ref{sec:flares} and for regions with strong magnetic fields in Sect.~\ref{sec:heatingmagreg}. 
Chromospheric waves and oscillations, which are potentially important sources of heating, are discussed in Sect.~\ref{sec:oscillwaves}.

\label{sec:bigquestflare}
\paragraph{Solar flares} are violent eruptions during which large amounts of energy are released in the form of radiation and high-energy particles. 
Major flares occur in active regions with strong magnetic {fields. 
The} strongest flares (X~class) can -- often in connection with coronal mass ejections (CME) -- directly influence heliosphere and Earth (see Sect.~\ref{sec:bqspaceweather}). 
However, there is a large spectrum of flare energies, which reaches down to so-called micro- and nano-flares. 
Such weaker events can occur on smaller spatial scales and potentially are candidates for heating the ``quiet'' solar corona even outside active regions.  
Despite their obvious importance, the basic physics of flares at all scales is not understood, especially regarding particle acceleration mechanisms, but also concerning the source of the still enigmatic emission component at sub-THz frequencies. 
{Furthermore, quasi-periodic} pulsations are common and possibly intrinsic features of flares. 
Understanding the nature of the pulsations puts important constrains on the interpretation of physical mechanisms responsible for the accumulation and release of magnetic energy and acceleration of charged particles in flares, and may become an important diagnostic tool for  the physical conditions in the flaring sites. 
The progress in this research direction requires spatial resolution superior than provided by available observational tools, as well as a high spectral resolution in the microwave and sub-THz bands.
ALMA's  spectral, spatial and temporal resolution for observations in the sub-THz range together with the ability to probe the thermal structure and magnetic fields in flaring regions promise therefore ground-breaking discoveries in this respect. 
See Sect.~\ref{sec:flares} for more details.

\label{sec:bigquestprom}
\paragraph{Solar prominences} are regions of cool and dense plasma lying in a much hotter and rarer solar corona. 
While active prominences exhibit large-scale motions and erupt in a matter of hours, quiescent prominences have large-scale structures that remain stable for days or weeks, although their fine structures change on time-scales of a few minutes. 
The existence of prominences is mainly due to the coronal magnetic fields that support prominence material against gravity and insulate it from the hot coronal environment. 
Prominences are referred to as (dark) filaments when observed in absorption against the (bright) solar disk. 

{
ALMA will provide us with unprecedented details on the temperature structure of the prominence material, which 
has direct implications for the  energy balance of prominences. 
}
ALMA's high spatial and temporal resolution will also be key in the understanding of the small-scale dynamics of the prominence fine structures and to the understanding of the origin of prominences. 
ALMA will provide invaluable insight into the role and impact of prominences on the evolution of the physical conditions in our space environment.
{See Sect.~\ref{sec:prominences} for more details.}

\label{sec:bqspaceweather}
{\paragraph{``Space Weather''} is a very dynamic area of research, which,  
broadly speaking, deals with the study of perturbations of the heliospheric environment, including around Earth, the planets and other solar-system bodies, originating from the Sun. 
ALMA observations addressing the aforementioned major research questions will have implications for the underlying solar drivers of space weather, including solar flares (Sect.~\ref{sec:flares}),  prominence (a.k.a. filament) eruptions (Sect.~\ref{sec:prominences}), and CMEs, including their triggers and internal structure right from the events' birth.}

\subsection{Quiet Sun regions}
\label{sec:quietsun}
\begin{figure}
\centering
\resizebox{\textwidth}{!}
{\includegraphics[]{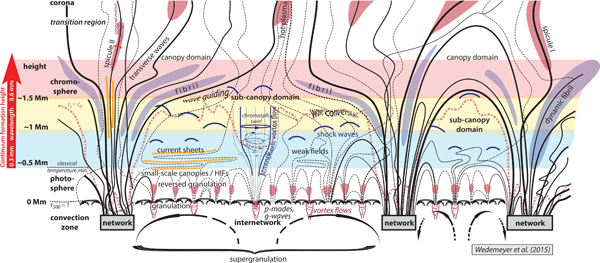}}
\caption{Schematic structure of Quiet Sun regions. 
This illustration is based on the cartoon by \citet{2009SSRv..144..317W} but has been updated and now contains vortex flows and torsional motions.
The horizontally aligned coloured areas represent the layers which are mapped by ALMA.
}
\label{fig:quietsunscheme}
\end{figure}

%
Our understanding of the chromosphere in quiet Sun regions has been improved substantially during the last decades owing to advances in observation techniques at ultraviolet (UV), visible and near-infrared (NIR) wavelengths and advances in numerical simulations. 
The cartoon in Fig.~\ref{fig:quietsunscheme} illustrates the complex structure of quiet Sun regions with a wealth of dynamic physical processes as it is known today.  
{Nevertheless,  quiet Sun internetwork regions are reasonably well understood so that 
continuum intensity observations of such regions may be used for radiometric calibrations of ALMA, which could be of potential interest for  radio astronomers in general.}

\subsubsection{The thermal structure and dynamics of Quiet Sun regions} 
\label{sec:quietsunthermal}
%
The classical semi-empirical reference atmospheres by \citet[][hereafter referred to as VAL]{1981ApJS...45..635V} and \citet[][hereafter FAL]{1993ApJ...406..319F} were constructed by adjusting a (hydrostatic) plane-parallel atmospheric stratification until the corresponding emergent spectra optimally matched a set of selected observations \citep[see also][]{2008ApJS..175..229A}. 
Among the selected target observations were EUV continuum intensities and the Ly$\alpha$ and Ly$\beta$ spectral profiles, although the spatial resolution of those observations was rather low compared to today's standards. 
While elaborate at the time, especially in capturing the NLTE subtleties of the optically-thick line formation, this class of static models cannot account for the pronounced small-structure and dynamics of the solar atmosphere so impressively documented by modern high-resolution instruments. 
The hope of the VAL and FAL-type models was that the assumption of a one-dimensional, plane-parallel atmospheric stratification and the usage of temporally and spatially averaged observations, would lead to a stratification that reproduced the atmospheric properties in the mean.  
However, the true structuring and dynamical properties of the atmosphere, especially in the chromosphere, are so severe that an average model cannot  capture a meaningful picture of the actual complexity. 
For example, the observation of dark carbon monoxide (CO) lines at the extreme limb implied the presence of cold gas, 1000\,K cooler than the minimum temperatures of the VAL- and FAL-type models \citep[e.g.][]{noyes72b,ayres81b}.  
In fact, the  EUV and UV intensities that form the basis of the 1D models in the chromosphere have a highly non-linear relation to the (local) gas temperature, which renders the average stratifications essentially meaningless in the presence of unresolved significant temperature variations.  
Now, however, ALMA provides an unprecedented measure of the small-scale thermal structure of the chromosphere and at the same time potentially exploiting the CO molecules in a new light as an additional thermal probe (see also Sect.~\ref{sec:carbonmonoxide}). 
Using IRIS, \citet{Schmitt2014} observed absorption features which indicate temperatures below 5000\,K 
existing above the transition region. 
The authors suggested that this lends credence to numerical studies that pointed towards elevated pockets of cool gas in the chromosphere -- a hypothesis that can be tested with ALMA.

An early attempt to account for atmospheric dynamics was the 1D time-dependent simulations of Carlsson \& Stein \citep{1994chdy.conf...47C,1995ApJ...440L..29C} of chromospheric Ca\,II excitation by propagating shock waves. 
These waves were generated from initial disturbances in the photosphere, which then propagate upwards into layers with lower density, and ultimately steepen into shock waves in the chromosphere. 
The resulting strong variations in quantities, especially the gas temperature, can successfully reproduce spatially and temporally averaged near-UV observations (of the Ca\,II ``flashes'' or ``K~grains'') and at the same time allow for cooler, less disturbed intervals in-between shocks, as implied by observations of carbon monoxide in the low chromosphere \citep{asensio03}. 
The result emphasised the ambiguities introduced by relying solely on UV diagnostics to empirically deduce properties of the solar chromosphere, again owing to the highly non-linear response of the UV to local conditions in a thermally diverse outer atmosphere.

The calculations of the emergent thermal free-free radiation at submillimeter and millimeter wavelengths for the 
{\citet{1994chdy.conf...47C}}
simulations produced brightness temperatures that varied strongly in time due to the propagating shock waves and oscillation modes, and thus provide a much more reliable proxy for the local gas temperature in the continuum forming layers \citep{2004A&A...419..747L}. 
The resulting synthetic (sub)millimeter maps were compared to interferometric observations of quiet Sun regions by the Berkeley-Illinois-Maryland Array (BIMA) at a wavelength of 3.5\,mm \citep{2006A&A...456..713L, 2008Ap&SS.313..197L}, although the pattern seen on the smallest spatial scales in the simulations were not resolvable in these observations (see also Fig.~\ref{fig:bimaimages} for examples of BIMA images). 
The dynamic picture was further supported by studies based on 3D models \citep[e.g.,][]{2007A&A...471..977W,2015A&A...575A..15L}, which in addition exhibited an intermittent and fast evolving pattern in the synthetic millimeter continuum intensity maps with structures on scales down to 0.1\,arcsec and time scales on the order of a few 10\,s. 
This small-scale pattern correlates closely with the thermal pattern at chromospheric heights as it was already known from earlier simulations \citep{2000ApJ...541..468S,wedemeyer03c,wedemeyer03a,2004A&A...414.1121W,2005ESASP.592E..87H,2010MmSAI..81..582C}. 
The pattern is produced by propagating shock waves like those seen in 1D simulations, although, crucially, in 3D the waves travel in all directions, not just vertically.  
The dynamic co-existence of hot gas in shock fronts and cool gas in post-shock regions finally provides a way to understand the presence of cold carbon monoxide in the chromosphere, which otherwise would be completely at odds with the high temperatures in VAL-type models  \citep{ayres96,asensio03,2005A&A...438.1043W,2007A&A...462L..31W}.

However, because the molecular formation rates are very slow in the post-shock gas, due to the low densities, even considering multi-step chemical pathways \citep{ayres96}, and because the post-shock regions are expected to be spatially intermittent, it is somewhat puzzling that the off-limb emissions of the CO fundamental rovibrational bands at 5~$\mu$m are so strong and spatially prevalent at the limb.  
These emissions, originally noted by \citet{1994Sci...263...64S}, occur because cold gas apparently exists in the chromosphere. 
It can appear in emission above the limb despite the low temperatures of the gas ($\sim$3600~K), because thermally emitting molecules are seen against the very dark IR sky at heights above the limb where the 5~$\mu$m photospheric continuum has become transparent.  
\citet{ayres02} carried out over-the-limb 2-D spherical  transport simulations of the CO 5~$\mu$m rovibrational spectrum in a variety of models including: (1)~the VAL laterally homogeneous reference model; (2)~multiple slices from the  1D time-dependent simulation by {\citet{1994chdy.conf...47C}}, with the 1D snapshots arrayed at random across the surface with a lateral dimension compatible with the dynamic ``K-grains'' observed prolifically in the quiet Sun ($\sim$1\,Mm); and (3)~various spatial combinations of the warm VAL thermal profile with cold temperature components in the low chromosphere region up to about 1000\,km, where the hot chromospheric canopy naturally destroys the molecules.  
In these simulations, the homogenous VAL temperature profile produces only very minor limb extensions of the CO~bands, coming entirely from the $T_{\rm min}$ region of that model.  
The {\citet{1994chdy.conf...47C}} variant did much better, because of cold regions in the higher altitude range due to adiabatic cooling associated with the acoustic pulses, producing significant CO limb emissions, although not as high above the limb as seen in observations.  
Models with VAL-like temperature profiles interspersed with small-scale cold regions, with temperatures below 3000\,K for most of the 500-1000\,km height range but with only 20\,\% covering fraction, did about as well as the 
{\citet{1994chdy.conf...47C}} 
model, but still did not reproduce the full spatial extension of the observed strong CO lines above the limb.  
Only a model in which the hot part of the VAL in the chromosphere was reduced to 20\,\% coverage, with the remaining 80\,\% allocated to a temperature profile with a cool dip in the chromospheric layers, although with minimum temperatures above 3000\,K, was able to fully reproduce the details of the CO off-limb emissions observed under conditions of very good seeing.  
These simulations seem to suggest that molecule-friendly conditions are pervasive in the low chromosphere, with hot gas being a relatively minor constituent.

The simulations, moreover, predict a very different appearance of the CO lines when observed on the disk, say in a heliogram taken in the core of a strong 5~$\mu$m CO line.  
The homogeneous VAL model of course would show a smooth featureless CO image across the surface, aside from large-scale slowly varying center-to-limb effects.  
The dynamic model by {\citet{1994chdy.conf...47C}} predicts a pattern of light and dark features across the surface, about equally mixed, with Doppler shifting of the CO lines in each coherent patch.  
Near the limb, however, the intermittent dark patches, tied together by the nearly horizontal light paths, dominate the scene because they are optically thicker in CO than the warmer regions.  
For the two VAL\,+\,COOLC (Òcool componentÓ) simulations, the 20\,\% COOLC version delivers a disk center image that is dominated by the brighter emission from the warm VAL component, but has conspicuous dark points from the intermittent but very cold regions, which become optically thick in CO in the cold ``COOLC1'' lower chromosphere.  
On the other hand, the 80\,\% ``COOLC0'' version shows a more bland image, brighter regions sprinkled between larger dark regions.  
The latter, however, are not as dark as for the minority COOLC1 case because the less extreme thermal profile of COOL0 has the disk-center CO lines becoming optically thick near the classical $T_{\rm min}$, where the gas is warmer, rather than higher up where the very cool temperatures occur.

These models admittedly are quite simplistic compared to the 3D dynamical simulations of these complex layers that have become available in recent years (see Sect~\ref{sec:quietsunmodel}).  
Nevertheless, the two-component VAL\,+\,COOLC simulations of on-disk and off-limb CO are instructive, and have important implications for ALMA.  
Namely, both types of models (minority cool mixed with majority hot, and vice versa) predict the off-limb extensions of CO, and the behavior close to the limb, with varying degrees of success, but much better than traditional 1D reference models.  
However, the two general classes of two-component simulations make very different predictions for a disk-center quiet-Sun CO-core heliogram.  
Ironically, perhaps, the minority cool component has the most obvious contrasty appearance, with small-scale very dark spots superimposed on a warmer background; whereas the majority cool component heliogram would appear much blander, with a pervasive, somewhat dark background with a sprinkling of small-scale not-so-bright points.  
In short, the cool regions are more visible when they are less important.  
In principle, contemporary CO imaging could settle the question straightforwardly, and indeed the best quality CO heliograms available do not reveal any obvious small-scale very dark points needed in the situation where the cool component is in the minority.  
However, that conclusion must be tempered by the fact that the spatial resolution of existing solar telescopes at 5~$\mu$m is only somewhat less than 1~Mm so that smaller-scale very dark points could have easily remained unresolved and thus undetected.  
The 4-m DKIST solar telescope will help resolve the question when it comes on line later this decade, but its resolution at 5~$\mu$m only will be less than a factor of 3, or so, better than existing 1.5-m-class solar instruments.

On the other hand, ALMA not only can image the complex layers of the low chromosphere in the sub-mm region with a spatial resolution much better than today's largest solar instruments in the thermal-IR, but also tune through those layers by observing at different frequencies.  
A single high spatial resolution time series of surface maps by ALMA at its highest frequencies will immediately decide whether cool gas in the lower chromosphere is pervasive, or not; and whether its existence is owed to intermittent dynamical phenomena, or instead perhaps to more stable pockets of gas that are not strongly heated for a time and suffer a ``molecular cooling catastrophe'', which plunges their internal temperatures to very low values  
\citep{ayres81}.

\begin{figure}
\centering
\resizebox{\textwidth}{!}
{\includegraphics[]{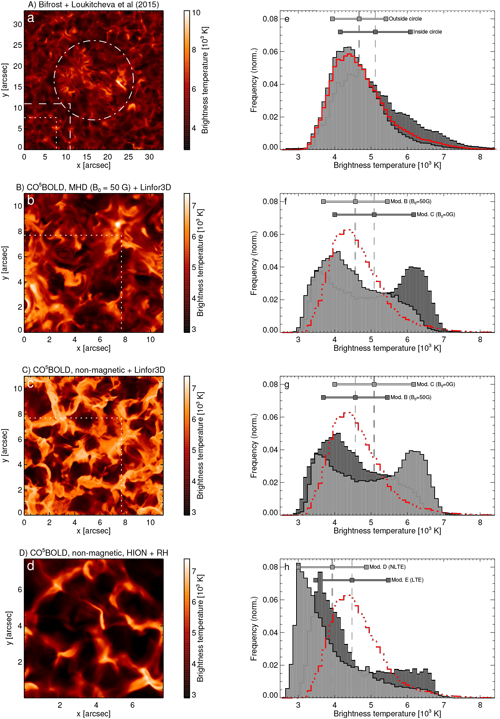}}
\caption{Synthetic brightness temperature at $\lambda = 1$\,mm for model 
snapshots from four different simulations (see Table~\ref{tab:quietsunmodels}). 
From top to bottom: A Bifrost simulation with an average magnetic field strength 
of \mbox{$<|\vec{B}|>_\mathrm{ph}\, = 50$\,G} in the photosphere (model~A), and CO$^5$BOLD 
simulations with an initial field strength of \mbox{$|\vec{B}|_0 = 50$\,G} (model~B), the 
corresponding non-magnetic model~C, and the smaller model~D, which takes into 
account time-dependent hydrogen ionisation. 
The upper left panel shows the primary beam size of an \mbox{12-m} ALMA antenna (i.e. the diffraction limit), which has a diameter of 21'' at $\lambda = 1$\,mm (dot-dashed circle) and the sizes of the smaller CO$^5$BOLD models (dashed rectangle for models~B and C~and dotted rectangle for model~D).
The corresponding brightness temperature histograms  are shown in the right column as light grey 
area. 
For comparison, a second distribution is displayed as a dark grey area. 
For model~A (panel~e), the histogram of the pixels inside the primary beam (the circle in panel~a, dark grey here) are compared to those outside (light grey), where the latter excludes the bright loops in the center. 
The distribution for the whole model is over-plotted as red thick line. 
In the panels~f and~g, the models~B and C are compared to each other. 
In panel~h, model~D (light grey, NLTE) is compared to model~E (dark grey, NLTE), which is equivalent to model~D but uses LTE electron densities instead of the results from the time-dependent hydrogen ionisation. 
The vertical lines mark the mean brightness temperature, whereas the horizontal lines with square symbols at the top of each panel represent the mean plus/minus one standard deviation. 
The histogram for the pixels outside the circle in model~A (i.e. without the magnetic loops in the center) is plotted as reference in panels~f-h (red triple-dot-dashed line). 
}
\label{fig:Tbquietsun}
\end{figure}

\begin{table}[bp!]
\caption{Properties of the considered numerical models and the resulting synthetic brightness temperature maps. 
Model~A has been computed with the MHD code Bifrost \citep{2011A&A...531A.154G} and 
models B-E with  CO$^5$BOLD \citep{2012JCoPh.231..919F}. 
The models have already been described in the following publications:  
L15 \citep{2015A&A...575A..15L}, 
W07 \citep{2007A&A...471..977W}, 
W12 \citep{2012Natur.486..505W}, 
W13 \citep{2013JPhCS.440a2005W}. 
}
The values in parentheses for model~A refer to the region outside the circle drawn in Fig.~\ref{fig:Tbquietsun}a. 
\begin{center}
\begin{tabular}{l|c|c|c|c|c}
\hline\noalign{\smallskip}
Model & A & B & C & D&E\\
\hline\noalign{\smallskip}
\hline\noalign{\smallskip}
Reference& L15  &  W12& W12, W13 & W07 & W07 \\
MHD code               &Bifrost  &
\hspace*{-1.5mm}CO$^5$BOLD\hspace*{-1.5mm}&
\hspace*{-1.5mm}CO$^5$BOLD\hspace*{-1.5mm}&
\hspace*{-1.5mm}CO$^5$BOLD\hspace*{-1.5mm}&
\hspace*{-1.5mm}CO$^5$BOLD\hspace*{-1.5mm}\\
Radiative transfer code&L15  &Linfor3D  &Linfor3D  &RH&RH\\
Horizontal cells                          &  504\,$\times$\,504& 286\,$\times$\,286&286\,$\times$\,286&140\,$\times$\,140&140\,$\times$\,140\\
Horizontal extent [Mm] & 24.1\,$\times$\,24.1 & 8.0\,$\times$\,8.0 & 8.0\,$\times$\,8.0 & 5.6\,$\times$\,5.6& 5.6\,$\times$\,5.6 \\ 
Horizontal extent [arcsec] & 33.3\,$\times$\,33.3 & 11.0\,$\times$\,11.0 & 11.0\,$\times$\,11.0 & 7.7\,$\times$\,7.7& 7.7\,$\times$\,7.7\\
Initial / avg. photospheric & & & & \\
magnetic field strength           &50 & 50& 0& 0& 0 \\
Non-equilibrium H                 & & & & & \\
ionisation                                      &yes&no&no&yes&(yes)\\
\hline\noalign{\smallskip}
\multicolumn{5}{l}{Synthetic brightness temperature at $\lambda = 1$\,mm}\\
\hline\noalign{\smallskip}
Average            [K] &  4785.9  (4673.6)& 4568.6  &  5086.9& 3935.9  & 4476.5\\
Standard deviation [K] &   822.9   (736.2)&  879.7  &  1085.5&  934.6  &  998.2\\
Minimum            [K] &  2705.8  (3158.4)& 2966.6  &  3044.2& 2686.7  & 3109.9\\
Maximum            [K] & 11093.8  (8928.7)& 7481.3  &  7257.4& 7272.8  & 7081.7\\
Contrast               & 17.2\,\% (15.8\,\%)& 19.3\,\%&21.3\,\%& 23.7\,\%& 22.3\,\%\\
\hline\noalign{\smallskip}
\end{tabular}
\end{center}
\label{tab:quietsunmodels}
\end{table}%

\subsubsection{Numerical predictions of quiet Sun ALMA observations}
\label{sec:quietsunmodel}
%
In Figure~\ref{fig:Tbquietsun} and Table~\ref{tab:quietsunmodels}, we compare different 3D numerical models. 
Model~A was computed with the Bifrost code \citep{2011A&A...531A.154G} and is classified as `enhanced network' whereas the other models (B-D) were calculated with CO$^5$BOLD  \citep{2012JCoPh.231..919F} and represent quiet Sun regions. 
The synthetic brightness was produced with the radiative transfer codes LINFOR3D (based on the Kiel code LINFOR/LINLTE\footnote{LINFOR3D manual at http://www.aip.de/$\sim$mst/linfor3D\_main.html.}) and RH \citep{uitenbroek00b}, whereas the results for model~A were calculated in detail by \citet{2015A&A...575A..15L}. 
Model~A has a horizontal extent of 24.1\,Mm\,$\times$\,24.1\,Mm (33''.3\,$\times$\,33''.3) and reaches into the corona. 
The magnetic field strength in the photosphere is on average $<|\vec{B}|>_\mathrm{ph}\, = 50$\,G with two major regions of opposite polarity, which are roughly 8\,Mm apart.  
The photospheric patches of opposite polarity are connected by magnetic loops, which extend into the low corona. 
These magnetic loops are nevertheless visible in the corresponding synthetic brightness temperature map for a wavelength of $\lambda = 1$\,mm (see upper left panel in Fig.~\ref{fig:Tbquietsun}). 
The brightness temperature in these loops is higher than in the surrounding (quiet) chromosphere but only 0.3\,\% of all pixels have a brightness temperature higher than 8000\,K. 
The corresponding histograms are displayed in the lower left panel for the whole simulation and the outer four corners, which thus exclude most of the bright loops in the centre.  
The brightness temperatures range from only $\sim 2700$\,K to $\sim 11100$\,K with an average of $\sim 4790$\,K and a standard deviation of $\sim 820$\,K. 
The 1\,mm map is compared to a corresponding 3\,mm map and Mg\,II\,k line wing and line core images in Fig.~\ref{fig:bifrostmmvsmg}. 
The numerical model~A is described in more detail in \citet[][see also IRIS technical note~33, http://sdc.uio.no/search/simulations]{carlssonbifrostmodel}.

The other models have a smaller horizontal and vertical extent and therefore do not include large-scale magnetic loops like those in model~A. 
The maximum brightness temperature at $\lambda = 1$\,mm is therefore $7\,500$\,K or less in models B-D without the high-temperature tail in the histograms, which is caused by the hotter loops in model~A. 
Models~B, C, and~D nevertheless show differences due to different magnetic field, spatial extent and inclusion of non-equilibrium hydrogen ionisation. 
Models~B \citep{2012Natur.486..505W,2013JPhCS.440a2005W} and C \citep{2012JCoPh.231..919F} have the same spatial extent but model~B has an initial magnetic field strength of $|\vec{B}|_0 = 50$\,G whereas model~C is non-magnetic. 
The histograms of the brightness temperature at $\lambda = 1$\,mm are over-plotted for both models in the middle bottom panels of Fig.~\ref{fig:Tbquietsun} so that the influence of magnetic fields becomes directly visible. 
The pronounced small-scale mesh-like pattern, which is produced by the interaction of propagating shock waves in model~C (see third brightness temperature map in panel~c), is somewhat suppressed by the magnetic field structures in model~B. 
These field structures are rooted in kilo-Gauss foot points in the photosphere and expand into chromospheric funnels, which then influence the propagation of waves in the upper layers. 
Consequently, the peak in the histograms at high brightness temperatures due to shock fronts is pronounced for model~C but ablated in model~B. 
On the other hand, chromospheric swirls appear in the map for model~B (panel~b) and indicate the presence of magnetic tornadoes (see Sect.~\ref{sec:vortexflows}).

In Fig.~\ref{fig:Tbquietsun}d, the corresponding brightness temperature for the non-magnetic model~D is shown, which has the smallest horizontal extent of the models considered here. 
For model~D, the hydrogen ionisation is treated in detail by solving the time-dependent rate equations for a six level model atom with fixed radiative rates \citep{2006ASPC..354..306L}. 
The resulting non-equilibrium (non-LTE) electron density varies much less in the chromosphere than the corresponding LTE electron densities, which instantaneously follow any change in local gas temperature. 
The finite recombination rates in the non-equilibrium case effectively weaken the variations caused by the hot propagating shock waves.  
The brightness temperature distribution of the non-LTE case is compared to the corresponding LTE case, i.e. the same model but with equilibrium electron densities, which is referred to as model~E in Fig.~\ref{fig:Tbquietsun}h and Table~\ref{tab:quietsunmodels}. 
An analysis of the contribution functions of the continuum intensity reveals that the variation of the effective formation height for different spatial locations at a given wavelength in model~E is reduced to 60\,\% compared to model~D, demonstrating the importance of a realistic treatment of the hydrogen ionisation. 
The contribution functions for model~E are shown {in Fig.~\ref{fig:contf}.}  
According to \citet{2007A&A...471..977W}, the narrower continuum forming layer in model~E results in the reduction of the average brightness temperature in the resulting maps from 4,476\,K in model~D (i.e. with LTE electron densities)  to 3,935\,K for model~E (i.e. with the more realistic non-equilibrium electron densities).   
The standard deviation also decreases somewhat from $\delta T_\mathrm{b, LTE} = 998$\,K to $\delta T_\mathrm{b, NLTE} = 933$\,K. 
In both cases, the histograms exhibit a very pronounced peak at low temperatures. 
The high-temperature peak due to shock waves is reduced when non-LTE electron densities are considered, suggesting that ALMA will detect a considerable amount of cool gas that goes unnoticed by atomic lines in the visible and UV.

\begin{figure}
\centering
\resizebox{6cm}{!}
{\includegraphics[]{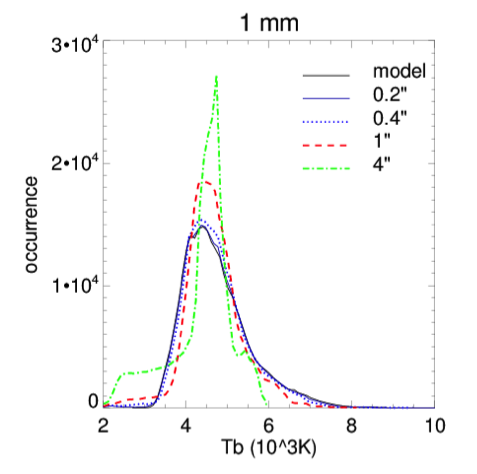}}
\caption{Spatial resolution effect on the intensity histogram for the wavelength of 1\,mm and four values of FWHM (spatial resolution): 0.''2 (solid), 0.''4 (dotted), 1'' (dashed), and 4'' (dot-dashed). }
\label{fig:spatresloukitcheva}
\end{figure}

Spatial resolution emerges as a critical problem for observations of the chromospheric fine-structure with ALMA. 
\citet{2015A&A...575A..15L} studied the effect of limited spatial resolution by applying artificial smearing to synthetic 
millimeter brightness temperature maps from a 3D snapshot of the aforementioned model~A. 
The minimum resolution in the model is defined by the horizontal extent of the grid cells (0.''064, cf.~Table~\ref{tab:quietsunmodels}). 
\citeauthor{2015A&A...575A..15L} found that the fine structure on the smallest spatial scales in the synthetic millimeter maps is increasingly lost when the resolution is reduced. 
At a resolution of 1'' the brightness temperature contrast decreased to 70-80\,\% of the original contrast value, depending on the wavelength. 
The fine structure on the smallest scales is already lost, while the larger-scale pattern is still distinguishable and closely reflects that in the original image. 
In the intensity histograms shown in Fig.~\ref{fig:spatresloukitcheva}, these effects are visible as narrowing of the histograms, disappearance of the high-brightness tail and growth of the low-brightness bulk. 
These findings are in line with the earlier study by \citet[][see their Fig.~10]{2007A&A...471..977W}. 
They state that the pattern on the smallest scales are hardly discernible at a resolution lower than 0.''3 whereas the larger mesh-like pattern due to shock waves can still be seen at a resolution of 0.''9. 
\citeauthor{2007A&A...471..977W} approximate the reduction of brightness temperature contrast $\delta T_\mathrm{b, rms} / \langle T_\mathrm{b} \rangle$ as function of angular resolution $\Delta \alpha$ with an exponential of the form  
\begin{equation} 
\frac{\delta T_\mathrm{b, rms}}{\langle T_\mathrm{b} \rangle } \propto  e^{-\Delta \alpha / D}\quad.
\end{equation}
The characteristic length scale $D$ of the chromospheric pattern is derived from fits to the contrast curves and depends on  wavelength and view angle. 
For a wavelength of 1\,mm at disk centre, for instance, the contrast reduces from 24\,\% to just 10\,\%, i.e. to about 40\,\% of the original value, when imposing artificial smearing with a spatial resolution corresponding to 1''.

\subsubsection{Magnetic fields in Quiet Sun regions}
\label{sec:qsmag}

\begin{figure}
\centering
\resizebox{\textwidth}{!}
{\includegraphics[]{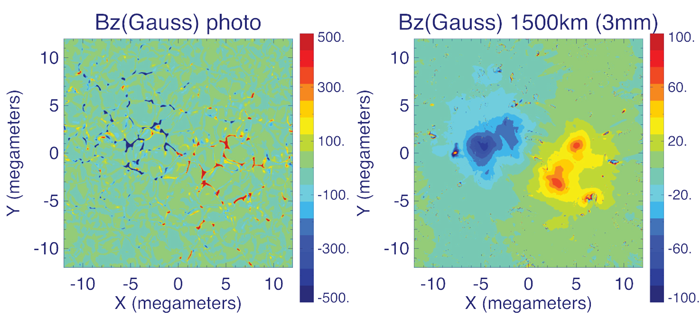}}
\caption{Longitudinal component of the magnetic field at the photospheric level (left) and as it is derived from synthetic millimeter brightness temperature maps at the chromospheric level (right).  
The latter roughly corresponds to a height of 1500\,km, which is where the emission at $\lambda = 3$\,mm mostly comes from in this case. 
The value range for photospheric magnetic field is limited to [$-500$\,G, $+500$\,G] for clarity. 
The employed brightness temperature maps are calculated based on a snapshot of the  3D RMHD simulations done with the Bifrost code 
\citep{2011A&A...531A.154G}. 
}
\label{fig:magneticfield}
\end{figure}

\citet{2015arXiv150805686L} carried out tests of the technique described in Sect.~\ref{sec:magneticfield} for ALMA frequencies by using a  3D simulation snapshot from the Bifrost code \citep{2011A&A...531A.154G} 
like the one described in Sect.~\ref{sec:quietsunmodel}. 
For a region of enhanced network with maximum magnetic field of 2\,kG at the photosphere (left panel in Fig.~\ref{fig:magneticfield}) and its rapid decrease with the height, the simulated circular polarisation at 3\,mm falls within $\pm 0.5$\,\%, which results in the restored longitudinal magnetic field within $\pm\,100$\,G at the effective formation height. 
{The (expected) polarisation is thus higher than the technical requirement of $0.1$\,\% for ALMA (see Sect.~\ref{sec:almamagneticfield}) and could thus be used to measure the magnetic field in the quiet Sun chromosphere. 
}

An example of the magnetic field strength at the height of approximately 1500\,km, as it is restored from the simulated polarised emission at $\lambda = 3$\,mm, is shown in the right panel of Fig.~\ref{fig:magneticfield}. 
For this example, the difference between the effective formation heights of the two normal modes does not exceed 2\,km at $\lambda = 3$\,mm and is only slightly greater than 3\,km for $\lambda = 10$\,mm.
The assumption of a constant magnetic field, which is implicitly made for this method, is valid for such small separations. 
Consequently, the method reproduces the longitudinal magnetic field with the ideal model data reasonably well. 
However, further analysis, e.g. involving realistic noise, is needed to fully validate the approach. 
Corresponding results for predicting magnetic fields in active regions are discussed in Sect.~\ref{sec:armodelling}.

\subsubsection{Vortex flows}
\label{sec:vortexflows}

A magnetic vortex flow is a fundamental physical process in the solar atmosphere, which forms when the photospheric footpoint of a magnetic field structure is swept into a vortex flow in an intergranular lanes at the solar surface
(see Fig.~\ref{fig:quietsunscheme}). 
The primary photospheric vortex is a direct consequence of the conservation of angular momentum of the downflowing plasma in the intergranular lanes (`bathtub effect') and thus an integral part of stellar surface convection \citep[see, e.g.,][and references therein]{2008ApJ...687L.131B,2011ApJ...727L..50K,2011A&A...533A.126M,2012A&A...541A..68M}. 
The rotation of the magnetic footpoint is mediated into the atmosphere above, where the plasma is forced to follow the rotating field structure \citep{2014PASJ...66S..10W}. 
There, it forms a observable signature, namely a `chromospheric swirl' \citep{2009A&A...507L...9W}
This kind of event, which is also referred to as a `magnetic tornado', may serve as a channel for energy and matter into the chromosphere and corona of the Sun and has therefore a potential contribution to heating the upper solar atmosphere \citep{2012Natur.486..505W}.
Based on measurements with the Swedish 1-m Solar Telescope, chromospheric swirls have lifetimes of ($12.7 \pm 4.0$)~min and diameters of $(4.0 \pm 1.4$~arcsec  (i.e, ($2900 \pm 1000$)\,km), whereas the width of substructure features (e.g., rings or spiral arms) can be as small as $<$\,0.''2. 
{ALMA observations of swirls (cf. Fig.~\ref{fig:Tbquietsun}b) would provide a direct measure of the gas temperatures within these structures and thus a mean to further investigate their role for the chromospheric energy transport and the possibly resulting heating of the outer solar layers.}

Rotating magnetic field structures are observed on an extended range of different spatial scales. 
On the Sun, it ranges from torsional motions in spicules \citep{2012ApJ...752L..12D} to very large vortices, which appear in connection with solar prominences \citep[e.g.][]{2012ApJ...752L..22L}. 
It is currently debated if the ``legs'' of prominences, also referred to as `giant solar tornadoes', are rotating, too \citep[e.g.,][]{2013ApJ...774..123W}. 
See Sect.~\ref{sec:prominences} for more information on solar prominences
{and Sect.~\ref{sec:magvortexscale} for potential implications for other fields of astrophysics.}

\subsubsection{Modelling of ion-neutral collisions} 
\label{sec:ionneutral}
%
State-of-the-art numerical simulations so far employ a single-fluid (ideal) MHD description, which is strictly speaking only valid if the relevant time scales are larger than the ion-neutral collision time. 
The latter is very short in the photosphere ($10^{-8} - 10^{-9}$\,s) and the lower  chromosphere ($10^{-5} - 10^{-7}$\,s). 
Therefore, in this part of the solar atmosphere the single-fluid description perfectly  describes the processes in the limits of spatial and temporal resolutions of ALMA. 
In middle and higher chromosphere, however, the chromospheric plasma is partially ionised with the ionisation degree varying significantly \citep{2002ApJ...572..626C}. 
The collision time between ions and neutral atoms (especially, neutral helium) is dramatically increased in the upper chromosphere and may reach the scales of ALMA observations ($\sim 1$\,s for proton-neutral helium collision). 
Therefore, the validity of the single-fluid approach has to be critically checked through comparisons of simulation results with ALMA observations. 
Such studies may motivate moving forward towards a multi-fluid MHD approach \citep[see, e.g.][]{2011A&A...529A..82Z}, which is ultimately wanted for a realistic modelling of the propagation and dissipation of Alfv{\'e}n waves in the solar atmosphere (see  Sect.~\ref{sec:alfvenwaves}).

\subsubsection{Center-to-limb variation and spicules}
%
Observations in the mm and cm range show less limb brightening than expected from the increase of the temperature with height \citep{1972A&A....21..119L,1993ApJ...415..364B,2013PASJ...65S..19K}
and this cannot be interpreted in terms of the instrumental resolution.  
It is usually explained as the effect of absorbing features, and spicules have been implicated to this end. In order for spicules to absorb, they have to be cooler than the ambient medium, which is certainly true for the transition region and the corona into which they penetrate, but not necessarily so in the chromosphere 
\citep[see, e.g.,][]{2012SSRv..169..181T}. 

{
In the 2D atmospheric model proposed by \citet{2005A&A...433..365S}, the authors included the presence of spicules to explain the low equatorial limb brightening ($\sim10$\%) observed at 17\,GHz. 
On the other hand, the high values observed at the poles were modeled by spicules-free regions associated with polar faculae \citep{2005A&A...440..367S}.
}

In order to measure the center-to-limb variation with ALMA the TP system must be fully implemented; then one can measure the brightness of both the network structures and the background at various positions on the disk and at the limb. 
Even without TP, the brightness difference between network and intranetwork  can be measured and this is an important diagnostic. Spicules themselves and the corresponding disk structures will be hard to observe, because their short lifetime \citep[$\sim15$ min for normal spicules, $\sim2$ min for type II spicules; see][]{2012SSRv..169..181T} will probably not allow sufficiently dense coverage of the {u-v} plane to properly image the complex structure 
{with the bright limb} 
in the field of view. 

\subsubsection{Polar brightenings}
\label{sec:polarbright} 
%
Radio observations have revealed the remarkable result that the polar regions of the Sun are brighter than the rest of the quiet Sun in a limited range of frequencies from 17 GHz to about 87 GHz 
{\citep[e.g.,][]{1980BCrAO..61...43E,1986PASJ...38....1K,1998ASPC..140..373S,1999ApJ...527..415N,2000A&AS..143..227P,2003A&A...401.1143S,2010A&A...509A..51S,2011ApJ...734...64S,2012ApJ...750L..42G,2014ApJ...780L..23N}. 
}
Most of the recent studies of the polar brightening have used data obtained by the Nobeyama  radioheliograph (NoRH) at 17\,GHz.
The NoRH images show that the polar brightening consists of two components: a diffuse component of 1500\,K excess brightness and patchy compact sources with localized excess brightness of about 3500\,K.  
The compact sources change significantly with time. 
While some weak short-lived features (lifetimes about 1 minute) are difficult to distinguish from side-lobes, some others live longer, do not move, and are probably real sources \citep{1999ApJ...527..415N}.

\citet{1998ASPC..140..373S} proposed that the polar brightening arises from a general limb brightening (due to an enhanced opacity in the line of sight) superposed on the bright features intrinsic to the polar region. 
However, the nature of the bright compact sources is largely unknown and they have no correspondence with small-scale bright regions in images from both the Extreme Ultraviolet Imaging Telescope (EIT) aboard the Solar and Heliospheric Observatory (SoHO) \citep{1999ApJ...527..415N} and the Atmospheric Imaging Assembly (AIA) aboard the Solar Dynamics Observatory (SDO) \citep{2014ApJ...780L..23N}.

{This lack of one-to-one correspondence between 17\,GHz and EUV bright regions could be due to the presence of radio optically thick features, like spicules, that are completely transparent at EUV lines 
\citep{2010A&A...509A..51S}
and this shadowing may extend onto the disk at the extreme limb \citep[e.g.,][]{1976ApJ...203..753L}.
}

The long-term variation of the polar brightening has also been studied \citep{1998ASPC..140..373S,2012ApJ...750L..42G,2014ApJ...780L..23N}. 
At 17\,GHz the polar brightening anti-correlates with the solar cycle as defined by the sunspot number and correlates with the polar magnetic field strength.   
In contrast, no polar brightening is apparent in the synoptic maps obtained by SSRT at 5.7\,GHz (Altyntsev 2015, private comm.). 
{\citet{2011ApJ...734...64S},} 
\citet{2012ApJ...750L..42G} and \citet{2014ApJ...780L..23N} have found that both the polar brightening and polar field strength were larger during the cycle 22/23 minimum than during the cycle 23/24 minimum. 
Concerning the occurrence of minimum polar brightening, they found that the north and south solar hemispheres were not tightly synchronized.

The study of the solar cycle dependence of the polar brightening is only meaningful using data from solar-dedicated instruments. 
However, ALMA  with its unprecedented spatial resolution and sensitivity has the potential to provide important new observations that will help us {understanding} the nature of polar brightening's compact sources.

\subsection{Active regions and sunspots} 
\label{sec:arasp}

\subsubsection{Active region modelling and predictions for ALMA}
\label{sec:armodelling}
\begin{figure}[t]
\centering
\resizebox{\textwidth}{!}
{\includegraphics[]{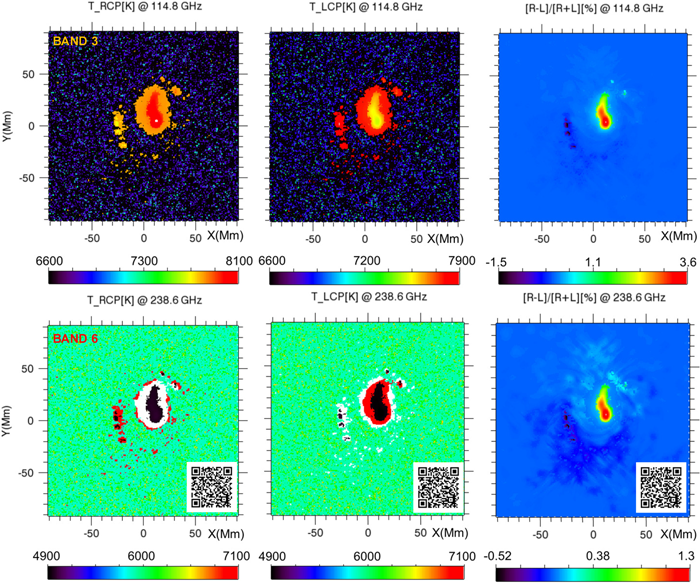}}
\caption{Examples of simulated images for AR~12158 in two ALMA frequency bands (identified at the top-left of the leftmost images). 
From left to right the plots display brightness temperature in right circularly polarized light, brightness temperature in left circularly polarized light and the relative difference between the two polarisations.
Note that the umbra appears bright (hot) at lower frequencies but dark (cool) at higher frequencies. 
The degree of polarisation along with the spatially resolved brightness temperature spectra is the key to obtaining the magnetic field map at a given frequency, which translates to a certain (range of) chromospheric heights. 
Multi-frequency data offer an elegant chromospheric tomography method \citep[see, e.g.,][]{2007A&A...471..977W,2014A&A...561A.133L,2015arXiv150805686L}. 
{The QR codes provide} links to on-line movies.
}
\label{fig:simactiveregion}
\end{figure}

As has been detailed in Sect.~\ref{sec:quietsun}, the chromosphere is extremely difficult to model in the quiet Sun, let alone in active regions where the magnetic field becomes even stronger and more highly structured. 
Detailed modeling of the microwave emission from active regions 
{\citep[see, e.g.,][and references therein]{1980A&A....82...30A,2015ApJ...805...93W}}
showed the importance of the gyroresonance mechanism  (see Sect.~\ref{sec:gyroemission}) and its potential role as an accurate diagnostic of the magnetic field. 
{In the ALMA spectral range, however, the gyroresonant emission is unimportant in the vast majority of the cases, while the free-free emission makes the dominant contribution (see Sect.~\ref{sec:mmradiation} for details).
}

To support solar ALMA observations of active regions, 
{\citet{2015arXiv150608395F}} 
developed an efficient algorithm integrated in the  3D modeling tool GX~Simulator {\citep[][available in SolarSoft]{2015ApJ...799..236N}}, facilitating the quick computation of  synthetic maps in the ALMA spectral range which give a rough idea of how a given active region will look at \mbox{(sub-)}millimeter wavelengths.  
Based on photospheric white-light images and magnetograms of a selected active region, an appropriate 1D model atmosphere is chosen for each pixel, i.e., for each spatial position in the image. 
The algorithm employs the set of 1D static model atmospheres by {\citet{2009ApJ...707..482F,2014SoPh..289..515F},}  which describe averaged atmospheric stratifications corresponding to different types of features in the photosphere {including} umbra, penumbra, network, internetwork, enhanced network, facula, or plage.
Each model provides the height stratification of {all relevant} physical quantities {including} 1D distributions of the electron temperature and density together with (non-LTE) ionized and neutral hydrogen densities. 
After the appropriate model is chosen, the emergent millimeter radiation is computed for each pixel accordingly. 
The algorithm accurately takes into account gyroresonance (GR) and free-free processes, {including} magnetized plasma dispersion and frequency-dependent mode coupling \citep{2014ApJ...781...77F},  non-LTE hydrogen ionization, all heavier singly-ionized atoms, the helium ionization state, and emission due to collisions of electrons with neutral hydrogen and helium (H$^-$ and He$^-$ opacities).

Some results of this modeling for AR~12158 (observed 10 Sep 2014) are shown in Fig.~\ref{fig:simactiveregion} for two frequencies representing the currently operating ALMA bands~3 and~6. 
This modeling suggests that different  features will be distinguishable at the ALMA frequencies with the brightness temperature contrast up to a few hundred~K. 
Interestingly, in line with the results of \citet{2014A&A...561A.133L}, the umbra, which is dark (cool) at high frequencies as well as in white light, appears bright (hot) at  lower frequencies. 
In addition, the brightness temperature noticeably depends on the magnetic field value due to the dependence of the free-free opacity on the magnetic field, which can be exploited for magnetic field measurements (see Sect.~\ref{sec:magneticfield}). 
Obviously, this simplified modeling is only a first step towards more realistic, eventually  3D models. 
In particular,  dynamic small-scale processes, which can have significant impact on the atmospheric structure, can by nature not be described with static 1D models (see also the similar discussion for quiet Sun regions in Sect.~\ref{sec:quietsunthermal}). 
On the other hand, a consistent 3D numerical model of a whole active region with sufficient spatial resolution is computationally not feasible with the resources available today. 
An alternative approximative approach would be to use a systematic grid of smaller 3D MHD models, which represent different sub-regions in an AR, ranging from weak-field regions to strong magnetic field concentrations. 
Based on magnetograms of real ARs, like those obtained with SDO/HMI, these different models can then be used as building blocks to fill a larger computational domain.

In a non-flaring active region the degree of circular polarisation reaches a few percent at millimeter wavelengths {\citep{2015arXiv150608395F}}, which allows reliable determination of the longitudinal component of chromospheric magnetic field from ALMA  observations. 
Examples of AR polarized emission maps in ALMA frequency bands (left circular polarisation (LCP) and right circular polarisation (RCP)) together with the degree of circular polarisation (P), reaching 3.6\,\% at 3\,mm, is shown in Fig.~\ref{fig:simactiveregion}. 
\label{sec:armagnetometryothertelescopes}
For more advanced AR magnetometry a combination of ALMA observational data with gyroresonant emission measurements in the microwave range available from the Very Large Array (VLA) and, in near future, from the Expanded Owens Valley Solar Array (EOVSA), the Siberian Solar Radio Telescope (SSRT), and the Chinese Solar Radio Telescope (CSRT), is desirable. 
The advantage of such a combination is that the gyroresonant emission yields the absolute value of the magnetic field at the transition region level {\citep[e.g.,][]{1994ApJ...420..903G,2011ApJ...728....1T,2015ApJ...805...93W},} which complements the line-of-sight magnetic field component measurements made with the free-free emission.
However, co-observing with some of these facilities might be limited given the geographic location of ALMA and resulting time differences and, in some cases, might be limited to studies of slowly varying features.

\subsubsection{Structure and dynamics of sunspot umbrae}
\label{sec:sunspotumbrae}

Although sunspot umbrae are well-studied, there are still a number of important issues that have defied attempts at understanding them \citep[see, e.g., the review by][]{2003A&ARv..11..153S}.  
Among these is the rather basic question of the temperature structure of the chromosphere above sunspot umbrae and the heating mechanism producing it. 
Models differ significantly from each other, and also the amount of heating required 
\citep{2014A&A...561A.133L}. 
Studies based on data gathered at visible wavelengths give controversial results regarding the height at which the chromospheric temperature rise starts in umbrae. 
\citet{2014A&A...561A.133L} have shown that it is possible using millimeter-wavelength observations to distinguish between the different models. 
This approach was so far hampered by the  low spatial resolution of available mm observations. 
This problem will be easily overcome by ALMA, even if not used in its highest resolution configuration. 
Furthermore, the ability of ALMA to observe at different wavelengths and thus to sample different atmospheric layers will facilitate exact measurements of the thermal structure above sunspot umbrae, {including} determinations of the exact height of the temperature minimum layer.

Sunspot umbrae have turned out to be far more complex than previously thought with a wealth of (often hidden) fine structure and dynamical phenomena \citep[see, e.g.,][]{2013ApJ...776...56R,2015ApJ...798..136Y}, such as umbral dots \citep{1969SoPh....7..351B,1969SoPh....7..366W,2013A&A...554A..53R}, umbral flashes \citep{1969SoPh....7..351B,2013A&A...556A.115D}, umbral oscillations \citep{2014A&A...569A..72S}, umbral microjets \citep{2013A&A...552L...1B}, light bridges of different types and their oscillations \citep[e.g.,][]{1997ApJ...490..458R,2014A&A...568A..60L,2014ApJ...792...41Y}, jets emanating from light bridges \citep{2011ApJ...738...83S} as well as umbral brightenings which are footpoints of hot coronal loops \citep{2013A&A...556A..79A}.  
Many of these phenomena (flashes, oscillations, microjets and light-bridge jets) are clearly chromospheric phenomena whose study will greatly profit from the ability of ALMA to trace different temperature regimes and heights as well as from the extremely high cadence that ALMA provides. 
The high cadence allows the evolution of these often short-lived phenomena to be followed in detail.

\subsubsection{Umbral oscillations and penumbral waves}
\label{sec:penumbralwaves}
%
Umbral oscillations \citep[which are actually waves, see][]{1992SoPh..138...93A}  were detected in the microwave range by \citet{1999SoPh..185..177G}  in Nobeyama circular polarisation data at 17\,GHz; they were spatially resolved in VLA observations both in Stokes~I and V at 8.5\,GHz and 5\,GHz 
\citep{2002A&A...386..658N} where they appeared as localized, intermittent oscillations.  
\citet{2012ApJ...746..119R} analyzed a long time series of simultaneous NoRH and AIA observations and determined time lags and spectral characteristics of umbral oscillations, which were consistent with those of upward propagating magnetoacoustic gravity waves, with the magnetic field acting as a waveguide up to the $\sim$1\,MK level.

Penumbral waves, which may \citep{2000A&A...355..375T} or may not \citep{2001A&A...375..617C}  be an extension of umbral waves have not been detected in the radio range, probably due to  the very intense sunspot-associated emission and/or the scarcity of high resolution observations. 
Recent observations \citep[e.g.][]{2015ApJ...800..129M} are consistent with the picture that umbral and penumbral waves have a common origin and propagate upwards along field lines the inclination of which varies between the sunspot center to the penumbra.

Both umbral oscillations and penumbral waves are expected to be associated with density variations that produce changes in the {free-free} emission detectable by ALMA; having a relatively simple spatial structure, they should be readily observable in snapshot mode.
Such observations will offer accurate diagnostics of the physical condition in a range of heights that will be extremely useful in modeling their behavior.

\subsubsection{Small-scale dynamic events in sunspot penumbrae}
\label{sec:eventspenum}

\begin{figure}[t]
\centering
\includegraphics[width=11cm]{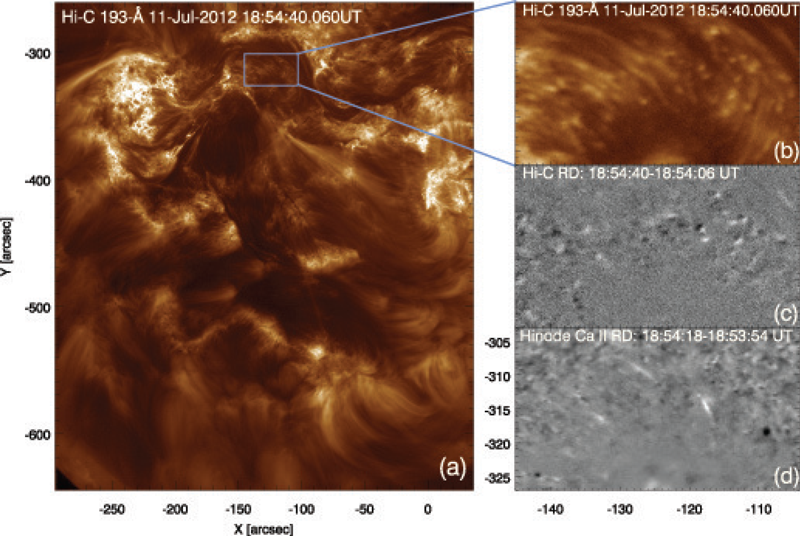}
\caption{Hi-C observation at a wavelength of 19.3\,nm taken on July~11, 2012.  
(a)~Full field-of-view. 
(b)~Close-up region (blue box in panel~a) showing an enlarged part of a sunspot penumbra. 
(c)~Corresponding running difference (RD) image of the close-up region showing many penumbral bright dots \citep{alpert15}, which also can be identified in panel~b. 
(d)~A running difference image in the Ca\,II\,H line obtained by Hinode's filtergraph for the same field of view, exhibiting a few penumbral microjets (seen as white streaks). 
These jets are studied in detail by \citet{2015arXiv151107900T}.
}
\label{fig:hicfg}
\end{figure}

Modern ground- and space-based telescopes have revealed the presence of many dynamic events in and above sunspot penumbrae, namely, running penumbral waves, localized up- and downflows, inward motion of bright penumbral grains, chromospheric microjets \citep{2007Sci...318.1594K}, and moving bright dots in the transition region \citep{2014ApJ...790L..29T} and in the corona \citep{alpert15}. 
Here we focus mainly on the two latest discoveries of small-scale dynamic events in penumbrae: penumbral microjets (PMJs), and penumbral bright dots (BDs). 
Examples of both types of events can be seen in Fig.~\ref{fig:hicfg}b-d.

It was thought earlier that penumbral microjets could be generated by the reconnection between the two magnetic components of penumbra, namely  spines (with more vertical field) and interspines (with more horizontal field). 
Recently, \citet{2015arXiv151107900T} proposed a modified formation mechanism of penumbral microjets based on the investigation of the small-scale structure of penumbral filaments by \citet{2013A&A...557A..25T}, who performed a spatially coupled inversion \citep{2012A&A...548A...5V,2013A&A...557A..24V}
of  spectropolarimetric data of a sunspot observed with the Spectropolarimeter (SP) of the Solar Optical Telescope \citep[SOT,][]{2008SoPh..249..233I,2008SoPh..249..167T} onboard the Hinode spacecraft. 
\citet{2015arXiv151107900T} also searched for coronal and transition region signatures of penumbral microjets but hardly found such signatures except for some larger penumbral jets, which appeared repeatedly at the same locations in chromospheric Ca\,II\,H line images.
This difference and the origin of both, the `normal-sized' and the larger jets, remains elusive. 
Based on the results of \citet{2015arXiv151107900T} and in line with the earlier proposal by \citet{2010ApJ...715L..40M}, it seems reasonable to assume that both types of jet are produced by magnetic reconnection events occurring between spines and opposite polarity fields, which are observed near the heads of penumbral filaments. 
However, only one third of the filaments studied by \citet{2013A&A...557A..25T} displayed opposite polarity fields near their heads. 
On the other hand, much of the internal structure of penumbral filaments might have remained undetected due to the limited spatial resolution of the employed instruments.   
To explore the small-scale internal structure of sunspot penumbrae, observations with very high spatial resolution are therefore crucial. 
In this respect, tracking dynamic features at high temporal cadence at (subarcsecond) angular resolution with ALMA could provide important clues on the formation mechanism and evolution of penumbral microjets.  
Ideally, such observations should be carried out simultaneously with observations with other space-borne telescopes such as, e.g., Hinode and the proposed \mbox{Solar-C}.

Penumbral bright dots, which are another kind of dynamic event in sunspot penumbrae, move inward or outward along  penumbral filaments. 
Such events were first observed with the Interface Region Imaging Spectrograph (IRIS) and then also with the High-resolution Coronal Imager \citep[Hi-C,][]{2014SoPh..289.4393K} in a narrow passband range centered at a wavelength of 19.3\,nm \citep{alpert15}.   
See the Hi-C image in Fig.~\ref{fig:hicfg} for some examples. 
The fact that the bright dots are detected by IRIS and in the 19.3\,nm passband of Hi-C, which includes some transition region emission, implies that these events originate in the transition region or possibly even lower in the atmosphere, i.e. in the chromosphere. 
The light curves of some of the bright dots observed with Hi-C indeed clearly indicate their origin in the transition region, while the light curves for many other events remain inconclusive. 
Among the various possible explanations of the origin of penumbral bright dots are, e.g., magnetic reconnection in inclined penumbral magnetic fields, shocks produced by downward flowing gas, and reconnection events caused by penumbral waves \citep{2014ApJ...790L..29T,alpert15}. 
Higher cadence and spatial resolution observations with ALMA can certainly provide important constraints on the so far elusive origin of  penumbral bright dots.

\subsubsection{Ellerman Bombs}
%
Ellerman Bombs (EBs) are usually identified as prominent small-scale brightenings in the wings of the H$\alpha$ line \citep{1917ApJ....46..298E}. 
They have mean lifetimes of \mbox{$\sim 10-15$~min} and are mainly observed in the vicinity of active regions \citep{2002ApJ...575..506G} or regions of emerging magnetic flux \citep{2008ApJ...684..736W} and do not produce any observable effects in the transition region or corona \citep{2013ApJ...774...32V,2015ApJ...798...19N}. 
These events are generally considered to be due to reconnection in the upper photosphere, which leads to increased temperatures therefore explaining the enhanced H$\alpha$~wings \citep[see e.g.][]{2013JPhCS.440a2007R}. 
High cadence coupled with high spatial resolution data from ALMA, coupled with its ability to observe different chromospheric layers can help to answer the question regarding the contribution to the chromosphere.
For example, do these small-scale, explosive phenomena in the lower solar atmosphere act as agents for triggering local instabilities in the magnetic environment of the solar surface?

\subsubsection{Explosive Events}

\begin{figure}
\centering
\resizebox{11cm}{!}
{\includegraphics[]{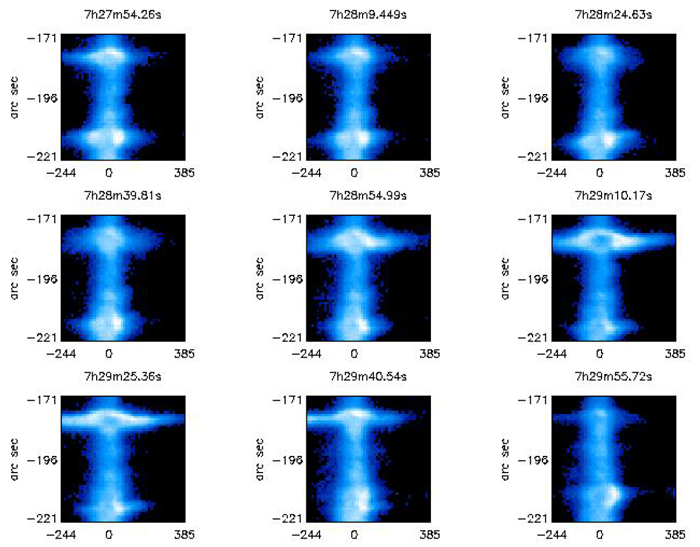}}
\caption{SUMER O\,VI~103.7\,nm slit data showing both blue and red shifts events (commonly termed explosive events) over a two minute interval. 
From \citet{1999A&A...351.1139P}.
}
\label{fig:explosiveevent}
\end{figure}

Another feature that ALMA is ideally suited for are the so-called ``explosive events'' (EEs, see Fig.~\ref{fig:explosiveevent}). 
These are characterized by non-Gaussian line profiles showing Doppler velocities of \mbox{$50\,-\,150$\,km\,s$^{-1}$} \citep{1983ApJ...272..329B}. 
However, despite many attempts their true nature remains unknown, although there is strong evidence that they are produced by magnetic reconnection \citep{1997Natur.386..811I}. 
Recently, \citet{2014ApJ...797...88H} using the Interface Region Imaging Spectrograph (IRIS) found that EEs are associated with very small-scale plasma ejection followed by retraction in the chromosphere. 
They observed a jet in a single IRIS pixel and conclude that it is thus only $\lesssim 120$\,km wide. 
These small-scale jets originate from a compact bright-point-like structure of $\sim 1.''5$ size as seen in the IRIS 133.0\,nm images. 
With ALMA's spatial resolution and cadence we may at last be able to resolve and identify the nature of these events and possibly answer some of the many open questions in this respect:  
Which physical processes generate the observed up- and down-flows and rotation in Explosive Events and which processes do produce these events in the first place? 
And do Explosive Events contribute directly or indirectly to the mass and energy transfer in the solar atmosphere?

\subsection{Solar flares} 
\label{sec:flares}

Solar flare radio emission represents transient enhancements  on top of a more slowly varying 
background  \citep[e.g.][]{1965sra..book.....K,2004ApJ...603L.121K}. 
The great sensitivity and precision of ALMA means that we can study even the weakest events, and probe them throughout the corona and lower atmosphere. 
We discuss the possibilities in separate sections here for major events, the microflare-nanoflare domain, and the flaring lower atmosphere. 
See \citet{2013A&ARv..21...58K} for a recent review of observations and theory.

\subsubsection{Major events}
\label{sec:majorflares}

\begin{figure}
\centering
\resizebox{\textwidth}{!}{\includegraphics[]{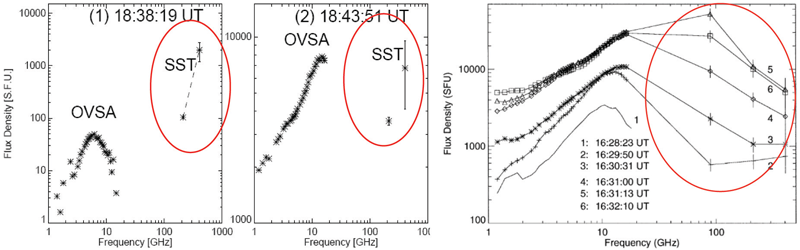}}
\caption{\textit{Left and middle panel:} broadband (1-405\,GHz) spectra of a solar flare on 2006~Dec~6 for two different time frames \citep{2009SoPh..255..131K}. 
The data were obtained with the Owens Valley Solar Array (OVSA) and the Solar Submillimeter Telescope (SST).   
A mysterious distinct sub-THz spectral component is apparent at 212\,GHz and 405\,GHz. 
\textit{Rightmost panel:}  broadband (1-405\,GHz) spectra of another solar flare on 2001~Aug~25 for a few consecutive time frames  \citep{2004SoPh..223..181R}. 
In a great contrast with the left panels, the radio spectrum of this flare represents a single, presumably, gyrosynchrotron spectrum with unusually high spectral peak frequency, around 100\,GHz at the peak time, while the spectrum decreases at 212\,GHz and 405\,GHz.
}
\label{fig:flare_subTHz}
\end{figure}

\edt{The microwave emission during flares comes primarily from mildly relativistic} electrons
via the gyrosynchrotron emission mechanism, with some role for thermal free-free emission.
For a typical flaring loop filled with accelerated electrons, the standard gyrosynchrotron spectrum
(see Sect.~\ref{sec:gyroemission})  reaches a peak around 5-20\,GHz \citep{1975SoPh...44..155G,2004ApJ...605..528N} and decreases towards THz frequencies with a roughly power-law spectrum, whose spectral index is specified by the fast electron energy spectrum and their pitch-angle anisotropy \citep{1968Ap&SS...2..171M,1969ApJ...158..753R,2003ApJ...587..823F,2011ApJ...742...87K}. 
In contrast, dense thermal plasma, which, for example, can be created as a result of chromospheric evaporation in response on the flare energy deposition, produces a flat flux-density  \edt{continuum} in its optically thin region and a Rayleigh-Jeans  \edt{continuum}  at longer wavelengths.
\edt{In some cases, available sub-THz observations do show either a decreasing spectrum, consistent with the non-thermal gyrosynchrotron mechanism, or an increasing continuum, consistent with the optically thick radiation.}  
From this perspective, the discovery \citep{2001ApJ...548L..95K,2002A&A...381..694T,2004A&A...420..361L,2004ApJ...603L.121K,2008A&A...492..215C,2009ApJ...697..420K} of a  \edt{continuum increasing with frequency and an apparently \textit{non-thermal} component in the sub-THz range, turned out to be entirely unexpected
(as in Fig.~\ref{fig:flare_subTHz}, left panel) even though entirely reasonable for a thermal one 
\citep{1975SoPh...43..405O}.} 
\edt{On a non-thermal hypothesis,} it is currently unclear what particles (electrons, positrons, or maybe protons) are responsible for this spectral component, nor what emission mechanism is at work. 
Accordingly many ideas have been proposed recently \citep{2010ApJ...709L.127F,2013A&ARv..21...58K,2014SoPh..289.3017Z,2014ApJ...791...31K}.  
Nevertheless, no consensus has been reached concerning the relevant emission process yet since the currently available observations are made at only a few (in most cases, only two) widely separated frequencies with very limited spatial information.
In order to distinguish the thermal and non-thermal emissions, the brightness temperature has to be measured across the whole source.
The source sizes are estimated to be about 10-20" 
\citep[see, e.g.,][for observations in different wavelength ranges]{1996ApJ...458L..49S,1999ApJ...522..547R,2014ApJ...787...15A,2015JGRA..120.4155K} and thus are on the order of the ALMA primary beam width. 
Based on the existing data, including the limited infrared observations now available \citep[e.g.,][]{2012ApJ...750L...7X,2015JGRA..120.4155K,2015SoPh..290.2809T} we expect to observe structure on all scales available to ALMA, and at all wavelengths. 
Interferometric observations with ALMA are therefore expected to greatly clarify the situation by adding the much-needed spectral and spatial information in the sub-THz range and thus enabling distinguishing between the thermal and non-thermal emission processes and the context association of the sub-THz source with other existing structures. 
\edt{As has been said, at least three distinct emission mechanisms are expected to make a contribution in the sub-THz emission from solar flares: thermal free-free, non-thermal gyrosynchrotron, and a non-thermal ÔmysteriousÕ rising component. 
In principle, all these components may appear together, but presumably from distinct spatial locations, which calls for  ALMA imaging of potentially complex sources. 
}
This complexity requires also the development of comprehensive modeling in three dimensions \citep{2015ApJ...799..236N}.

\begin{figure}
\centering
\resizebox{10cm}{!}
{\includegraphics[]{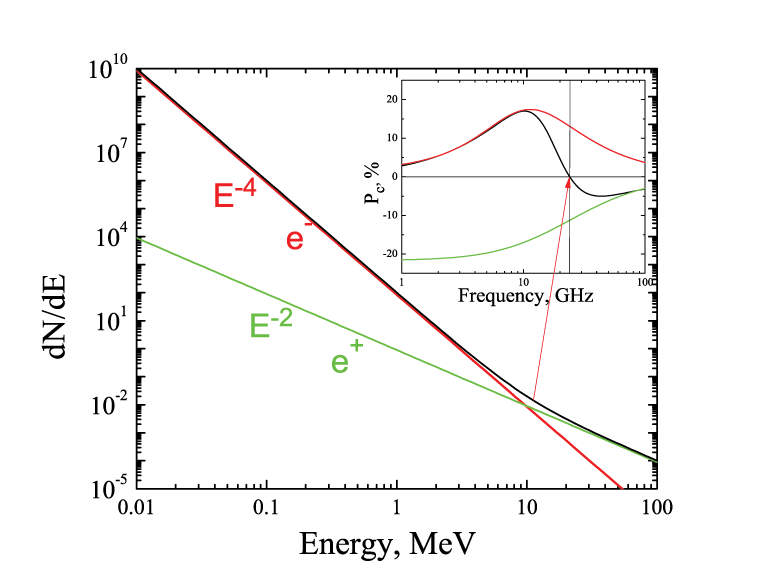}}
\caption{Illustration of the positron contribution effect on the polarisation of the {gyrosynchrotron} emission. Schematic energy spectra of fast electrons (red) and positrons (green) favourable for the high-energy positron detection are shown. The black line shows the composite spectrum dominated by electrons at low energies, but by positrons at high energies. 
The behaviour of the circular polarisation of the emission produced by these two components and by their composition is shown in the inset. 
A distinctive signature of the positron presence is the polarisation reversal at a high frequency shown by the red arrow, which happens because the positrons preferably produce the ordinary mode emission in contrast to the electrons, which produce extraordinary mode emission.
{Figure from \citet{2013PASJ...65S...7F}.} 
}
\label{fig:positron_cartoon}
\end{figure}

Interestingly, the yet-to-be-explained distinct sub-THz component does not happen in each and every flare. 
It might not be present even in the most powerful events, like that shown in Fig.~\ref{fig:flare_subTHz}, right. 
The emission is very strong and the spectrum peak frequency is very high (around 100\,GHz), but  no extra component is present in the sub-THz range. 
In such cases, ALMA will provide invaluable spectral information, which is the key to recovering the fast particle spectrum and other physical parameters of the flaring loops. 
\citet{2013SoPh..288..549G} employed a microwave spectrum in the range 1-18\,GHz to develop a forward fit diagnostics of the flaring loop parameters including the accelerated electron spectral slope. 
But they could neither recover the electron energy spectrum shape at the highest available energy nor determine the high-energy cut-off in that spectrum. 
However, this information is vitally needed to investigate the efficiency of the particle acceleration in flares and, thus, determine the responsible acceleration mechanism(s).
At those high frequencies the gyrosynchrotron emission is produced by relativistic particles, which can be either accelerated electrons or secondary positrons (created in nuclear interactions). 
In order to distinguish the positron and electron contributions, one has to measure the sense of circular polarisation along with direction of the magnetic field vector to determine the magneto-ionic mode dominating the emission-ordinary or extraordinary-to conclude in favor of positrons or electrons, respectively \citep[see Fig.~\ref{fig:positron_cartoon}; from][]{2013PASJ...65S...7F}.

So far, only indirect identifications of the wave mode are available with the Nobeyama data \citep{2013PASJ...65S...7F}.
ALMA with its broad spectral range, high spatial resolution, polarisation measurement purity (combined with the superior SDO vector magnetic measurements) will be an ideal tool for detecting the relativistic positron (and, thus, nuclear) component in solar flares --- powerful, but almost unexplored diagnostics.
\edt{In addition, after the positrons have become thermalized, they may form positronium, which might be detectable by ALMA. 
Thus, the evolution of the positrons from the original relativistic state to the final thermal state, when they interact with thermal electrons to form the positronium and eventually annihilate, can be nailed down. 
Then, detection of the thermal component from the flare can significantly constrain the energy balance, evolution, and partitions in flares. 
Although we cannot yet foresee what exactly will add the routine detection of the `mysterious' components, there is no doubt that ALMA observations of the flare emission will directly contribute to our understanding of the key questions about the solar flares.
}

\subsubsection{Microflares and nanoflares}
\label{sec:flaresmicronano}
%
The magnitudes of solar flares follow a power-law distribution over many decades \citep[e.g.][and references therein]{2012RSPTA.370.3217P}.
In this range the flare properties do not vary substantially \citep[e.g.][]{2004ApJ...612..530Q}, although there is a weak tendency for more energetic events to produce a harder (flatter) hard X-ray spectrum \citep{2005A&A...439..737B}. 
In moderately active regions, microflares with non-thermal energies in excess of $10^{26}$\,erg are found about once per hour \citep[e.g.,][]{2002SoPh..210..431B,2002SoPh..210..445K}. 
At microwaves, non-thermal gyrosynchrotron emission has been suggested for a number of events \citep[e.g.][]{1982A&A...107..178F,1997ApJ...477..958G,1999ApJ...513..983N,2006A&A...451..691K}.

Thermal flare signatures in the EUV down to a few times $10^{23}$\,ergs are found in quiet regions \citep{2000ApJ...529..554P}.
\citet{2000SoPh..191..341K} reported frequent enhancements related to such quiet region flares in radio waves.
Half of these events have a thermal radio spectrum (increasing with frequency). 
\citet{1999A&A...341..286B} found thermal signatures of quiet region flares at a wavelength of 2\,cm, originating mostly in the chromosphere.

The great sensitivity of ALMA, even in single-dish operation, guarantees abundant observations of these weaker events.
At the bottom of the energy scale, with energy releases of order a few $\times\ 10^{-9}$ those of major flares, we find the ``nanoflare'' concept.
This interesting idea plausibly explains the entirety of coronal heating via tiny episodic flare-like processes \citep{1988ApJ...330..474P}.
At high occurrence rates the nanoflares would be individually undetectable \citep[e.g.][]{1991SoPh..133..357H}, but indirect tests for their presence might reveal them; for lower rates of occurrence one could possibly observe particular events \citep[e.g.][]{2014Sci...346B.315T}.

\begin{figure}
\centering
\resizebox{10cm}{!}
{\includegraphics[]{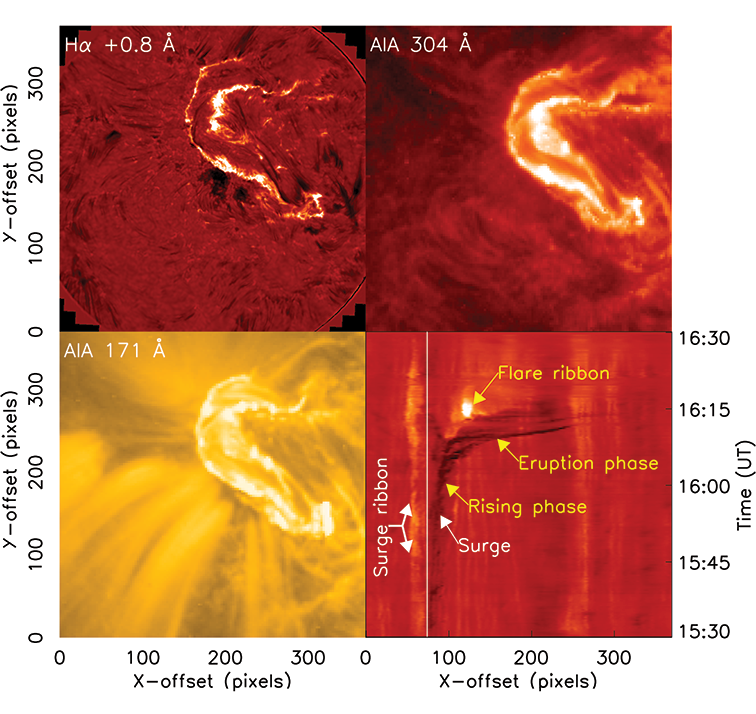}}
\vspace*{-3mm}
\caption{Simultaneous imaging of a C4.3-class flare from   11 Nov 2010 at different wavelengths. 
\textit{(Upper~left)}~Image taken in the wing of the H$\alpha$ line ($\Delta \lambda = 0.08$\,nm) at the Dunn Solar Telescope (NSO) with the IBIS instrument.
\textit{(Upper~right)}~AIA/SDO passband at 30.4\,nm. 
\textit{(Bottom~left)}~AIA/SDO passband at 17.1\,nm. 
\textit{(Bottom~right)}~A time-slice image in the H$\alpha$ showing the ribbon associated with the surge-like event, the rising filament and its eruption as a dark feature. 
All images comprise 400\,$\times$\,400~pixels with a pixel size of 0.''1. 
The FOV of each image is therefore 40''\,$\times$\,40''.
The images are based on the work by \citet{2014A&A...566A.148H}. 
}
\label{fig:halphaflare}
\end{figure}

\subsubsection{The lower atmosphere}
\label{sec:flarelowatmos}
%
The presence of broad H$\alpha$ line profiles at the outer edges of expanding flare ribbons provided early clues to the development of our current flare models, for example, and modern high-resolution observations reveal the presence of intense flare emission in white light in compact photospheric flare kernels \citep{2011ApJ...739...96K,2012ApJ...753L..26M}.
An H$\alpha$ image of a C4.3-class flare, which occurred on 11~Nov~2010, is shown in Fig.~\ref{fig:halphaflare} next to simultaneous images in the 17.1\,nm and 30.4\,nm passbands recorded with AIA/SDO. 
At ALMA frequencies one can observe the development of flare effects in the lower atmosphere with an almost entirely new capability.

From the modeling point of view, flares or microflares offer a new challenge, but in principle a welcome one because of their ``impulse response'' excitation of physical processes.
For example, \citet{1992ApJ...384..656W} reported observations of an extraordinarily compact event
\citep[see also, e.g.,][]{2007ApJ...666.1256B,2013PASJ...65S...1M}, and \citet{2008ApJ...688L.119J} reported white-light emission from a GOES class C1.2~flare on a 300\,km spatial scale.
The IRIS mission has begun to provide imaging spectroscopy of the chromosphere, including of course flares \citep[e.g.][]{2014arXiv1409.8603Y}, and thus has encouraged the development of extensive modeling tools as a routine part of the analysis of data.
The scientific potential of simultaneous multi-wavelength observations of flares was already demonstrated during the Solar Maximum Mission (SMM) in the early 1980's, during which several flares were observed simultaneously in the UV at selected individual spectral lines, e.g. O\,V~137.1\,nm, Si\,IV~139.4\,nm etc, and at hard X-ray energies of several keV. 
Several papers reported simultaneous (to within a fraction of a second) increases in the UV lines and at hard X-ray energies \citep[e.g.,][]{1988ApJ...330..480C}. 
It has now been shown \citep{2012SoPh..280..111D} that this is due to transient ionization, with the O\,V~137.1\,nm line being enhanced in the first fraction of a second with the peak in the line contribution function occurring initially at a higher electron temperature than in ionization equilibrium. 
The rise time and enhancement factor depend mostly on the electron density. 
The high cadence capability of ALMA will be essential due to these transient ionisation effects.

The appearance of strong impulsive-phase infrared sources (\citeauthor{2004ApJ...607L.131X}~\citeyear{2004ApJ...607L.131X} at  1,560\,nm and \citeauthor{2013ApJ...768..134K}~\citeyear{2013ApJ...768..134K} at 10,000\,nm) establishes the major significance of the lower solar atmosphere in flare physics. 
These findings, together with the absence of evidence for strong beaming \citep{1998ApJ...500.1003K,2006ApJ...653L.149K,2012ApJ...753L..26M,2013SoPh..284..405D} strongly suggest that ALMA {will significantly contribute} to answering important open questions about the nature of solar flares.

\subsubsection{Quasi-periodic pulsations}
\label{sec:qpp}
%
An intriguing and intensively debated feature of flaring energy releases are quasi-periodic pulsations (QPP) which are often seen in light curves of solar flares as well-pronounced periodic modulation of the emission intensity \citep{2009SSRv..149..119N}. 
These periodic variations were first detected by \citet{1969ApJ...155L.117P} in hard X-rays and have been consequently observed in all wavelength bands including microwaves, white light, soft and hard X-rays, and gamma-rays.
In a number of flares quasi-periodic pulsations  are seen simultaneously with different instruments, ground-based and space-borne.  
For example, quasi-periodic pulsations  have been simultaneously detected in the microwave band from the ground and in hard X-ray and gamma-rays from space 
{\citep[e.g.][]{1969ApJ...155L.117P,1983ApJ...271..376K,1983Natur.305..292N,2001ApJ...562L.103A,2003ApJ...588.1163G,2008ApJ...684.1433F,2010ApJ...708L..47N},}
and in soft X-rays with broadband filters on the GOES and PROBA-2 {spacecrafts \citep{2012ApJ...749L..16D,2015SoPh..tmp...50S}.}

Typical periods of quasi-periodic pulsations  range from a fraction of a second to several minutes, and sometimes are of a very high quality. The modulation depth can reach 100\%. 
Statistical studies in the microwave band show that quasi-periodic pulsations  are a common and perhaps intrinsic feature of flares \citep{2010SoPh..267..329K}. 
The term \lq\lq quasi\rq\rq-periodic reflects that the variations do not always have a clear harmonic pattern, and often are amplitude- and period-modulated. 
This modulation can be connected with noise, or it can be associated with the gradual variation of the physical parameters of the flaring active region and hence carry useful information.

It is commonly accepted that quasi-periodic pulsations  can be produced by several non-exclusive mechanisms, such as modulation of the plasma parameters by MHD waves and oscillations \citep[e.g.][]{2006A&A...446.1151N}, modulation of the non-thermal electron and positron kinematics by MHD oscillations \citep[e.g.][]{1982SvAL....8..132Z}, periodic triggering of energy releases by external fast and slow magnetoacoustic waves \citep[][]{2006A&A...452..343N,2006SoPh..238..313C}, alternate currents (torsional Alfv\'en oscillations) in current-carrying loops \citep{1998A&A...337..887Z}, and self-oscillatory regimes of spontaneous magnetic reconnection \citep[e.g.][]{2000A&A...360..715K}, including the coalescence over-stability of two current-carrying loops \citep{1987ApJ...321.1031T}. 
Quasi-periodic pulsations  in the sub-second period range could be associated with wave-particle interactions \citep{1987SoPh..111..113A}. 
Occasional observations of multi-period quasi-periodic pulsations  \citep[e.g.][]{2009A&A...493..259I, 2011ApJ...740...90V} might indicate that several different mechanisms could operate simultaneously, or that quasi-periodic pulsations  are induced by several different co-existing MHD waves. 
Indeed, in some cases, multi-period quasi-periodic pulsations  have been identified with standing slow and fast sausage modes of flaring coronal loops \citep{2011ApJ...740...90V}.

The possible association of quasi-periodic pulsations  with coronal MHD oscillations is based not only on the coincidence of the period ranges. 
In a number of cases quasi-periodic pulsations  have the time signatures typical for the MHD waves detected in the corona in the EUV band. 
For example an exponentially decaying harmonic dependence was observed simultaneously in the microwave quasi-periodic pulsations  and in the EUV emission intensity oscillation caused by a standing slow magnetoacoustic wave \citep{2012ApJ...756L..36K}. 
However, in the majority of cases, it has been difficult to establish the specific mechanism responsible for a quasi-periodic pulsation because of insufficient spatial resolution. 
For example, the 10''~resolution of the Nobeyama Radioheliograph (NoRH) does not  allow resolving the width of a typical flaring loop and hence the spatial transverse structure of the quasi-periodic pulsations's microwave source. 
ALMA's high spatial resolution will certainly dramatically improve our understanding of this phenomenon.

Observations of solar flares in the sub-THz band have shown high quality second and sub-second quasi-periodic pulsations  \cite[e.g][]{2003SoPh..218..211M, 2009ApJ...697..420K}. 
Thus, revealing the nature of the dependence of the flaring emission intensity on the frequency in the GHz and sub-THz band would be incomplete without quantifying them in detail and explaining the nature of quasi-periodic pulsations  in this band.

Moreover, observations of quasi-periodic pulsations  and their structure at different frequencies would provide us with additional important constraints necessary for the discrimination between possible physical mechanisms responsible for the peculiarity of the sub-THz spectrum of flaring emission. 
One necessary step forward in this direction is the development and generalisation of forward modelling of microwave emission quasi-periodic pulsations  performed by \citet{2014ApJ...785...86R,2015REZNIKOVA}, including different sources of the periodicity, e.g. different modes of MHD oscillations, quasi-periodically varying parameters, such as the spectral, spatial and pitch angle distribution of the injected non-thermal particles. 
A preliminary analytical study of sub-THz quasi-periodic pulsations  in terms of the alternate current model has recently been performed by \citet{2014SoPh..289.3017Z} and showed consistency with the available observations performed without the spectral and spatial resolution.  
In particular, this model could reproduce the observed dependence of the pulsation repetition rate on the signal intensity, which is different at the two separated frequencies of the observation, 212\,GHz and 405\,GHz. 
The confident identification of the physical mechanism responsible for quasi-periodic pulsations  and, in the case of modulation by MHD oscillations, of the modulating MHD mode, requires information about the spatial and spectral structure of quasi-periodic pulsations  at different frequencies. 
For example, a decisive signature of the sausage oscillation are the anti-phase variations in different parts of the microwave band \citep{2012ApJ...748..140M}. 
Also, the fine spectral structure of microwave quasi-periodic pulsations  may contain unique signatures of different MHD modes \citep{2013ApJ...777..159Y}. Forward modelling of similar time-spatial-spectral signatures of other possible mechanisms for quasi-periodic pulsations  needs to be performed.

Thus, high-precision imaging spectroscopy and polarimetry with ALMA with a spatial resolution of at least an order of magnitude better than NoRH's and its high spectral resolution, will revolutionise our understanding of flaring quasi-periodic pulsations.
Furthermore, the quasi-periodic injection and acceleration of  fast electrons \citep[e.g.][]{1969ApJ...155L.117P,1983ApJ...271..376K,1983Natur.305..292N,2008ApJ...684.1433F} results in a corresponding measurable quasi-periodic impact of the electrons precipitating into the chromosphere (see Sect.~\ref{sec:particlebeam}). 
ALMA can measure the corresponding thermal chromospheric response and, thus, provide a much needed chromospheric diagnostics for the processes causing  quasi-periodic pulsations.

\subsubsection{Particle beam heating of the chromosphere}
\label{sec:particlebeam} 
%
From the phenomenon of type~III radio emission it is well known that particle beams form in the solar corona.
Reverse drift bursts \citep[e.g.][]{1995ApJ...455..347A} reveal electron beams moving downwards to the chromosphere, but it is still unclear how far down they go.  
Downward electron beams have a long theoretical development as part of the so-called thick-target model as an explanation for hard X-ray and white-light footpoint sources, and hence for flare energetics in general \citep{1971ApJ...164..151K,1971SoPh...18..489B,1972SoPh...24..414H}.
These electron beams are thought to hit the plasma at the loop footpoints in the lower (dense) solar atmosphere, where the bulk of hard X-ray (HXR) emission is produced. 
Recent observations with RHESSI have allowed to look into the structure of HXR footpoints \citep[see, e.g., the review by][]{2011SSRv..159..301K}.
The observations \citep{2002SoPh..210..383A,2010ApJ...717..250K,2011A&A...533L...2B} established that higher energy HXR emission is indeed produced lower in the atmosphere. 
RHESSI observations \citep{2008A&A...489L..57K,2010ApJ...721.1933S,2010ApJ...717..250K,2011A&A...533L...2B,2012ApJ...760..142B} of hard X-ray foot-point sources suggest heights in the range $700\,-\,1200$\,km.  
\citet{2012ApJ...753L..26M}  found significantly lower heights, at the highest $(305 \pm 170)$\,km, but a further study {of three (partially occulted)} events \citep{2015ApJ...802...19K} suggests that the white light and HXR centroids are observed to be co-spatial and appear about 800\,km above the photosphere.
ALMA observations {will} provide crucial constraints on the formation heights.

Beam transport is also very attractive for explaining weaker events such as micro- and nanoflares \citep[e.g.,][]{2014Sci...346B.315T}. 
It is commonly accepted that during solar flares particle beams, which are accelerated by magnetic reconnection processes in the low corona, bombard the denser chromospheric layer. 
Owing to this bombardment the chromosphere is rapidly heated and the chromospheric plasma is evaporated into coronal parts of flare loops.
Processes, which involve heating of the chromosphere by particle beams, have been studied intensively by means of 1D radiation hydrodynamic modeling \citep{1984ApJ...279..896N,1984SoPh...90..357M}.

Furthermore, it was also found that these processes are strongly influenced by the so-called return current \citep{1990A&amp;A...234..496V,1992A&A...264..679K}, which is formed due to electromagnetic effects during the beam propagation through the flare atmosphere. 
The return current fully compensates the electric current of the beam and influences the location in the chromosphere, where the beam energy is deposited.  
It was proposed that the return current produces the impact polarisation of optical chromospheric lines \citep{2002A&A...383..713K}. 
As detailed in Sect.~\ref{sec:mmsunalma}, ALMA is expected to serve as a thermometer for the chromospheric plasma, which facilitates the study of all the processes connected with particle beam heating.
However, ALMA alone will not be able to directly recognize particle beams but indirectly through changes in the emission due to beam heating of the chromosphere. 
In any case, ALMA observations need to be supported by further observations in a broad range of the electromagnetic spectrum including X-ray, gamma-ray and the THz range.

\subsubsection{Triggering mechanism of subflares in active regions} 
%
In a recent work, \citet{2014ApJ...795L..24T} reported external triggering of solar subflares in a braided coronal magnetic structure observed by Hi-C \citep{2013Natur.493..501C,2014SoPh..289.4393K}. 
Magnetic reconnection between two chromospheric/transition region loops below the braided coronal magnetic structure resulted in both a shorter and a longer loop. 
The shorter one presumably submerged, as implied by the flux cancellation seen at that site, while the longer loop erupted and interacted with the overlying braided coronal loop, thus triggering a subflare with an estimated thermal energy release of 10$^{28}$\,erg. 
At least ten such events were detected in four hours of observation using different channels of the Atmospheric Imaging Assembly (AIA) aboard the Solar Dynamics Observatory (SDO). 
This observation suggests a new possible mechanism for the external triggering of subflares in addition to the one most commonly thought, namely the spontaneous internal triggering as a consequence of excessive energy accumulation by footpoint shuffling of magnetic field lines \citep{1988ApJ...330..474P}. 
It remains to be seen, however, how common this external triggering is and how important it is for coronal heating in general. 
High resolution chromospheric images obtained by ALMA, {in particular when co-observing with future Hi-C flights,  would provide important constraints on the magnetic and thermal topology of the active region chromosphere before, during and after subflares, and the external triggering of these events.}

\subsection{Chromospheric heating in regions with strong magnetic field} 
\label{sec:heatingmagreg}
%
Maintaining the high temperatures in the magnetic chromosphere, i.e. above network and plage regions, requires a heating rate that is larger than that of the corona \citep{1977ARA&A..15..363W}, yet the dominant mechanism for heating the magnetic chromosphere remains unknown. 
A variety of observations have revealed evidence for potential candidate heating mechanisms from spicules \citep{2007PASJ...59S.655D, 2006ASPC..354..276R}, Alfv{\'e}n waves \citep{2007Sci...318.1574D,2009Sci...323.1582J,2012Natur.486..505W,2014Sci...346D.315D}, magneto-acoustic shocks \citep{2006ApJ...648L.151J, 2006ApJ...647L..73H,2008A&A...479..213B} to gravity waves \citep{2008ApJ...681L.125S}. 
Similarly, a variety of theoretical models for chromospheric heating in magnetic regions have been studied, including high frequency waves \citep{2008ApJ...680.1542H}, Alfv{\'e}n waves \citep{2010ApJ...711..164V,2014ApJ...796L..23S}, transverse kink waves \citep{2008ApJ...676L..73V}, multi-fluid effects \citep{2010ApJ...724.1542K,2012ApJ...753..161M}, and plasma instabilities \citep[e.g.,][]{2008A&A...480..839F}. 
Waves and oscillations and related implications for atmospheric heating are discussed in more detail in Sect.~\ref{sec:oscillwaves}.

This large number of possible sources of heating has made it difficult to determine which of these many processes actually dominate(s). 
This is in part because many of the models lack the necessary physical complexity, including multi-fluid effects, non-LTE radiative transfer, or time dependent ionization. 
However, equally importantly the requirements on the heating (i.e., the ``heating rate'') have often been derived from simplified, semi-empirical 1D models \citep[e.g.,][]{1993ApJ...406..319F} that differ significantly from the time-dependent, highly spatially structured solar chromosphere. 
In other words, a model's ``success'' has often been determined by comparing the energy dissipation from a mechanism to an average heating rate derived from a 1D static atmosphere.

To make further progress it is critical that more quantitative and realistic comparisons are made between observations and models, e.g., through comparisons of observations (e.g., in the Mg\,II and Ca\,II lines) and synthetic observables from models \citep{2013ApJ...764L..11D,2013ApJ...772...89L,2013ApJ...778..143P}. 
While many of the chromospheric spectral lines require advanced calculations that include non-LTE radiative transfer, the availability of ALMA observations (which largely avoid non-LTE calculations) will provide a critical new component to constrain the spatio-temporal properties of chromospheric heating in and around magnetic network and plage regions for a wide range of heights.

ALMA observations have the potential to provide more direct constraints for theoretical modelers that, when combined with sub-arcsecond resolution observations of chromospheric lines observed from the ground or space (e.g., with IRIS and/or SST/CRISP/ CHROMIS), will lead to new insights concerning which of the many forms of mechanical energy that have been invoked by models actually dominate the deposition of non-radiative energy in the magnetized chromosphere. 
Because coronal loops are rooted in the network and plage regions, a better understanding of chromospheric heating has direct implications for coronal heating models.

\subsection{Chromospheric oscillations and waves}
\label{sec:oscillwaves}

\subsubsection{Wave propagation in the solar atmosphere } 
\label{sec:waveprop}
%
Waves and oscillations are interesting not only from the point of view that they can propagate energy into the chromosphere and dissipate that energy to produce non-radiative heating, but they also carry information about the structure of the atmosphere in which they propagate. 
Since the sound speed is a slowly varying function of temperature in the chromosphere, pressure disturbances with frequencies above the acoustic cut-off  propagating upwards in the chromosphere provide information about the difference of the formation height of the accompanying temperature and velocity perturbations, or, reversely, if the difference of formation heights is known, allow inferences about the phase speed of the waves and hence about the wave mode. 

Several authors \citep{Mein1971,Mein1976,Lites1982,Staiger1984,Fleck1989} who studied the wave propagation behavior in the chromosphere by analyzing time series of the Ca\,II infrared triplet lines at 854.2\,nm and 849.8\,nm found zero phase differences between the Doppler shifts of the two lines. 
The vanishing phase lag between the two lines was interpreted as evidence of a standing wave pattern in the chromosphere caused by total reflection of the waves at the transition region \citep{Mein1971,Mein1976,Staiger1984,Fleck1989}. 
This conclusion was based on the assumption that the Doppler signal of the two Ca\,II infrared lines is formed at different heights in the chromosphere, an assumption that was later demonstrated to be wrong \citep{Skartlien1994}. 
\citet{Fleck1994a}, \citet{Fleck1994b} and \citet{Hofmann1995} extended the earlier work of \citet{Fleck1989} and included the Ca\,II\,K and He\,I\,1083\,nm lines together with the Ca\,II\,854.2\,nm line in their observations. 
It is generally believed that the He\,I\,1083\,nm line is formed high up in the chromosphere just below the transition region \citep[e.g.,][and references therein]{1996SoPh..163...79B,2007SoPh..245..167B,2009CEAB...33..337J} and that the formation height of Ca\,II\,K$_3$  also is different from that of the Ca\,II infrared triplet \citep{Skartlien1994}. 
The phase difference spectra between these three lines, however, still showed a nearly uniform value close to 0$^\circ$, especially at frequencies above the cutoff frequency, providing further evidence of a dominating non-propagating component of the chromospheric wave field.

{Quasi-simultaneous observations of diagnostics with different formation heights, 
as will be possible in a linear manner with ALMA, have been carried out by \citet{1999A&A...347..335D}, who observed the two lower-chromospheric lines N\,I\,131.9\,nm and C\,II\,133.5\,nm  with the SUMER instrument on the SoHO spacecraft. 
The differences between (intensity and velocity) power spectra of network and internetwork regions are in line with results based on Ca\,II power spectra, whereas it seems that C\,II intensity lags that of N\,I by up to 16\,sec near 3\,mHz  when phase spectra are averaged along the slit. 
}

According to the dispersion relation of acoustic-gravity waves in a stratified atmosphere, propagating sound waves display a linear increase of the phase difference with frequency. 
Interestingly, the analysis of recent IRIS observations of the Mg\,II\,h and k~line also shows vanishing phase differences between the k$_3$, k$_{2v}$  and k$_{2r}$ Ca\,II\,K line core features, which are believed to form at different heights.  
Why is the apparent phase speed of high frequency acoustic waves so high? 
Are these results misleading because of complex radiation transfer effects in these optically thick lines?

\citet{2004A&A...419..747L,2006A&A...456..713L,2008Ap&SS.313..197L}, and \citet{2006A&A...456..697W} have demonstrated the feasibility of measuring chromospheric oscillations in the mm range. 
Multi-wavelength time series of ALMA observations of the temperature fluctuations of inter-network oscillations should allow travel time measurements between different heights as these disturbances propagate through the chromosphere and thus should finally settle the long-standing question about the propagation characteristics of acoustic waves in the chromosphere. 

\subsubsection{Alfv{\'e}n waves} 
\label{sec:alfvenwaves}
%
In addition to acoustic waves, Alfv\'en waves may play an important role in chromospheric heating. 
In the chromosphere, collisions between ions and neutrals (see also Sect.~\ref{sec:ionneutral}) are an efficient dissipative mechanism for Alfv\'en waves with frequencies near the ion-neutral collision frequency \citep{2001ApJ...558..859D,2004A&A...422.1073K,2005A&A...442.1091L,2011A&A...529A..82Z,2013ApJ...765...81R,2013ApJ...767..171S,2013ApJS..209...16S}. 
Estimations of the heating rate due to Alfv\'en waves damped by ion-neutral collisions \citep{2011JGRA..116.9104S,2011ApJ...735...45G,2013ApJ...777...53T} suggest that this mechanism may generate sufficient heat to compensate the radiative losses at low altitudes in the solar atmosphere. 
Recently, \citet{2015A&A...573A..79S} investigated the damping of Alfv\'en waves  as a function of height in a static chromospheric model based on the semi-empirical  model~F of \citet{1993ApJ...406..319F}. 
They found a critical interval of wavelengths for which impulsively excited Alfv\'en waves are overdamped as a result of the  strong ion-neutral dissipation. 
Equivalently, for periodically driven Alfv\'en waves there is an optimal frequency for which the damping is most effective. 
The critical wavelengths and optimal frequencies vary with height and are very sensitive to the ion-neutral collision cross section for momentum transfer used in the theoretical {computations.
Uncertainties} 
 in the collision frequency and cross section  cause uncertainties in the various transport coefficients that govern basic collisional phenomena in the plasma \citep{2012ApJ...753..161M}. 
On the other hand, \citet{2013A&A...549A.113Z} investigated the role of neutral helium and stratification on the propagation of torsional Alfv\'en waves through the chromosphere. 
While high-frequency waves are efficiently damped by ion-neutral collisions, \citet{2013A&A...549A.113Z}  concluded that low-frequency waves may not reach the transition region because they may become evanescent at lower heights owing to gravitational stratification. 
Hence, propagation of  Alfv\'en waves through the chromosphere into the solar corona should be considered with caution. 
The ability of ALMA to observe different layers in the chromosphere by using different bands, along with the high temporal and spatial resolutions,  may be crucial to understand how Alfv\'en waves actually propagate and damp in the chromosphere and thus how they contribute to chromospheric heating. ALMA offers the unprecedented opportunity to study high-frequency waves in the chromosphere.

Magnetohydrodynamic waves are also {ubiquitous} in solar prominences \citep{2002SoPh..206...45O,2009ApJ...704..870L,2013ApJ...779L..16H}. 
Since the properties of the prominence plasma are akin to those in the chromosphere, dissipation of wave energy due to ion-neutral collisions may also play a role in the energy balance of prominences. 
In the same way as in the chromosphere, ALMA observations may shed light on the impact of high-frequency waves on prominence plasmas.

\subsubsection{Resonant absorption and associated heating}
\label{sec:resonantabs}
%
Transverse MHD waves have been shown to be ubiquitous in the solar atmosphere \citep{2007Sci...318.1574D,2011Natur.475..477M,2007Sci...318.1577O,2007Sci...317.1192T}. 
Such waves constitute important heating agents due to their relatively high energy flux. 
Numerous theoretical and numerical studies have shown that resonant absorption is an expected phenomenon accompanying the propagation of transverse MHD waves along chromospheric or coronal waveguides \citep{1978ApJ...226..650I,1991SoPh..133..227S,2008ApJ...682L.141A,2008ApJ...679.1611T,2008ApJ...676L..73V,2010ApJ...718L.102V}. 
An observational consequence of this mechanism is the damping of transverse motions of the waveguides (such as flux tubes) in the plane-of-the-sky (POS) due to the conversion of energy between the kink mode and Alfv\'en waves. 
This conversion occurs at the resonance points, where the phase speed of the kink wave is equal to that of the Alfv\'en waves, and may not involve significant dissipation. 
It has been therefore difficult to associate the observed wave damping with heating and to estimate the exact energy contribution from these waves to chromospheric and coronal heating.

High-resolution ALMA observations at chromospheric and coronal heights ({including} prominences and coronal rain) can contribute in several ways towards narrowing down the heating contribution of transverse MHD waves. 
For instance, the high spatial resolution of ALMA allows to refine the upper limits for the energy contents of these waves, available prior to dissipation or mode conversion \citep{2013ApJ...768..191G,2014ApJ...795...18V}. 
Another important way is by clarifying the dissipation mechanisms of these Alfv{\'e}nic waves, which must occur at presently unresolved spatial scales. As explained earlier, resonant absorption can lead to strong damping of the POS transverse motion and a corresponding increase of the azimuthal flows in the boundary layer, which behave like \mbox{$m=1$} torsional Alfv{\'e}n modes. 
Through coordinated observations between ALMA and spectrometers such as IRIS, both motions associated with transverse MHD waves can be detected (POS and azimuthal motions along the line-of-sight), allowing precise MHD mode identification and correct application of MHD seismology techniques \citep{2011ApJ...733L..15V,2015ApJ...809...2015ApJ...809...71O72A,2015ApJ...809...71O}.

Also, resonant absorption is expected to generate small spatial scales in which, through the process of phase mixing, heating is produced. 
Furthermore, analytical and numerical modelling demonstrates that low amplitude transverse MHD modes can have a significant impact in the morphology of coronal or prominence flux tubes \citep{Ofman_1994GeoRL..21.2259O,Terradas_2008ApJ...687L.115T,Soler_2010ApJ...712..875S}. 
The strong velocity shear obtained through resonant absorption can lead to Kelvin-Helmholtz instabilities~(KHI), which extract the energy from the resonance layer and dump it into heat through viscous and ohmic processes in the generated vortices and current sheets. 
The Kelvin-Helmholtz instability vortices that result show-up as strand-like or thread-like structures in coronal or prominence flux tubes, respectively \citep{2014ApJ...787L..22A,2015ApJ...809...72A,2015ApJ...809...71O}. 
{These vortices should be detectable at a resolution of 20\,\% of the flux tube's radius, which would be $\sim 0.3''$ for a flux tube with 1000\,km radius and thus an achievable goal with ALMA.} 
Detection of such Kelvin-Helmholtz instability vortices and the ensuing turbulence would allow to correctly estimate the wave energy being dissipated.
	
\subsection{Magnetic loops in the upper atmosphere} 
\label{sec:magloops}

\subsubsection{The fine-structure of coronal loops} 
\label{sec:coronalloops}
For a long time solar coronal loops have been considered to consist of bundles of thin strands that remain undetected with the current instrumental capabilities \citep[see for e.g.,][and references therein]{1993ApJ...405..767G, 2010LRSP....7....5R}. 
Today, that perspective continues to remain as prevalent as ever. 
Coronal loops were first detected in coronagraphic observations in the 1940s \citep[and references therein]{1991plsc.book.....B, 2004psci.book.....A}. 
Within coronal loops, neighbouring field lines are considered to be thermally isolated and substructures are referred to as ``strands''. 
{It is likely} that most observations represent superpositions of hundreds of unresolved strands at various stages of heating and cooling \citep{2006SoPh..234...41K}. 
Other studies suggest that elementary loop components should be even finer, with typical cross-sections of the strands on the order of 10-100\,km \citep{2003SoPh..216...27B, 2004ApJ...605..911C, 2007ApJ...661..532D}. 
See also the results obtained from imaging spectroscopy at radio wavelengths by \citet{2013ApJ...763L..21C} with the Karl G.~Jansky Very Large Array (VLA), which support loop diameters of less than 100\,km. 
{In order to fully} describe the sub-structures of coronal loops one fundamental issue needs to be addressed, namely the unknown spatial scale of the coronal heating mechanism \citep{2010LRSP....7....5R}.
The multi-stranded interpretation of a coronal loop cross-section predicts a broad temperature distribution of unresolved threads \citep{2011ApJ...732...81A}. 
{\citet{2013A&A...556A.104P} concluded} that the fundamental reason for the constant cross-section is due to the presence of a small temperature gradient across the structure, hence, a simpler spatial structure would 
{suffice. 
\citet{2011ApJ...739...33S}} studied cool loops that were chosen in the 17.1\,nm channel from SDO/AIA and found 12~loops with narrow temperature distributions, consistent with isothermal plasma that could be modelled independently as a monolithic loop. 
In~2012 the High-resolution Coronal Imager \citep[Hi-C: ][]{2013Natur.493..501C} imaged the 1.5~MK solar corona at the highest ever resolution (in that passband) of 216\,-\,288\,km \citep{2014ApJ...787L..10W}.          
From this data, \citet{2013ApJ...772L..19B} measured the Gaussian widths of 91~coronal loops and the resulting distribution had a peak at a cross-sectional width of 270\,km.

{Coronal loops} in active regions can easily be subject to the thermal instability mechanism, through which the plasma undergoes catastrophic cooling from coronal temperatures down to chromospheric temperatures  \citep{1999ApJ...512..985A,2001ApJ...550.1036A}. 
The end result of this process is coronal rain, chromospheric material 
{observable by ALMA.}  
As we will see in the next sections, coronal rain provides unique advantages for solar atmospheric research, such as being a high resolution tracer of the local and global magnetic field topology of loops.

\begin{figure*}
\centering
\resizebox{\textwidth}{!}{\includegraphics{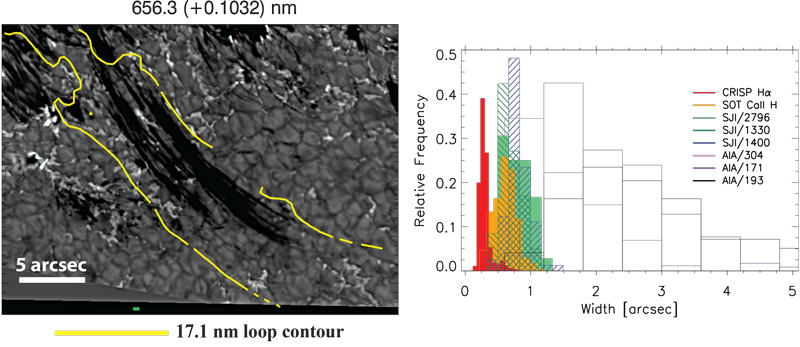}}
\vspace*{-10mm}
\caption{{\it Left}: 
A high-resolution (grey-scale) image of a post-flare loop taken with the CRisp Spectro-Polarimeter (CRISP) at the Swedish 1-m Solar Telescope (SST) in the H$\alpha$ far red-wing (656.3\,nm +0.1032\,nm) \citep{2014ApJ...797...36S}. 
The appearance of coronal rain (dark strands in H$\alpha$) is co-spatial with high temperature emission in the SDO/AIA\,17.1\,nm channel (yellow contours, \mbox{$\log T=5.8$}) from the coronal loop. 
{\it Right}: 
Histogram of coronal rain widths for two datasets combining various wavelengths: H$\alpha$ with the SST (in red, average measured logarithmic temperature $\log~T \approx 3.8$), Ca\,II\,H with Hinode/SOT (orange, $\log~T \approx$ 4), Mg\,II\,k~279.6\,nm with IRIS/SJI (light green, $\log~T \approx 4.2$), C\,II\,133.0\,nm with IRIS/SJI (dark green, $\log~T \approx 4.3$), Si\,IV\,140.0\,nm with IRIS/SJI (blue, $\log~T \approx 4.8$), He\,II\,30.4\,nm with SDO/AIA (light purple, $\log~T \approx 5$), Fe\,IX\,171 with SDO/AIA (dark purple, $\log~T \approx 5.9$), and Fe\,XII\,19.3\,nm with SDO/AIA (black, $\log~T \approx 6.2$). 
Explanation for these datasets can be found in \citet{2015ApJ...806...81A}. 
The bin size in each wavelength is set to the spatial resolution of each instrument. 
Strand-like structure in these loops extends from chromospheric to at least transition region temperatures. 
The shape of the distributions strongly suggest that the bulk of the coronal rain widths remains undetected at present day resolution.
Also compare to the statistical analysis of strands in this coronal loop by \citet{2014ApJ...797...36S}.
}
\label{fig:finescale}
\end{figure*}

\subsubsection{{Coronal rain - a proxy for the spatial and thermal inhomogeneity of the solar corona}}
\label{sec:coronalrain}
%
{A major question in coronal heating concerns the spatial distribution of the heating mechanism in loops. 
There is substantial evidence for most of the heating in active region loops being concentrated at the footpoints, which implies that thermal conduction is transporting the energy from energetic events at the footpoint upwards along the loops and is thus the main mechanism maintaining the corona in active regions \citep{2006SoPh..234...41K, 2010LRSP....7....5R}.  
Numerical studies have shown that such spatial distribution of heating often sets the loops into a state of thermal non-equilibrium \citep{1991ApJ...378..372A,1999ApJ...512..985A,2003A&A...411..605M,2004A&A...424..289M,2005A&A...436.1067M,2005ApJ...624.1080M,2010ApJ...716..154A,2010ApJ...709..499S,2013ApJ...773...94M}, in which cycles of heating (where loops become over-dense) are followed by cooling (in which loops deplete) due to the thermal instability mechanism \citep{1953ApJ...117..431P, Field_1965ApJ...142..531F}. 
The lowest temperatures in these cycles depend on a broad parameter space, and range from cool chromospheric temperatures to transition region temperatures.
During the cooling stage the plasma rapidly condenses into cold and dense clumps termed coronal rain, often observed falling along loop-like trajectories in a large number of chromospheric and transition region lines \citep{1970PASJ...22..405K,Leroy_1972SoPh...25..413L,1998SoPh..182...73K,Schrijver_2001SoPh..198..325S,2005ESASP.600E..30M,2006ApJ...643.1245U,2007A&A...475L..25O,2009ApJ...694.1256T,2012ApJ...745..152A,2015ApJ...806...81A}.}

Recent studies suggest that coronal rain is a fairly common phenomenon of active regions playing an important role in the chromosphere-corona mass cycle \citep{2012ApJ...745..152A, 2012ApJ...745L..21L, 2015ApJ...806...81A}. 
While these studies suggest that the temperatures in coronal rain can go down to 2000\,K or so, it is not yet clear whether such low temperatures are common in the rain. 
{Through the thermostat and high resolution capability of ALMA, the atmosphere above an active region can be probed for thermal bremsstrahlung from the cool material. 
Also, coronal rain is expected to have strong emission in recombination lines that are detectable with ALMA. 
Determining the presence of such cold material in the solar atmosphere is of large importance since it would imply that coronal heating is an extremely inhomogeneous process, with a very cold and frequent counterpart.}

Combining both a chromospheric and coronal nature, coronal rain constitutes the highest resolution window  into the substructure of loops. 
At the highest spatial resolution achievable today, the coronal rain clumps appear organised in strand-like structures with ever decreasing widths on the order of a few 100\,km or less (see Fig.~\ref{fig:finescale}, right panel), suggesting a tip-of-the-iceberg scenario in which loop substructure may not be fully resolved yet \citep{2012SoPh..280..457A,2012ApJ...745..152A,2014ApJ...797...36S,2015ApJ...806...81A}. 
These results have been supported by numerical simulations of coronal rain formation, through which a higher percentage of coronal rain clumps are obtained at sub-resolution scales \citep{2013ApJ...771L..29F}.
The strand-like structure of coronal rain is not only limited to chromospheric temperatures but extends to at least transition region temperatures \citep{2015ApJ...806...81A}. Such results point to the existence of fundamental substructure within coronal loops, consistent with the theoretically predicted strands (see Sect.~\ref{sec:coronalloops}). By providing high spatial resolution maps of sizes and temperature of coronal rain, ALMA can help determining the bulk of the size distribution of such fundamental substructure and its role in the chromosphere-corona mass cycle.

The nature of coronal rain strands and prominence threads remains elusive \citep[][see also Sect.~\ref{sec:prominences}]{Tandberg-Hansen_1995ASSL..199.....T, Lin_2011SSRv..158..237L, 2015ASSL..415...31E}. The formation and evolution of such structure may be strongly linked to coronal heating mechanisms. Recent work suggests the generation of such fine-scale structure and associated heating through dynamic instabilities \citep[such as KHI][]{2014ApJ...787L..22A,2015ApJ...809...72A,2015ApJ...809...71O}. With high spatial and temporal resolution, and multi-passband observations, ALMA can clarify the generation of such fine-scale structure by tracking the heating events in these cool structures and by determining the thermal and non-thermal processes involved.

Perhaps the most famous example of coronal rain is post-flare loops, in which the strong basal heating from the flare produces large chromospheric evaporation leading to over-dense strongly radiating loops which end up cooling catastrophically and falling back to the chromosphere \citep{1976ApJ...210..575F,1978ApJ...223.1046F}. 
This multi-thermal aspect of the plasma is often invoked to explain the usually large flux densities observed in the millimetre and sub-millimetre range \citep{2011SoPh..273..339T} but is not well constrained during the flare event.  
Also, the spatial distribution of the main thermal component during a flare is not well known. 
ALMA can help determining the amount and spatial distribution of the chromospheric temperature plasma such as coronal rain contributing to the thermal radiation.
Observations of coronal rain with ALMA will help determining the thermal and spatial inhomogeneity of the solar corona, and thus establish fundamental aspects of coronal heating and thermal instability mechanisms.

\subsection{Prominences and filaments}
\label{sec:prominences}
%
As already mentioned in Sect.~\ref{sec:bigquestprom}, prominences (aka filaments\footnote{Prominences are referred to as (dark) filaments when observed in absorption against the (bright) solar disk.}) are cool, dense regions of plasma lying in a much hotter and rarified coronal environment, which are supported by coronal magnetic fields against gravity.  
Comprehensive reviews of prominence physics can be found in \citet{Tandberg-Hansen_1995ASSL..199.....T}, \citet{2010SSRv..151..243L}, \citet{2010SSRv..151..333M}, \citet{2014LRSP...11....1P} or in the recent book by \citet{2015ASSL..415.....V}.

The primary contribution of ALMA to the understanding of solar prominences is its unique capability to measure the temperature of the cool prominence plasma with high spatial and temporal resolution \citep[see, e.g.,][]{2004A&A...419..747L,2012SoPh..277...31H,2013RMxAA..49....3P}.  
Previous studies based on mm and sub-mm radio emission \citep[see, e.g.,][]{1992SoPh..137...67V,1993ApJ...418..510B,1993A&A...274L...9H,1995SoPh..156..363I} had to rely on low-resolution observations available at the time. 
Now ALMA will be able to resolve the thermal fine-structure of prominences on far smaller spatial scales, perhaps reaching the physical dimensions of the prominence fine structures. 
Moreover, the high temporal resolution of ALMA will facilitate studying the evolution and dynamics of the (thermal) the small-scale structures, including, e.g.,  localized plasma flows, oscillations and passing waves.

\begin{figure}[t]
\centering
\resizebox{\textwidth}{!}
{\includegraphics[]{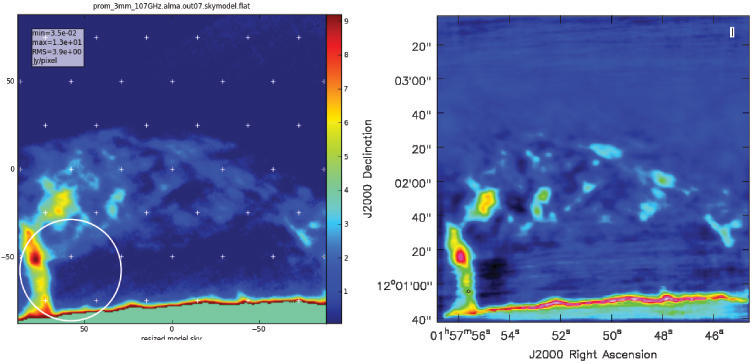}}
\caption{\textit{Left:} Modelled (ideal) brightness temperature map at 100\,GHz. 
  The white circle shows ALMA field-of-view for single pointing, the crosses are
  the centres of pointing for interferometric mosaic. 
  Fourty-nine pointings are necessary for Nyquist sampling of the displayed part of the
  prominence. 
  \textit{Right:}~Corresponding image as seen by ALMA at 3\,mm in Band~3 simulated with the CASA::simobserve() task. 
  {Images from \citet{2015SolPh...accept...H}. 
  We note that the dynamic range of images obtained by interferometric observing at the limb may be challenging.}
} 
\label{fig:promvis}
\end{figure}

Basic plasma parameters can easily be obtained, assuming that the prominence plasma is optically thin, that the background signal can be safely removed, and that the radiometric calibration is correct \citep{2013RMxAA..49....3P}.
As the thermal structure of the cool prominence plasma is still not well understood, it will be of primary interest to probe the plasma under different viewing angles (e.g. filament on the disk vs. prominence above the limb), and across the wavelength range offered by {ALMA.
Due to the increase of opacity of the prominence plasma and thus the variation of the formation region of the continuum with increasing wavelength}  \citep[see, e.g.,][]{2015SolPh...accept...H}, 
ALMA will be able to effectively scan the prominence thermal structure from the coolest fine-structure cores towards the hotter layers surrounding them. 
This will provide new insights into the structure of the {lower} temperature prominence-corona transition region (PCTR), which surrounds a whole prominence or individual prominence fine structures, and resulting in direct implications for the energy balance problem in prominences.

ALMA observations of the thermal structure of prominences can be tested and cross-calibrated using synthetic temperature maps obtained by forward modelling. 
For this purpose, multi-dimensional {prominence models} with realistic temperature distribution including the PCTR can be used, e.g.,  the 2D models of \citet{2009A&A...503..663G} and \citet{2001A&A...375.1082H}, or the  3D whole-prominence fine structure models of \citet{2015Apj...accept...G}.  
\citet{2015SolPh...accept...H} simulated the visibility of solar prominences and the detectability of their internal structures in future ALMA observations. 
Unlike in the case of the quiet solar chromosphere (see Sect.~\ref{sec:quietsun}), the brightness temperature model is based on the relation between the kinetic plasma temperature and the derived integral intensity of the H$\alpha$ line \citep[see][and references therein]{2015SolPh...accept...H}. 
Figure~\ref{fig:promvis} shows the {modeled} (i.e. ideal) brightness temperature map calculated from relations (10) and (14) in \citet{2015SolPh...accept...H} together with pointings for an ALMA interferometric mosaic at 3\,mm (ALMA Band~3) and the corresponding image reconstructed from the ALMA visibilities.

{According to previous radio observations of filaments \citep{1979SoPh...61..335R}, the intensity contrast with respect to the quiet sun decreases with frequency, being $\sim$5 \% at 3.5\,mm \citep[see also][]{2001SoPh..199..115C}. 
}
At higher frequencies, \citet{1993ApJ...418..510B} found negligible contrast at and near the location of H$\alpha$ filaments at 0.85\,mm. 
This finding demonstrates that filaments can expected to be observable in the low frequency ALMA bands, with the TP system being necessary due to their large extent. 
\edt{As a matter of fact, a filament corridor is visible in the SW quadrant in the ALMA map shown in Fig.~\ref{fig:almafulldisk}.}
In addition, dense patches of filamentary material might be observable at higher frequencies. 
Such observations will provide information on the density and temperature structure and, possibly, on the 
magnetic field of filaments, complementary to the information obtainable from prominences.

In summary, the unique capabilities of ALMA will result in significant contributions to our understanding of solar prominences. 
For instance, the anticipated results concern 
(i)~detailed information on the thermodynamical parameters the prominence fine structure plasma, 
(ii)~the nature of the PCTR and its role in conducting energy from the corona to the prominence, 
(iii)~the differences in 3D magnetic structures between prominences, and 
(iv)~how the prominence plasma collects and evolves 
(see also the associated issue of coronal rain in Sect.~\ref{sec:coronalrain}). 
Furthermore, ALMA will help advance the understanding of wave propagation (cf.~Sect.~\ref{sec:waveprop}) and damping in  prominences and with it more details of the prominence plasma and magnetic field properties. 
ALMA will also help to investigate currently debated `giant' solar tornadoes \citep[e.g.,][and references therein]{2012ApJ...761L..25O,2012ApJ...756L..41S,2012ApJ...752L..22L,2013A&A...549A.105P,2013ApJ...774..123W}, which are often associated with solar prominences. 
While recent observations allowed more information to be gathered on the state of the plasma \citep{Levens}, open questions remain as to the exact nature of the magnetic structure supporting this kind of phenomenon.

\subsection{{Potential implications} for other fields of astrophysics} 
\label{sec:otherfields}

\begin{figure}
\centering
\resizebox{9cm}{!}
{\includegraphics[]{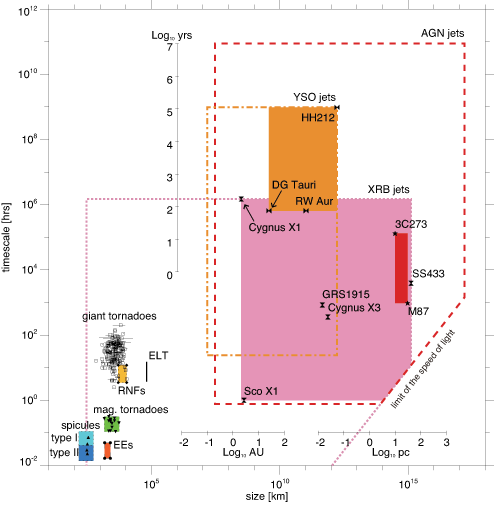}}
\caption{
Characteristic sizes and timescales of different rotating magnetic phenomena in all scales: giant tornadoes (squares), rotating network fields (RNFs: diamonds), magnetic tornadoes (triangles, tip downward), explosive events (EEs: circles), spicules (triangles, tip upward), and an extremely large tornado (ELT) on solar atmosphere and, in addition, astrophysical jets in young stellar objects (YSOs: bowties), in X-ray binaries (XRBs: sandglasses), and in active galactic nuclei (AGNs: stars).  
The individual data points for objects in the solar atmosphere are explained in \citep{2013ApJ...774..123W} and for objects in YSOs, XRBs, and AGNs are explained in the text in Section~\ref{sec:magvortexscale}.  
Filled boxes indicate the areas of observed data points whereas unfilled boxes indicate the area of theoretical possibility.
}
\label{fig:vortexscale}
\end{figure}

The Sun serves {as} a natural plasma laboratory, which hosts a multitude of physical processes that can also occur in other astrophysical objects either under similar conditions or on other scales. 
Many of the results that are expected from ALMA observations of the Sun, therefore have  potential implications for other fields of astrophysics.

Observations of the \mbox{(sub-)millimeter} emission from the solar atmosphere can be compared to (spatially unresolved) ALMA observations of other stars, thus opening up a much larger parameter space. 
Given its high sensitivity, ALMA is well-suited for such stellar observations as demonstrated by, e.g.,  
 \citet{2015A&A...573L...4L}. 
They managed to resolve the stellar components of the binary $\alpha$~Centauri with ALMA and measured the flux density in three different bands, thus providing important constraints for the thermal structure of the atmospheres of these two sun-like stars. 
Comparing such stellar ALMA observations to solar results promises important clues for the heating and activity of stellar atmospheres in general, including, e.g., the source of the so-called ``basal flux'' \citep[e.g.,][]{Schrijver1987,Perez2011} and flares (see Sect.~\ref{sec:flares}). 
For example, quasi-periodic pulsations (see Sec.~\ref{sec:qpp}) are found both for solar and stellar flares. 
With dedicated campaigns \citep{anfinogentov2013} and with the Kepler mission \citep{walkowicz2011,balona2012,shibayama2013}, flares and quasi-periodic pulsations in them are regularly detected. 
However, seismology with these pulsations is even less straightforward, since the spatial scales of the flare region cannot be directly measured on the stars. 
The correct identification of the mechanism responsible for quasi-periodic pulsations in solar flares will enable a more targeted seismology of the stellar flares as well. 
{As in the solar case, ALMA will also be an excellent instrument for the study of stellar flare oscillations.  
For instance,}  
\citet{2006A&A...458..921W} detected oscillations of 30 to 40\,sec in flares observed with the GALEX satellite in several stars.  
These oscillations were interpreted as due to magneto-hydrodynamic waves in active region coronal loops \citep[cf., e.g.,][]{2011ApJ...740...90V}.

\label{sec:magvortexscale}
Observations of magnetic tornadoes and vortex flows in the upper solar atmosphere (see Sect.~\ref{sec:vortexflows}) seem to suggest a relation between the characteristic size and the lifetime of different kinds of rotating magnetic field structures on the Sun (Fig.~\ref{fig:vortexscale}). 
This finding gives rise to the question if such a relation could be extrapolated to larger scales on the Sun and elsewhere in the universe and if there is a physical mechanism which would explain such a scalability. 
Large-scale, {long-lived rotating magnetic fields} are indeed observed as part of astrophysical jets in a wide variety of physical systems, from protostars \citep{2000MNRAS.314..241D,2002ApJ...576..222B,2004Ap&SS.292..553C,2010Sci...330.1209C}, through galactic compact objects \citep{1998Natur.392..673M,2001ApJ...558..283F,2005Natur.436..819G,2006MNRAS.370..399B} to active galactic nuclei \citep{1999Natur.401..891J,2001Sci...294..128L,2002A&A...388.1106B,2009ApJ...700.1690G,2013MNRAS.431..695M} and a galactic centre molecular tornado \citep{2007PASJ...59..189S} 
(see Fig.~\ref{fig:vortexscale}).    
Although the basic physical mechanism that creates and drives these jets is yet to be understood, rotating magnetic fields play a critical role in creating, driving, and directing jets across all scales.  
Actually, successful theoretical models, which have been developed for these different systems, incorporate magnetic fields threading the central objects such as accretion discs or rapidly spinning stars and compact objects. 
These models are tested observationally by measuring polarised radio emission \citep[see][for a review and references therein]{2012SSRv..169...27P} and compared to laboratory experiments \citep[for an overview see][and references therein]{2009PhPl...16d1005B}. 
ALMA will enable magnetic field measurements as well as vortex velocities, which are indispensable for understanding this type of phenomena and for testing if the suggested scalability and potential similarities between the individual phenomena outlined above are true or not. 

\section{Conclusions and Outlook} 
\label{sec:concoutlook}
%
While ALMA has already produced some remarkable results for a wide range of astrophysical topics, its potential for solar observations is yet to be revealed. 
Observing the Sun requires the development of observing modes, which are quite different from the usual modes. 
In return, such observations will produce some fascinating results. 
Essential contributions are anticipated for the science cases outlined in this review article. 
In particular, important steps towards advancing our understanding for central questions in solar physics are expected, for instance, concerning the heating of the outer solar atmosphere, solar flares, prominences and their implications for stellar activity and space weather.

It is important to remember that ALMA observations will be even more valuable when put into context with other observations of the chromosphere, e.g. obtained with IRIS (Mg\,II lines), SDO (magnetograms, continuum images, He\,II line), other ground-based solar observations (H$\alpha$, Ca\,II\,H and K lines, 
He\,I\,1083\,nm, magnetic field maps, G-band, etc), as well as forthcoming observatories (e.g. Solar-C). 
Taken all together, new advances can be expected to address fundamental questions about the Sun. 
We expect solar observing proposals to be accepted beginning with ALMA cycle~4 in 2016. 
ALMA is expected to remain operational for several decades, and, as most other ground-based observatories, will further improve its already impressive capabilities with time, thus extending the range of possible scientific applications in the future.

\begin{acknowledgements} 
This review was written in connection with the Solar Simulations for the Atacama Large Millimeter Observatory Network (SSALMON, http://www.ssalmon.uio.no).   
{We are grateful to the many colleagues who are working actively on realising the solar observing modes for ALMA as part of CSV/EOC activities carried out by the solar development teams of the NA/EU/EA-ARCs and JAO. 
In particular, we would like to thank A.~Remijan and R.~Hills for helpful comments.} 
S.~Wedemeyer acknowledges support (UiO-PES2020) by the Faculty of Mathematics and Natural Sciences of the University of Oslo, Norway, and the Research Council of Norway (grant 221767/F20).
R.~Braj\v sa was supported  by the European Commission FP7 with the projects eHEROES (284461, 2012-2015) and SOLARNET  (312495, 2013-2017), as well as by the Croatian Science Foundation (project~6212 ``Solar and Stellar Variability'').
G.~Fleishman is supported by NSF grants AGS-1250374 and AGS-1262772 and NASA grant NNX14AC87G to the New Jersey Institute of Technology.
M. Loukitcheva acknowledges Saint-Petersburg State University for research grants 6.0.26.2010, {6.37.343.2015 and 6.42.1428.2015,} and grant RFBR 15-02-03835. 
M.~B\'{a}rta thanks for the support of the European Commission through the CIG grant PCIG-GA-2011-304265 (SERAF) and GACR grant 13-24782S. 
M.~B\'{a}rta, M.~Karlicky, E.~Kontar and V.~M.~Nakariakov acknowledge the Marie Curie PIRSES-GA-2011-295272 RadioSun project.
R.~Soler acknowledges support from MINECO and FEDER funds (AYA2011-22846), from a ``Juan de la Cierva'' grant (JCI-2012-13594), from MECD (CEF11-0012), from CAIB (``Grups Competitius'' program), and UIB (``Vicerectorat dÕInvestigacio i Postgrau'').
S.~K.~Tiwari is supported by appointment to the NASA Postdoctoral Program at the NASA/MSFC, administered by ORAU through a contract with NASA.
E.~Scullion is a Government of Ireland Post-doctoral Research Fellow supported by the Irish Research Council.
N.~Labrosse acknowledges support from STFC grant ST/L000741/1, and, together with H.~Hudson, funding from the European CommunityÕs Seventh Framework Programme (FP7/2007-2013) under grant agreement no. 606862 (F-CHROMA).
Research at the Armagh Observatory is grant-aided by the N. Ireland Dept. of Culture, Arts and Leisure.
V.~M.~Nakariakov acknowledges the STFC consolidated grant ST/L000733/1. 
T.~Van~Doorsselaere was supported by an Odysseus grant of the FWO Vlaanderen, the IAP~P7/08 CHARM (Belspo) and the GOA-2015-014 (KU~Leuven).
{
A~Nindos' work has been partly co-financed by the European Union (European Social Fund -ESF) and Greek national funds through the Operational Program ``Education and Lifelong Learning'' of the National Strategic Reference Framework (NSRF) - Research Funding Program: ``Thales. Investing in knowledge society through the European Social Fund''.
}
T.~V.~Zaqarashvili acknowledges  FWF (the Austrian Fonds zur F\"orderung der Wissenschaftlichen Forschung) project P26181-N27.
{C.~L.~Selhorst acknowledges financial support from the S‹o Paulo Research Foundation (FAPESP), grant number 2014/10489-0.} 
We acknowledge usage of  the SAO/NASA Astrophysics Data System (ADS).  
\end{acknowledgements}
%
\small
\bibliographystyle{aa} 

\begin{thebibliography}{483}
\expandafter\ifx\csname natexlab\endcsname\relax\def\natexlab#1{#1}\fi

\bibitem[{{Ackermann} {et~al.}(2014){Ackermann}, {Ajello}, {Albert},
  {Allafort}, {Baldini}, {Barbiellini}, {Bastieri}, {Bechtol}, {Bellazzini},
  {Bissaldi}, {Bonamente}, {Bottacini}, {Bouvier}, {Brandt}, {Bregeon},
  {Brigida}, {Bruel}, {Buehler}, {Buson}, {Caliandro}, {Cameron}, {Caraveo},
  {Cecchi}, {Charles}, {Chekhtman}, {Chen}, {Chiang}, {Chiaro}, {Ciprini},
  {Claus}, {Cohen-Tanugi}, {Conrad}, {Cutini}, {D'Ammando}, {de Angelis}, {de
  Palma}, {Dermer}, {Desiante}, {Digel}, {Di Venere}, {Silva}, {Drell},
  {Drlica-Wagner}, {Favuzzi}, {Fegan}, {Focke}, {Franckowiak}, {Fukazawa},
  {Funk}, {Fusco}, {Gargano}, {Gasparrini}, {Germani}, {Giglietto}, {Giordano},
  {Giroletti}, {Glanzman}, {Godfrey}, {Grenier}, {Grove}, {Guiriec}, {Hadasch},
  {Hayashida}, {Hays}, {Horan}, {Hughes}, {Inoue}, {Jackson}, {Jogler},
  {J{\'o}hannesson}, {Johnson}, {Kamae}, {Kawano}, {Kn{\"o}dlseder}, {Kuss},
  {Lande}, {Larsson}, {Latronico}, {Lemoine-Goumard}, {Longo}, {Loparco},
  {Lott}, {Lovellette}, {Lubrano}, {Mayer}, {Mazziotta}, {McEnery},
  {Michelson}, {Mizuno}, {Moiseev}, {Monte}, {Monzani}, {Moretti}, {Morselli},
  {Moskalenko}, {Murgia}, {Murphy}, {Nemmen}, {Nuss}, {Ohno}, {Ohsugi},
  {Okumura}, {Omodei}, {Orienti}, {Orlando}, {Ormes}, {Paneque}, {Panetta},
  {Perkins}, {Pesce-Rollins}, {Petrosian}, {Piron}, {Pivato}, {Porter},
  {Rain{\`o}}, {Rando}, {Razzano}, {Reimer}, {Reimer}, {Ritz}, {Schulz},
  {Sgr{\`o}}, {Siskind}, {Spandre}, {Spinelli}, {Takahashi}, {Takeuchi},
  {Tanaka}, {Thayer}, {Thayer}, {Thompson}, {Tibaldo}, {Tinivella}, {Tosti},
  {Troja}, {Tronconi}, {Usher}, {Vandenbroucke}, {Vasileiou}, {Vianello},
  {Vitale}, {Werner}, {Winer}, {Wood}, {Wood}, {Wood}, \&
  {Yang}}]{2014ApJ...787...15A}
{Ackermann}, M., {Ajello}, M., {Albert}, A., {et~al.} 2014, \apj, 787, 15

\bibitem[{{Alissandrakis} {et~al.}(1992){Alissandrakis}, {Georgakilas}, \&
  {Dialetis}}]{1992SoPh..138...93A}
{Alissandrakis}, C.~E., {Georgakilas}, A.~A., \& {Dialetis}, D. 1992, \solphys,
  138, 93

\bibitem[{{Alissandrakis} \& {Kundu}(1978)}]{1978ApJ...222..342A}
{Alissandrakis}, C.~E. \& {Kundu}, M.~R. 1978, \apj, 222, 342

\bibitem[{{Alissandrakis} {et~al.}(1980){Alissandrakis}, {Kundu}, \&
  {Lantos}}]{1980A&A....82...30A}
{Alissandrakis}, C.~E., {Kundu}, M.~R., \& {Lantos}, P. 1980, \aap, 82, 30

\bibitem[{{Alissandrakis} \& {Patsourakos}(2013)}]{2013A&A...556A..79A}
{Alissandrakis}, C.~E. \& {Patsourakos}, S. 2013, \aap, 556, A79

\bibitem[{{Alissandrakis} \& {Preka-Papadema}(1984)}]{1984A&A...139..507A}
{Alissandrakis}, C.~E. \& {Preka-Papadema}, P. 1984, \aap, 139, 507

\bibitem[{{Alpert} {et~al.}(2015){Alpert}, {Tiwari}, {Moore}, {Winebarger}, \&
  {Savage}}]{alpert15}
{Alpert}, S.~E., {Tiwari}, S.~K., {Moore}, R.~L., {Winebarger}, A.~R., \&
  {Savage}, S. 2015, \apjl, 00, to be submitted

\bibitem[{{Anfinogentov} {et~al.}(2013){Anfinogentov}, {Nakariakov},
  {Mathioudakis}, {Van Doorsselaere}, \& {Kowalski}}]{anfinogentov2013}
{Anfinogentov}, S., {Nakariakov}, V.~M., {Mathioudakis}, M., {Van
  Doorsselaere}, T., \& {Kowalski}, A.~F. 2013, \apj, 773, 156

\bibitem[{{Antiochos} \& {Klimchuk}(1991)}]{1991ApJ...378..372A}
{Antiochos}, S.~K. \& {Klimchuk}, J.~A. 1991, \apj, 378, 372

\bibitem[{{Antiochos} {et~al.}(1999){Antiochos}, {MacNeice}, {Spicer}, \&
  {Klimchuk}}]{1999ApJ...512..985A}
{Antiochos}, S.~K., {MacNeice}, P.~J., {Spicer}, D.~S., \& {Klimchuk}, J.~A.
  1999, \apj, 512, 985

\bibitem[{{Antolin} {et~al.}(2015{\natexlab{a}}){Antolin}, {Okamoto}, {De
  Pontieu}, {Uitenbroek}, {Van Doorsselaere}, \&
  {Yokoyama}}]{2015ApJ...809...72A}
{Antolin}, P., {Okamoto}, T.~J., {De Pontieu}, B., {et~al.} 2015{\natexlab{a}},
  \apj, 809, 72

\bibitem[{{Antolin} \& {Rouppe van der Voort}(2012)}]{2012ApJ...745..152A}
{Antolin}, P. \& {Rouppe van der Voort}, L. 2012, \apj, 745, 152

\bibitem[{{Antolin} {et~al.}(2010){Antolin}, {Shibata}, \&
  {Vissers}}]{2010ApJ...716..154A}
{Antolin}, P., {Shibata}, K., \& {Vissers}, G. 2010, \apj, 716, 154

\bibitem[{{Antolin} {et~al.}(2015{\natexlab{b}}){Antolin}, {Vissers},
  {Pereira}, {Rouppe van der Voort}, \& {Scullion}}]{2015ApJ...806...81A}
{Antolin}, P., {Vissers}, G., {Pereira}, T.~M.~D., {Rouppe van der Voort}, L.,
  \& {Scullion}, E. 2015{\natexlab{b}}, \apj, 806, 81

\bibitem[{{Antolin} {et~al.}(2012){Antolin}, {Vissers}, \& {Rouppe van der
  Voort}}]{2012SoPh..280..457A}
{Antolin}, P., {Vissers}, G., \& {Rouppe van der Voort}, L. 2012, \solphys,
  280, 457

\bibitem[{{Antolin} {et~al.}(2014){Antolin}, {Yokoyama}, \& {Van
  Doorsselaere}}]{2014ApJ...787L..22A}
{Antolin}, P., {Yokoyama}, T., \& {Van Doorsselaere}, T. 2014, \apjl, 787, L22

\bibitem[{{Arregui} {et~al.}(2008){Arregui}, {Terradas}, {Oliver}, \&
  {Ballester}}]{2008ApJ...682L.141A}
{Arregui}, I., {Terradas}, J., {Oliver}, R., \& {Ballester}, J.~L. 2008, \apjl,
  682, L141

\bibitem[{{Asai} {et~al.}(2001){Asai}, {Shimojo}, {Isobe}, {Morimoto},
  {Yokoyama}, {Shibasaki}, \& {Nakajima}}]{2001ApJ...562L.103A}
{Asai}, A., {Shimojo}, M., {Isobe}, H., {et~al.} 2001, \apjl, 562, L103

\bibitem[{{Aschwanden}(1987)}]{1987SoPh..111..113A}
{Aschwanden}, M.~J. 1987, Solar Physics, 111, 113

\bibitem[{{Aschwanden}(2004)}]{2004psci.book.....A}
{Aschwanden}, M.~J. 2004, {Physics of the Solar Corona. An Introduction}
  (Praxis Publishing Ltd)

\bibitem[{{Aschwanden} {et~al.}(1995){Aschwanden}, {Benz}, {Dennis}, \&
  {Schwartz}}]{1995ApJ...455..347A}
{Aschwanden}, M.~J., {Benz}, A.~O., {Dennis}, B.~R., \& {Schwartz}, R.~A. 1995,
  \apj, 455, 347

\bibitem[{{Aschwanden} \& {Boerner}(2011)}]{2011ApJ...732...81A}
{Aschwanden}, M.~J. \& {Boerner}, P. 2011, \apj, 732, 81

\bibitem[{{Aschwanden} {et~al.}(2002){Aschwanden}, {Brown}, \&
  {Kontar}}]{2002SoPh..210..383A}
{Aschwanden}, M.~J., {Brown}, J.~C., \& {Kontar}, E.~P. 2002, \solphys, 210,
  383

\bibitem[{{Aschwanden} {et~al.}(2001){Aschwanden}, {Schrijver}, \&
  {Alexander}}]{2001ApJ...550.1036A}
{Aschwanden}, M.~J., {Schrijver}, C.~J., \& {Alexander}, D. 2001, \apj, 550,
  1036

\bibitem[{{Asensio Ramos} {et~al.}(2003){Asensio Ramos}, {Trujillo Bueno},
  {Carlsson}, \& {Cernicharo}}]{asensio03}
{Asensio Ramos}, A., {Trujillo Bueno}, J., {Carlsson}, M., \& {Cernicharo}, J.
  2003, \apjl, 588, L61

\bibitem[{{Avrett} \& {Loeser}(2008)}]{2008ApJS..175..229A}
{Avrett}, E.~H. \& {Loeser}, R. 2008, \apjs, 175, 229

\bibitem[{{Ayres}(1981)}]{ayres81}
{Ayres}, T.~R. 1981, \apj, 244, 1064

\bibitem[{{Ayres}(2002)}]{ayres02}
{Ayres}, T.~R. 2002, \apj, 575, 1104

\bibitem[{{Ayres} {et~al.}(2013){Ayres}, {Lyons}, {Ludwig}, {Caffau}, \&
  {Wedemeyer-B{\"o}hm}}]{2013ApJ...765...46A}
{Ayres}, T.~R., {Lyons}, J.~R., {Ludwig}, H.-G., {Caffau}, E., \&
  {Wedemeyer-B{\"o}hm}, S. 2013, \apj, 765, 46

\bibitem[{{Ayres} {et~al.}(2006){Ayres}, {Plymate}, \&
  {Keller}}]{2006ApJS..165..618A}
{Ayres}, T.~R., {Plymate}, C., \& {Keller}, C.~U. 2006, \apjs, 165, 618

\bibitem[{{Ayres} \& {Rabin}(1996)}]{ayres96}
{Ayres}, T.~R. \& {Rabin}, D. 1996, \apj, 460, 1042

\bibitem[{{Ayres} \& {Testerman}(1981)}]{ayres81b}
{Ayres}, T.~R. \& {Testerman}, L. 1981, \apj, 245, 1124

\bibitem[{Bacciotti {et~al.}(2002)Bacciotti, Ray, Mundt, Eisl{\"o}ffel, \&
  Solf}]{2002ApJ...576..222B}
Bacciotti, F., Ray, T.~P., Mundt, R., Eisl{\"o}ffel, J., \& Solf, J. 2002, The
  Astrophysical Journal, 576, 222

\bibitem[{{Balona}(2012)}]{balona2012}
{Balona}, L.~A. 2012, \mnras, 423, 3420

\bibitem[{{Bastian}(2002)}]{bastian02}
{Bastian}, T.~S. 2002, Astron.~Nachr., 323, 271

\bibitem[{{Bastian} {et~al.}(1998){Bastian}, {Benz}, \&
  {Gary}}]{1998ARA&A..36..131B}
{Bastian}, T.~S., {Benz}, A.~O., \& {Gary}, D.~E. 1998, \araa, 36, 131

\bibitem[{{Bastian} {et~al.}(1993{\natexlab{a}}){Bastian}, {Ewell}, \&
  {Zirin}}]{1993ApJ...418..510B}
{Bastian}, T.~S., {Ewell}, Jr., M.~W., \& {Zirin}, H. 1993{\natexlab{a}}, \apj,
  418, 510

\bibitem[{{Bastian} {et~al.}(1993{\natexlab{b}}){Bastian}, {Ewell}, \&
  {Zirin}}]{1993ApJ...415..364B}
{Bastian}, T.~S., {Ewell}, Jr., M.~W., \& {Zirin}, H. 1993{\natexlab{b}}, \apj,
  415, 364

\bibitem[{{Bastian} {et~al.}(2007){Bastian}, {Fleishman}, \&
  {Gary}}]{2007ApJ...666.1256B}
{Bastian}, T.~S., {Fleishman}, G.~D., \& {Gary}, D.~E. 2007, \apj, 666, 1256

\bibitem[{{Battaglia} {et~al.}(2005){Battaglia}, {Grigis}, \&
  {Benz}}]{2005A&A...439..737B}
{Battaglia}, M., {Grigis}, P.~C., \& {Benz}, A.~O. 2005, \aap, 439, 737

\bibitem[{{Battaglia} \& {Kontar}(2011)}]{2011A&A...533L...2B}
{Battaglia}, M. \& {Kontar}, E.~P. 2011, \aap, 533, L2

\bibitem[{{Battaglia} \& {Kontar}(2012)}]{2012ApJ...760..142B}
{Battaglia}, M. \& {Kontar}, E.~P. 2012, \apj, 760, 142

\bibitem[{{Beck} {et~al.}(2008){Beck}, {Schmidt}, {Rezaei}, \&
  {Rammacher}}]{2008A&A...479..213B}
{Beck}, C., {Schmidt}, W., {Rezaei}, R., \& {Rammacher}, W. 2008, \aap, 479,
  213

\bibitem[{{Beckers} \& {Tallant}(1969)}]{1969SoPh....7..351B}
{Beckers}, J.~M. \& {Tallant}, P.~E. 1969, \solphys, 7, 351

\bibitem[{Beckert \& Falcke(2002)}]{2002A&A...388.1106B}
Beckert, T. \& Falcke, H. 2002, Astronomy and Astrophysics, 388, 1106

\bibitem[{Begelman {et~al.}(2006)Begelman, King, \&
  Pringle}]{2006MNRAS.370..399B}
Begelman, M.~C., King, A.~R., \& Pringle, J.~E. 2006, Monthly Notices of the
  Royal Astronomical Society, 370, 399

\bibitem[{Bellan {et~al.}(2009)Bellan, Livio, Kato, Lebedev, Ray, Ferrari,
  Hartigan, Frank, Foster, \& Nicola{\"\i}}]{2009PhPl...16d1005B}
Bellan, P.~M., Livio, M., Kato, Y., {et~al.} 2009, Physics of Plasmas, 16, 1005

\bibitem[{{Benz} {et~al.}(2012{\natexlab{a}}){Benz}, {Brajsa}, {Shimojo},
  {Karlicky}, \& {Testi}}]{2012IAUSS...6E.205B}
{Benz}, A.~O., {Brajsa}, R., {Shimojo}, M., {Karlicky}, M., \& {Testi}, L.
  2012{\natexlab{a}}, IAU Special Session, 6, 205

\bibitem[{{Benz} \& {Grigis}(2002)}]{2002SoPh..210..431B}
{Benz}, A.~O. \& {Grigis}, P.~C. 2002, \solphys, 210, 431

\bibitem[{{Benz} \& {Krucker}(1999)}]{1999A&A...341..286B}
{Benz}, A.~O. \& {Krucker}, S. 1999, \aap, 341, 286

\bibitem[{{Benz} {et~al.}(2012{\natexlab{b}}){Benz}, {Krucker}, {Hurford},
  {Arnold}, {Orleanski}, {Gr{\"o}belbauer}, {Klober}, {Iseli}, {Wiehl},
  {Csillaghy}, {Etesi}, {Hochmuth}, {Battaglia}, {Bednarzik}, {Resanovic},
  {Grimm}, {Viertel}, {Commichau}, {Meuris}, {Limousin}, {Brun}, {Vilmer},
  {Skup}, {Graczyk}, {Stolarski}, {Michalska}, {Nowosielski}, {Cichocki},
  {Mosdorf}, {Seweryn}, {Przepi{\'o}rka}, {Sylwester}, {Kowalinski}, {Mrozek},
  {Podgorski}, {Mann}, {Aurass}, {Popow}, {Onel}, {Dionies}, {Bauer},
  {Rendtel}, {Warmuth}, {Woche}, {Pl{\"u}schke}, {Bittner}, {Paschke},
  {Wolker}, {Van Beek}, {Farnik}, {Kasparova}, {Veronig}, {Kienreich},
  {Gallagher}, {Bloomfield}, {Piana}, {Massone}, {Dennis}, {Schwarz}, \&
  {Lin}}]{2012SPIE.8443E..3LB}
{Benz}, A.~O., {Krucker}, S., {Hurford}, G.~J., {et~al.} 2012{\natexlab{b}}, in
  Society of Photo-Optical Instrumentation Engineers (SPIE) Conference Series,
  Vol. 8443, Society of Photo-Optical Instrumentation Engineers (SPIE)
  Conference Series, 3

\bibitem[{{Berger} \& {Simon}(1972)}]{berger-simon1972}
{Berger}, P.~S. \& {Simon}, M. 1972, \apj, 171, 191

\bibitem[{{Berkner} {et~al.}(2013){Berkner}, {Hauschildt}, \&
  {Baron}}]{2013A&A...550A.104B}
{Berkner}, A., {Hauschildt}, P.~H., \& {Baron}, E. 2013, \aap, 550, A104

\bibitem[{{Beveridge} {et~al.}(2003){Beveridge}, {Longcope}, \&
  {Priest}}]{2003SoPh..216...27B}
{Beveridge}, C., {Longcope}, D.~W., \& {Priest}, E.~R. 2003, \solphys, 216, 27

\bibitem[{{Bharti} {et~al.}(2013){Bharti}, {Hirzberger}, \&
  {Solanki}}]{2013A&A...552L...1B}
{Bharti}, L., {Hirzberger}, J., \& {Solanki}, S.~K. 2013, \aap, 552, L1

\bibitem[{{Bian} {et~al.}(2014){Bian}, {Emslie}, {Stackhouse}, \&
  {Kontar}}]{2014ApJ...796..142B}
{Bian}, N.~H., {Emslie}, A.~G., {Stackhouse}, D.~J., \& {Kontar}, E.~P. 2014,
  \apj, 796, 142

\bibitem[{{Bogod}(2011)}]{2011AstBu..66..190B}
{Bogod}, V.~M. 2011, Astrophysical Bulletin, 66, 190

\bibitem[{{Bogod} {et~al.}(2011){Bogod}, {Alesin}, \&
  {Pervakov}}]{2011AstBu..66..205B}
{Bogod}, V.~M., {Alesin}, A.~M., \& {Pervakov}, A.~A. 2011, Astrophysical
  Bulletin, 66, 205

\bibitem[{{Bogod} \& {Gelfreikh}(1980)}]{1980SoPh...67...29B}
{Bogod}, V.~M. \& {Gelfreikh}, G.~B. 1980, \solphys, 67, 29

\bibitem[{{Bonet} {et~al.}(2008){Bonet}, {M{\'a}rquez}, {S{\'a}nchez Almeida},
  {Cabello}, \& {Domingo}}]{2008ApJ...687L.131B}
{Bonet}, J.~A., {M{\'a}rquez}, I., {S{\'a}nchez Almeida}, J., {Cabello}, I., \&
  {Domingo}, V. 2008, \apjl, 687, L131

\bibitem[{{Boreiko} \& {Clark}(1986)}]{boreiko-clark1986}
{Boreiko}, R.~T. \& {Clark}, T.~A. 1986, \aap, 157, 353

\bibitem[{{Braj{\v s}a} {et~al.}(2007){Braj{\v s}a}, {Benz}, {Temmer},
  {Jurdana-{\v S}epi{\'c}}, {{\v S}aina}, \& {W{\"o}hl}}]{2007SoPh..245..167B}
{Braj{\v s}a}, R., {Benz}, A.~O., {Temmer}, M., {et~al.} 2007, \solphys, 245,
  167

\bibitem[{{Braj{\v s}a} \& {et al.}({2015})}]{Brajsa2015}
{Braj{\v s}a}, R. \& {et al.} {2015}, to be submitted to \aap

\bibitem[{{Braj{\v s}a} {et~al.}(1996){Braj{\v s}a}, {Pohjolainen}, {Ru{\v
  z}djak}, {Sakurai}, {Urpo}, {Vr{\v s}nak}, \&
  {W{\"o}hl}}]{1996SoPh..163...79B}
{Braj{\v s}a}, R., {Pohjolainen}, S., {Ru{\v z}djak}, V., {et~al.} 1996,
  \solphys, 163, 79

\bibitem[{{Braj{\v s}a} {et~al.}(2009){Braj{\v s}a}, {Rom{\v s}tajn},
  {W{\"o}hl}, {Benz}, {Temmer}, \& {Ro{\v s}a}}]{2009A&A...493..613B}
{Braj{\v s}a}, R., {Rom{\v s}tajn}, I., {W{\"o}hl}, H., {et~al.} 2009, \aap,
  493, 613

\bibitem[{{Brault} \& {Noyes}(1983)}]{brault-noyes1983}
{Brault}, J. \& {Noyes}, R. 1983, \apjl, 269, L61

\bibitem[{{Bray} {et~al.}(1991){Bray}, {Cram}, {Durrant}, \&
  {Loughhead}}]{1991plsc.book.....B}
{Bray}, R.~J., {Cram}, L.~E., {Durrant}, C., \& {Loughhead}, R.~E. 1991,
  {Plasma Loops in the Solar Corona} (Cambridge University Press)

\bibitem[{{Brooks} {et~al.}(2013){Brooks}, {Warren}, {Ugarte-Urra}, \&
  {Winebarger}}]{2013ApJ...772L..19B}
{Brooks}, D.~H., {Warren}, H.~P., {Ugarte-Urra}, I., \& {Winebarger}, A.~R.
  2013, \apjl, 772, L19

\bibitem[{{Brown}(1971)}]{1971SoPh...18..489B}
{Brown}, J.~C. 1971, \solphys, 18, 489

\bibitem[{{Brueckner} \& {Bartoe}(1983)}]{1983ApJ...272..329B}
{Brueckner}, G.~E. \& {Bartoe}, J.-D.~F. 1983, \apj, 272, 329

\bibitem[{{Bruls} {et~al.}(1995){Bruls}, {Solanki}, {Rutten}, \&
  {Carlsson}}]{1995A&A...293..225B}
{Bruls}, J.~H.~M.~J., {Solanki}, S.~K., {Rutten}, R.~J., \& {Carlsson}, M.
  1995, \aap, 293, 225

\bibitem[{{Buchholz} {et~al.}(1998){Buchholz}, {Ulmschneider}, \&
  {Cuntz}}]{Buchholz1998}
{Buchholz}, B., {Ulmschneider}, P., \& {Cuntz}, M. 1998, \apj, 494, 700

\bibitem[{{Cao} {et~al.}(2012){Cao}, {Goode}, {Ahn}, {Gorceix}, {Schmidt}, \&
  {Lin}}]{2012ASPC..463..291C}
{Cao}, W., {Goode}, P.~R., {Ahn}, K., {et~al.} 2012, in Astronomical Society of
  the Pacific Conference Series, Vol. 463, Second ATST-EAST Meeting: Magnetic
  Fields from the Photosphere to the Corona., ed. T.~R. {Rimmele},
  A.~{Tritschler}, F.~{W{\"o}ger}, M.~{Collados Vera}, H.~{Socas-Navarro},
  R.~{Schlichenmaier}, M.~{Carlsson}, T.~{Berger}, A.~{Cadavid}, P.~R.
  {Gilbert}, P.~R. {Goode}, \& M.~{Kn{\"o}lker}, 291

\bibitem[{{Cargill} \& {Klimchuk}(2004)}]{2004ApJ...605..911C}
{Cargill}, P.~J. \& {Klimchuk}, J.~A. 2004, \apj, 605, 911

\bibitem[{{Carlin} \& {Asensio Ramos}(2014)}]{2014arXiv1412.5386C}
{Carlin}, E.~S. \& {Asensio Ramos}, A. 2014, ArXiv e-prints

\bibitem[{{Carlsson} {et al.}({2016})}]{carlssonbifrostmodel}
{Carlsson}, M. \& {Hansteen}, B. \& {Gudiksen}, B. \& {et al.} 2016, \aap, 585, id.A4, 10 (2016)

\bibitem[{{Carlsson} {et~al.}(2010){Carlsson}, {Hansteen}, \&
  {Gudiksen}}]{2010MmSAI..81..582C}
{Carlsson}, M., {Hansteen}, V.~H., \& {Gudiksen}, B.~V. 2010, \memsai, 81, 582

\bibitem[{{Carlsson} {et~al.}(1992){Carlsson}, {Rutten}, \&
  {Shchukina}}]{carlsson1992}
{Carlsson}, M., {Rutten}, R.~J., \& {Shchukina}, N.~G. 1992, \aap, 253, 567

\bibitem[{{Carlsson} \& {Stein}(1994)}]{1994chdy.conf...47C}
{Carlsson}, M. \& {Stein}, R.~F. 1994, in Chromospheric Dynamics, ed.
  M.~{Carlsson} (University of Oslo, Oslo, Norway), 47, iSBN 82-7121-013-0
  (converted 978-82-7121-013-7)

\bibitem[{{Carlsson} \& {Stein}(1995)}]{1995ApJ...440L..29C}
{Carlsson}, M. \& {Stein}, R.~F. 1995, \apjl, 440, L29

\bibitem[{{Carlsson} \& {Stein}(2002)}]{2002ApJ...572..626C}
{Carlsson}, M. \& {Stein}, R.~F. 2002, \apj, 572, 626

\bibitem[{Carrasco-Gonz{\'a}lez {et~al.}(2010)Carrasco-Gonz{\'a}lez, Rodriguez,
  Anglada, Marti, Torrelles, \& Osorio}]{2010Sci...330.1209C}
Carrasco-Gonz{\'a}lez, C., Rodriguez, L.~F., Anglada, G., {et~al.} 2010,
  Science, 330, 1209

\bibitem[{{Chang} \& {Noyes}(1983)}]{chang-noyes1983}
{Chang}, E.~S. \& {Noyes}, R.~W. 1983, \apjl, 275, L11

\bibitem[{{Chen} {et~al.}(2013){Chen}, {Bastian}, {White}, {Gary}, {Perley},
  {Rupen}, \& {Carlson}}]{2013ApJ...763L..21C}
{Chen}, B., {Bastian}, T.~S., {White}, S.~M., {et~al.} 2013, \apjl, 763, L21

\bibitem[{{Chen} \& {Priest}(2006)}]{2006SoPh..238..313C}
{Chen}, P.~F. \& {Priest}, E.~R. 2006, \solphys, 238, 313

\bibitem[{{Cheng} {et~al.}(1988){Cheng}, {Vanderveen}, {Orwig}, \&
  {Tandberg-Hanssen}}]{1988ApJ...330..480C}
{Cheng}, C.-C., {Vanderveen}, K., {Orwig}, L.~E., \& {Tandberg-Hanssen}, E.
  1988, \apj, 330, 480

\bibitem[{{Chiuderi Drago} {et~al.}(2001){Chiuderi Drago}, {Alissandrakis},
  {Bastian}, {Bocchialini}, \& {Harrison}}]{2001SoPh..199..115C}
{Chiuderi Drago}, F., {Alissandrakis}, C.~E., {Bastian}, T., {Bocchialini}, K.,
  \& {Harrison}, R.~A. 2001, \solphys, 199, 115

\bibitem[{{Chiuderi Drago} {et~al.}(1977){Chiuderi Drago}, {Felli}, \&
  {Tofani}}]{1977A&A....61...79C}
{Chiuderi Drago}, F., {Felli}, M., \& {Tofani}, G. 1977, \aap, 61, 79

\bibitem[{{Christopoulou} {et~al.}(2001){Christopoulou}, {Georgakilas}, \&
  {Koutchmy}}]{2001A&A...375..617C}
{Christopoulou}, E.~B., {Georgakilas}, A.~A., \& {Koutchmy}, S. 2001, \aap,
  375, 617

\bibitem[{{Cirtain} {et~al.}(2013){Cirtain}, {Golub}, {Winebarger}, {de
  Pontieu}, {Kobayashi}, {Moore}, {Walsh}, {Korreck}, {Weber}, {McCauley},
  {Title}, {Kuzin}, \& {Deforest}}]{2013Natur.493..501C}
{Cirtain}, J.~W., {Golub}, L., {Winebarger}, A.~R., {et~al.} 2013, \nat, 493,
  501

\bibitem[{{Clark} {et~al.}(2000{\natexlab{a}}){Clark}, {Naylor}, \&
  {Davis}}]{clark2000a}
{Clark}, T.~A., {Naylor}, D.~A., \& {Davis}, G.~R. 2000{\natexlab{a}}, \aap,
  357, 757

\bibitem[{{Clark} {et~al.}(2000{\natexlab{b}}){Clark}, {Naylor}, \&
  {Davis}}]{clark2000b}
{Clark}, T.~A., {Naylor}, D.~A., \& {Davis}, G.~R. 2000{\natexlab{b}}, \aap,
  361, L60

\bibitem[{Coffey {et~al.}(2004)Coffey, Bacciotti, Woitas, Ray, \&
  Eisl{\"o}ffel}]{2004Ap&SS.292..553C}
Coffey, D., Bacciotti, F., Woitas, J., Ray, T.~P., \& Eisl{\"o}ffel, J. 2004,
  Astrophys Space Sci, 292, 553

\bibitem[{{Collados} {et~al.}(2010){Collados}, {Bettonvil}, {Cavaller},
  {Ermolli}, {Gelly}, {P{\'e}rez}, {Socas-Navarro}, {Soltau}, {Volkmer}, \&
  {EST Team}}]{2010AN....331..615C}
{Collados}, M., {Bettonvil}, F., {Cavaller}, L., {et~al.} 2010, Astronomische
  Nachrichten, 331, 615

\bibitem[{{Collados} {et~al.}(2012){Collados}, {L{\'o}pez}, {P{\'a}ez},
  {Hern{\'a}ndez}, {Reyes}, {Calcines}, {Ballesteros}, {D{\'{\i}}az}, {Denker},
  {Lagg}, {Schlichenmaier}, {Schmidt}, {Solanki}, {Strassmeier}, {von der
  L{\"u}he}, \& {Volkmer}}]{2012AN....333..872C}
{Collados}, M., {L{\'o}pez}, R., {P{\'a}ez}, E., {et~al.} 2012, Astronomische
  Nachrichten, 333, 872

\bibitem[{{Cristiani} {et~al.}(2008){Cristiani}, {Gim{\'e}nez de Castro},
  {Mandrini}, {Machado}, {Silva}, {Kaufmann}, \&
  {Rovira}}]{2008A&A...492..215C}
{Cristiani}, G., {Gim{\'e}nez de Castro}, C.~G., {Mandrini}, C.~H., {et~al.}
  2008, \aap, 492, 215

\bibitem[{{Cuntz} {et~al.}(2007){Cuntz}, {Rammacher}, \&
  {Musielak}}]{Cuntz2007}
{Cuntz}, M., {Rammacher}, W., \& {Musielak}, Z.~E. 2007, \apjl, 657, L57

\bibitem[{{Danilovic} {et~al.}(2010){Danilovic}, {Sch{\"u}ssler}, \&
  {Solanki}}]{2010A&A...513A...1D}
{Danilovic}, S., {Sch{\"u}ssler}, M., \& {Solanki}, S.~K. 2010, \aap, 513, A1

\bibitem[{Davis {et~al.}(2000)Davis, Berndsen, Smith, Chrysostomou, \&
  Hobson}]{2000MNRAS.314..241D}
Davis, C.~J., Berndsen, A., Smith, M.~D., Chrysostomou, A., \& Hobson, J. 2000,
  Monthly Notices of the Royal Astronomical Society, 314, 241

\bibitem[{{de la Cruz Rodr{\'{\i}}guez} {et~al.}(2013{\natexlab{a}}){de la Cruz
  Rodr{\'{\i}}guez}, {De Pontieu}, {Carlsson}, \& {Rouppe van der
  Voort}}]{2013ApJ...764L..11D}
{de la Cruz Rodr{\'{\i}}guez}, J., {De Pontieu}, B., {Carlsson}, M., \& {Rouppe
  van der Voort}, L.~H.~M. 2013{\natexlab{a}}, \apjl, 764, L11

\bibitem[{{de la Cruz Rodr{\'{\i}}guez} {et~al.}(2013{\natexlab{b}}){de la Cruz
  Rodr{\'{\i}}guez}, {Rouppe van der Voort}, {Socas-Navarro}, \& {van
  Noort}}]{2013A&A...556A.115D}
{de la Cruz Rodr{\'{\i}}guez}, J., {Rouppe van der Voort}, L., {Socas-Navarro},
  H., \& {van Noort}, M. 2013{\natexlab{b}}, \aap, 556, A115

\bibitem[{{de la Cruz Rodr{\'{\i}}guez} {et~al.}(2010){de la Cruz
  Rodr{\'{\i}}guez}, {Socas-Navarro}, {van Noort}, \& {Rouppe van der
  Voort}}]{2010MmSAI..81..716D}
{de la Cruz Rodr{\'{\i}}guez}, J., {Socas-Navarro}, H., {van Noort}, M., \&
  {Rouppe van der Voort}, L. 2010, \memsai, 81, 716

\bibitem[{{De Pontieu} {et~al.}(2012){De Pontieu}, {Carlsson}, {Rouppe van der
  Voort}, {Rutten}, {Hansteen}, \& {Watanabe}}]{2012ApJ...752L..12D}
{De Pontieu}, B., {Carlsson}, M., {Rouppe van der Voort}, L.~H.~M., {et~al.}
  2012, \apjl, 752, L12

\bibitem[{{De Pontieu} {et~al.}(2007){De Pontieu}, {Hansteen}, {Rouppe van der
  Voort}, {van Noort}, \& {Carlsson}}]{2007ApJ...655..624D}
{De Pontieu}, B., {Hansteen}, V.~H., {Rouppe van der Voort}, L., {van Noort},
  M., \& {Carlsson}, M. 2007, \apj, 655, 624

\bibitem[{{De Pontieu} {et~al.}(2001){De Pontieu}, {Martens}, \&
  {Hudson}}]{2001ApJ...558..859D}
{De Pontieu}, B., {Martens}, P.~C.~H., \& {Hudson}, H.~S. 2001, \apj, 558, 859

\bibitem[{{de Pontieu} {et~al.}(2007){de Pontieu}, {McIntosh}, {Hansteen},
  {Carlsson}, {Schrijver}, {Tarbell}, {Title}, {Shine}, {Suematsu}, {Tsuneta},
  {Katsukawa}, {Ichimoto}, {Shimizu}, \& {Nagata}}]{2007PASJ...59S.655D}
{de Pontieu}, B., {McIntosh}, S., {Hansteen}, V.~H., {et~al.} 2007, \pasj, 59,
  655

\bibitem[{{De Pontieu} {et~al.}(2007){De Pontieu}, {McIntosh}, {Carlsson},
  {Hansteen}, {Tarbell}, {Schrijver}, {Title}, {Shine}, {Tsuneta}, {Katsukawa},
  {Ichimoto}, {Suematsu}, {Shimizu}, \& {Nagata}}]{2007Sci...318.1574D}
{De Pontieu}, B., {McIntosh}, S.~W., {Carlsson}, M., {et~al.} 2007, Science,
  318, 1574

\bibitem[{{De Pontieu} {et~al.}(2014{\natexlab{a}}){De Pontieu}, {Rouppe van
  der Voort}, {McIntosh}, {Pereira}, {Carlsson}, {Hansteen}, {Skogsrud},
  {Lemen}, {Title}, {Boerner}, {Hurlburt}, {Tarbell}, {Wuelser}, {De Luca},
  {Golub}, {McKillop}, {Reeves}, {Saar}, {Testa}, {Tian}, {Kankelborg},
  {Jaeggli}, {Kleint}, \& {Martinez-Sykora}}]{2014Sci...346D.315D}
{De Pontieu}, B., {Rouppe van der Voort}, L., {McIntosh}, S.~W., {et~al.}
  2014{\natexlab{a}}, Science, 346, D315 (1255732

\bibitem[{{De Pontieu} {et~al.}(2014{\natexlab{b}}){De Pontieu}, {Title},
  {Lemen}, {Kushner}, {Akin}, {Allard}, {Berger}, {Boerner}, {Cheung}, {Chou},
  {Drake}, {Duncan}, {Freeland}, {Heyman}, {Hoffman}, {Hurlburt}, {Lindgren},
  {Mathur}, {Rehse}, {Sabolish}, {Seguin}, {Schrijver}, {Tarbell},
  {W{\"u}lser}, {Wolfson}, {Yanari}, {Mudge}, {Nguyen-Phuc}, {Timmons}, {van
  Bezooijen}, {Weingrod}, {Brookner}, {Butcher}, {Dougherty}, {Eder},
  {Knagenhjelm}, {Larsen}, {Mansir}, {Phan}, {Boyle}, {Cheimets}, {DeLuca},
  {Golub}, {Gates}, {Hertz}, {McKillop}, {Park}, {Perry}, {Podgorski},
  {Reeves}, {Saar}, {Testa}, {Tian}, {Weber}, {Dunn}, {Eccles}, {Jaeggli},
  {Kankelborg}, {Mashburn}, {Pust}, {Springer}, {Carvalho}, {Kleint}, {Marmie},
  {Mazmanian}, {Pereira}, {Sawyer}, {Strong}, {Worden}, {Carlsson}, {Hansteen},
  {Leenaarts}, {Wiesmann}, {Aloise}, {Chu}, {Bush}, {Scherrer}, {Brekke},
  {Martinez-Sykora}, {Lites}, {McIntosh}, {Uitenbroek}, {Okamoto}, {Gummin},
  {Auker}, {Jerram}, {Pool}, \& {Waltham}}]{2014SoPh..289.2733D}
{De Pontieu}, B., {Title}, A.~M., {Lemen}, J.~R., {et~al.} 2014{\natexlab{b}},
  \solphys, 289, 2733

\bibitem[{{DeForest}(2007)}]{2007ApJ...661..532D}
{DeForest}, C.~E. 2007, \apj, 661, 532

\bibitem[{{Dickson} \& {Kontar}(2013)}]{2013SoPh..284..405D}
{Dickson}, E.~C.~M. \& {Kontar}, E.~P. 2013, \solphys, 284, 405

\bibitem[{{Dolginov} \& {Yakovlev}(1974)}]{1974SvA....17..634D}
{Dolginov}, A.~Z. \& {Yakovlev}, D.~G. 1974, Sov. Ast., 17, 634

\bibitem[{{Dolla} {et~al.}(2012){Dolla}, {Marqu{\'e}}, {Seaton}, {Van
  Doorsselaere}, {Dominique}, {Berghmans}, {Cabanas}, {De Groof}, {Schmutz},
  {Verdini}, {West}, {Zender}, \& {Zhukov}}]{2012ApJ...749L..16D}
{Dolla}, L., {Marqu{\'e}}, C., {Seaton}, D.~B., {et~al.} 2012, \apjl, 749, L16

\bibitem[{{Domingo} {et~al.}(1995{\natexlab{a}}){Domingo}, {Fleck}, \&
  {Poland}}]{1995SSRv...72...81D}
{Domingo}, V., {Fleck}, B., \& {Poland}, A.~I. 1995{\natexlab{a}}, \ssr, 72, 81

\bibitem[{{Domingo} {et~al.}(1995{\natexlab{b}}){Domingo}, {Fleck}, \&
  {Poland}}]{1995SoPh..162....1D}
{Domingo}, V., {Fleck}, B., \& {Poland}, A.~I. 1995{\natexlab{b}}, \solphys,
  162, 1

\bibitem[{{Doyle} {et~al.}(2012){Doyle}, {Giunta}, {Singh}, {Madjarska},
  {Summers}, {Kellett}, \& {O'Mullane}}]{2012SoPh..280..111D}
{Doyle}, J.~G., {Giunta}, A., {Singh}, A., {et~al.} 2012, \solphys, 280, 111

\bibitem[{{Doyle} {et~al.}(1994){Doyle}, {Houdebine}, {Mathioudakis}, \&
  {Panagi}}]{1994A&A...285..233D}
{Doyle}, J.~G., {Houdebine}, E.~R., {Mathioudakis}, M., \& {Panagi}, P.~M.
  1994, \aap, 285, 233

\bibitem[{{Doyle} {et~al.}(1999){Doyle}, {van den Oord}, {O'Shea}, \&
  {Banerjee}}]{1999A&A...347..335D}
{Doyle}, J.~G., {van den Oord}, G.~H.~J., {O'Shea}, E., \& {Banerjee}, D. 1999,
  \aap, 347, 335

\bibitem[{{Dulk}(1985)}]{1985ARA&A..23..169D}
{Dulk}, G.~A. 1985, \araa, 23, 169

\bibitem[{{Dupree}(1968)}]{dupree1968}
{Dupree}, A.~K. 1968, \apjl, 152, L125

\bibitem[{{Efanov} {et~al.}(1980){Efanov}, {Labrum}, {Moiseev}, {Nesterov}, \&
  {Stewart}}]{1980BCrAO..61...43E}
{Efanov}, V.~A., {Labrum}, N., {Moiseev}, I.~G., {Nesterov}, N.~S., \&
  {Stewart}, R. 1980, Bulletin Crimean Astrophysical Observatory, 61, 43

\bibitem[{{Ekers} \& {Rots}(1979)}]{1979ASSL...76...61E}
{Ekers}, R.~D. \& {Rots}, A.~H. 1979, in Astrophysics and Space Science
  Library, Vol.~76, IAU Colloq. 49: Image Formation from Coherence Functions in
  Astronomy, ed. C.~{van Schooneveld}, 61

\bibitem[{{Ellerman}(1917)}]{1917ApJ....46..298E}
{Ellerman}, F. 1917, \apj, 46, 298

\bibitem[{{Ellis} \& {Bland-Hawthorn}(2009)}]{2009ApJ...707..457E}
{Ellis}, S.~C. \& {Bland-Hawthorn}, J. 2009, \apj, 707, 457

\bibitem[{{Engvold}(2015)}]{2015ASSL..415...31E}
{Engvold}, O. 2015, in Astrophysics and Space Science Library, Vol. 415,
  Astrophysics and Space Science Library, ed. J.-C. {Vial} \& O.~{Engvold}, 31

\bibitem[{{Fang} {et~al.}(2013){Fang}, {Xia}, \&
  {Keppens}}]{2013ApJ...771L..29F}
{Fang}, X., {Xia}, C., \& {Keppens}, R. 2013, \apjl, 771, L29

\bibitem[{{Fern{\'a}ndez-Menchero} {et~al.}(2014){Fern{\'a}ndez-Menchero}, {Del
  Zanna}, \& {Badnell}}]{fernandez2014}
{Fern{\'a}ndez-Menchero}, L., {Del Zanna}, G., \& {Badnell}, N.~R. 2014, \aap,
  566, A104

\bibitem[{{Field}(1965)}]{Field_1965ApJ...142..531F}
{Field}, G.~B. 1965, \apj, 142, 531

\bibitem[{{Fleck} \& {Deubner}(1989)}]{Fleck1989}
{Fleck}, B. \& {Deubner}, F.-L. 1989, \aap, 224, 245

\bibitem[{{Fleck} {et~al.}(1994{\natexlab{a}}){Fleck}, {Deubner}, {Hofmann}, \&
  {Steffens}}]{Fleck1994a}
{Fleck}, B., {Deubner}, F.-L., {Hofmann}, J., \& {Steffens}, S.
  1994{\natexlab{a}}, in Chromospheric Dynamics, ed. M.~{Carlsson}, 103

\bibitem[{{Fleck} {et~al.}(1994{\natexlab{b}}){Fleck}, {Deubner}, {Maier}, \&
  {Schmidt}}]{Fleck1994b}
{Fleck}, B., {Deubner}, F.-L., {Maier}, D., \& {Schmidt}, W.
  1994{\natexlab{b}}, in IAU Symposium, Vol. 154, Infrared Solar Physics, ed.
  D.~M. {Rabin}, J.~T. {Jefferies}, \& C.~{Lindsey}, 65

\bibitem[{{Fleishman} {et~al.}(2015){Fleishman}, {Loukitcheva}, \&
  {Nita}}]{2015arXiv150608395F}
{Fleishman}, G., {Loukitcheva}, M., \& {Nita}, G. 2015, ArXiv e-prints

\bibitem[{{Fleishman} {et~al.}(2013){Fleishman}, {Altyntsev}, \&
  {Meshalkina}}]{2013PASJ...65S...7F}
{Fleishman}, G.~D., {Altyntsev}, A.~T., \& {Meshalkina}, N.~S. 2013, \pasj, 65,
  S7 1

\bibitem[{{Fleishman} {et~al.}(2008){Fleishman}, {Bastian}, \&
  {Gary}}]{2008ApJ...684.1433F}
{Fleishman}, G.~D., {Bastian}, T.~S., \& {Gary}, D.~E. 2008, \apj, 684, 1433

\bibitem[{{Fleishman} \& {Kontar}(2010)}]{2010ApJ...709L.127F}
{Fleishman}, G.~D. \& {Kontar}, E.~P. 2010, \apjl, 709, L127

\bibitem[{{Fleishman} \& {Kuznetsov}(2010)}]{2010ApJ...721.1127F}
{Fleishman}, G.~D. \& {Kuznetsov}, A.~A. 2010, \apj, 721, 1127

\bibitem[{{Fleishman} \& {Kuznetsov}(2014)}]{2014ApJ...781...77F}
{Fleishman}, G.~D. \& {Kuznetsov}, A.~A. 2014, \apj, 781, 77

\bibitem[{{Fleishman} \& {Melnikov}(2003)}]{2003ApJ...587..823F}
{Fleishman}, G.~D. \& {Melnikov}, V.~F. 2003, \apj, 587, 823

\bibitem[{Fomalont {et~al.}(2001)Fomalont, Geldzahler, \&
  Bradshaw}]{2001ApJ...558..283F}
Fomalont, E.~B., Geldzahler, B.~J., \& Bradshaw, C.~F. 2001, The Astrophysical
  Journal, 558, 283

\bibitem[{{Fontenla} {et~al.}(1993){Fontenla}, {Avrett}, \&
  {Loeser}}]{1993ApJ...406..319F}
{Fontenla}, J.~M., {Avrett}, E.~H., \& {Loeser}, R. 1993, \apj, 406, 319

\bibitem[{{Fontenla} {et~al.}(2009){Fontenla}, {Curdt}, {Haberreiter},
  {Harder}, \& {Tian}}]{2009ApJ...707..482F}
{Fontenla}, J.~M., {Curdt}, W., {Haberreiter}, M., {Harder}, J., \& {Tian}, H.
  2009, \apj, 707, 482

\bibitem[{{Fontenla} {et~al.}(2014){Fontenla}, {Landi}, {Snow}, \&
  {Woods}}]{2014SoPh..289..515F}
{Fontenla}, J.~M., {Landi}, E., {Snow}, M., \& {Woods}, T. 2014, \solphys, 289,
  515

\bibitem[{{Fontenla} {et~al.}(2008){Fontenla}, {Peterson}, \&
  {Harder}}]{2008A&A...480..839F}
{Fontenla}, J.~M., {Peterson}, W.~K., \& {Harder}, J. 2008, \aap, 480, 839

\bibitem[{{Foukal}(1978)}]{1978ApJ...223.1046F}
{Foukal}, P. 1978, \apj, 223, 1046

\bibitem[{{Foukal}(1976)}]{1976ApJ...210..575F}
{Foukal}, P.~V. 1976, \apj, 210, 575

\bibitem[{{Freytag} {et~al.}(2012){Freytag}, {Steffen}, {Ludwig},
  {Wedemeyer-B{\"o}hm}, {Schaffenberger}, \& {Steiner}}]{2012JCoPh.231..919F}
{Freytag}, B., {Steffen}, M., {Ludwig}, H.-G., {et~al.} 2012, Journal of
  Computational Physics, 231, 919

\bibitem[{{Frolov}(2015)}]{2015EPJD...69..132F}
{Frolov}, A.~M. 2015, European Physical Journal D, 69, 132

\bibitem[{{Fuerst} {et~al.}(1982){Fuerst}, {Benz}, \&
  {Hirth}}]{1982A&A...107..178F}
{Fuerst}, E., {Benz}, A.~O., \& {Hirth}, W. 1982, \aap, 107, 178

\bibitem[{Gallo {et~al.}(2005)Gallo, Fender, Kaiser, Russell, Morganti,
  Oosterloo, \& Heinz}]{2005Natur.436..819G}
Gallo, E., Fender, R., Kaiser, C., {et~al.} 2005, Nature, 436, 819

\bibitem[{{Gary} {et~al.}(2013){Gary}, {Fleishman}, \&
  {Nita}}]{2013SoPh..288..549G}
{Gary}, D.~E., {Fleishman}, G.~D., \& {Nita}, G.~M. 2013, \solphys, 288, 549

\bibitem[{{Gary} {et~al.}(1997){Gary}, {Hartl}, \&
  {Shimizu}}]{1997ApJ...477..958G}
{Gary}, D.~E., {Hartl}, M.~D., \& {Shimizu}, T. 1997, \apj, 477, 958

\bibitem[{{Gary} \& {Hurford}(1994)}]{1994ApJ...420..903G}
{Gary}, D.~E. \& {Hurford}, G.~J. 1994, \apj, 420, 903

\bibitem[{{Gary} \& {Keller}(2004)}]{2004ASSL..314.....G}
{Gary}, D.~E. \& {Keller}, C.~U., eds. 2004, Astrophysics and Space Science
  Library, Vol. 314, {Solar and Space Weather Radiophysics - Current Status and
  Future Developments}

\bibitem[{Gebhardt \& Thomas(2009)}]{2009ApJ...700.1690G}
Gebhardt, K. \& Thomas, J. 2009, The Astrophysical Journal, 700, 1690

\bibitem[{{Gelfreikh} {et~al.}(1999){Gelfreikh}, {Grechnev}, {Kosugi}, \&
  {Shibasaki}}]{1999SoPh..185..177G}
{Gelfreikh}, G.~B., {Grechnev}, V., {Kosugi}, T., \& {Shibasaki}, K. 1999,
  \solphys, 185, 177

\bibitem[{{Georgoulis} {et~al.}(2002){Georgoulis}, {Rust}, {Bernasconi}, \&
  {Schmieder}}]{2002ApJ...575..506G}
{Georgoulis}, M.~K., {Rust}, D.~M., {Bernasconi}, P.~N., \& {Schmieder}, B.
  2002, \apj, 575, 506

\bibitem[{{Gomez} {et~al.}(1993){Gomez}, {Martens}, \&
  {Golub}}]{1993ApJ...405..767G}
{Gomez}, D.~O., {Martens}, P.~C.~H., \& {Golub}, L. 1993, \apj, 405, 767

\bibitem[{{Goode} \& {Cao}(2012)}]{2012SPIE.8444E..03G}
{Goode}, P.~R. \& {Cao}, W. 2012, in Society of Photo-Optical Instrumentation
  Engineers (SPIE) Conference Series, Vol. 8444, Society of Photo-Optical
  Instrumentation Engineers (SPIE) Conference Series, 3

\bibitem[{{Goode} \& {Cao}(2013)}]{2013SoPh..287..315G}
{Goode}, P.~R. \& {Cao}, W. 2013, \solphys, 287, 315

\bibitem[{{Goodman}(2011)}]{2011ApJ...735...45G}
{Goodman}, M.~L. 2011, \apj, 735, 45 (1

\bibitem[{{Goossens} {et~al.}(2013){Goossens}, {Van Doorsselaere}, {Soler}, \&
  {Verth}}]{2013ApJ...768..191G}
{Goossens}, M., {Van Doorsselaere}, T., {Soler}, R., \& {Verth}, G. 2013, \apj,
  768, 191

\bibitem[{{Gopalswamy} {et~al.}(2012){Gopalswamy}, {Yashiro}, {M{\"a}kel{\"a}},
  {Michalek}, {Shibasaki}, \& {Hathaway}}]{2012ApJ...750L..42G}
{Gopalswamy}, N., {Yashiro}, S., {M{\"a}kel{\"a}}, P., {et~al.} 2012, \apjl,
  750, L42

\bibitem[{{Gouttebroze} \& {Labrosse}(2009)}]{2009A&A...503..663G}
{Gouttebroze}, P. \& {Labrosse}, N. 2009, \aap, 503, 663

\bibitem[{{Grebinskij} {et~al.}(2000){Grebinskij}, {Bogod}, {Gelfreikh},
  {Urpo}, {Pohjolainen}, \& {Shibasaki}}]{2000A&AS..144..169G}
{Grebinskij}, A., {Bogod}, V., {Gelfreikh}, G., {et~al.} 2000, \aaps, 144, 169

\bibitem[{{Grechnev} {et~al.}(2003{\natexlab{a}}){Grechnev}, {Lesovoi},
  {Smolkov}, {Krissinel}, {Zandanov}, {Altyntsev}, {Kardapolova}, {Sergeev},
  {Uralov}, {Maksimov}, \& {Lubyshev}}]{2003SoPh..216..239G}
{Grechnev}, V.~V., {Lesovoi}, S.~V., {Smolkov}, G.~Y., {et~al.}
  2003{\natexlab{a}}, \solphys, 216, 239

\bibitem[{{Grechnev} {et~al.}(2003{\natexlab{b}}){Grechnev}, {White}, \&
  {Kundu}}]{2003ApJ...588.1163G}
{Grechnev}, V.~V., {White}, S.~M., \& {Kundu}, M.~R. 2003{\natexlab{b}}, \apj,
  588, 1163

\bibitem[{{Greve}(1977)}]{greve1977}
{Greve}, A. 1977, \solphys, 52, 423

\bibitem[{{Gudiksen} {et~al.}(2011){Gudiksen}, {Carlsson}, {Hansteen}, {Hayek},
  {Leenaarts}, \& {Mart{\'{\i}}nez-Sykora}}]{2011A&A...531A.154G}
{Gudiksen}, B.~V., {Carlsson}, M., {Hansteen}, V.~H., {et~al.} 2011, \aap, 531,
  A154+

\bibitem[{{Guidice} \& {Castelli}(1975)}]{1975SoPh...44..155G}
{Guidice}, D.~A. \& {Castelli}, J.~P. 1975, \solphys, 44, 155

\bibitem[{{Gun{\'a}r} \& {Mackay}(2015)}]{2015Apj...accept...G}
{Gun{\'a}r}, S. \& {Mackay}, D.~H. 2015, \apj, 803, 64

\bibitem[{{Hansteen} {et~al.}(2014){Hansteen}, {De Pontieu}, {Carlsson},
  {Lemen}, {Title}, {Boerner}, {Hurlburt}, {Tarbell}, {Wuelser}, {Pereira}, {De
  Luca}, {Golub}, {McKillop}, {Reeves}, {Saar}, {Testa}, {Tian}, {Kankelborg},
  {Jaeggli}, {Kleint}, \& {Mart{\'{\i}}nez-Sykora}}]{Hansteen2014}
{Hansteen}, V., {De Pontieu}, B., {Carlsson}, M., {et~al.} 2014, Science, 346,
  315

\bibitem[{{Hansteen} {et~al.}(2006){Hansteen}, {De Pontieu}, {Rouppe van der
  Voort}, {van Noort}, \& {Carlsson}}]{2006ApJ...647L..73H}
{Hansteen}, V.~H., {De Pontieu}, B., {Rouppe van der Voort}, L., {van Noort},
  M., \& {Carlsson}, M. 2006, \apjl, 647, L73

\bibitem[{{Hansteen} \& {Gudiksen}(2005)}]{2005ESASP.592E..87H}
{Hansteen}, V.~H. \& {Gudiksen}, B. 2005, in ESA Special Publication, Vol. 592,
  Solar Wind 11/SOHO 16, Connecting Sun and Heliosphere, 483--486

\bibitem[{{Harrison} {et~al.}(1993){Harrison}, {Carter}, {Clark}, {Lindsey},
  {Jefferies}, {Sime}, {Watt}, {Roellig}, {Becklin}, {Naylor}, {Tompkins}, \&
  {Braun}}]{1993A&A...274L...9H}
{Harrison}, R.~A., {Carter}, M.~K., {Clark}, T.~A., {et~al.} 1993, \aap, 274,
  L9

\bibitem[{{Harrison} {et~al.}(1995){Harrison}, {Sawyer}, {Carter}, {Cruise},
  {Cutler}, {Fludra}, {Hayes}, {Kent}, {Lang}, {Parker}, {Payne}, {Pike},
  {Peskett}, {Richards}, {Gulhane}, {Norman}, {Breeveld}, {Breeveld}, {Al
  Janabi}, {McCalden}, {Parkinson}, {Self}, {Thomas}, {Poland}, {Thomas},
  {Thompson}, {Kjeldseth-Moe}, {Brekke}, {Karud}, {Maltby}, {Aschenbach},
  {Br{\"a}uninger}, {K{\"u}hne}, {Hollandt}, {Siegmund}, {Huber}, {Gabriel},
  {Mason}, \& {Bromage}}]{1995SoPh..162..233H}
{Harrison}, R.~A., {Sawyer}, E.~C., {Carter}, M.~K., {et~al.} 1995, \solphys,
  162, 233

\bibitem[{{Hasan} \& {van Ballegooijen}(2008)}]{2008ApJ...680.1542H}
{Hasan}, S.~S. \& {van Ballegooijen}, A.~A. 2008, \apj, 680, 1542

\bibitem[{{Hauschildt} \& {Baron}(2010)}]{2010A&A...509A..36H}
{Hauschildt}, P.~H. \& {Baron}, E. 2010, \aap, 509, A36+

\bibitem[{{Hauschildt} \& {Baron}(2014)}]{2014A&A...566A..89H}
{Hauschildt}, P.~H. \& {Baron}, E. 2014, \aap, 566, A89

\bibitem[{{Heinzel} \& {Anzer}(2001)}]{2001A&A...375.1082H}
{Heinzel}, P. \& {Anzer}, U. 2001, \aap, 375, 1082

\bibitem[{{Heinzel} \& {Avrett}(2012)}]{2012SoPh..277...31H}
{Heinzel}, P. \& {Avrett}, E.~H. 2012, \solphys, 277, 31

\bibitem[{{Heinzel} {et~al.}(2015){Heinzel}, {Berlicki}, {B\'arta},
  {Karlick\'y}, \& {Rudawy}}]{2015SolPh...accept...H}
{Heinzel}, P., {Berlicki}, A., {B\'arta}, M., {Karlick\'y}, M., \& {Rudawy}, P.
  2015, \solphys, 290, Issue 7, 1981 

\bibitem[{{Hey}(2014)}]{hey2014}
{Hey}, J.~D. 2014, Journal of Physics B Atomic Molecular Physics, 47, 165701

\bibitem[{{Hillier} {et~al.}(2013){Hillier}, {Morton}, \&
  {Erd{\'e}lyi}}]{2013ApJ...779L..16H}
{Hillier}, A., {Morton}, R.~J., \& {Erd{\'e}lyi}, R. 2013, \apjl, 779, L16

\bibitem[{{Hoang-Binh}(1982)}]{hoang-binh1982}
{Hoang-Binh}, D. 1982, \aap, 112, L3

\bibitem[{{Hoang-Binh} {et~al.}(1987){Hoang-Binh}, {Brault}, {Picart},
  {Tran-Minh}, \& {Vallee}}]{hoang-binh1987}
{Hoang-Binh}, D., {Brault}, P., {Picart}, J., {Tran-Minh}, N., \& {Vallee}, O.
  1987, \aap, 181, 134

\bibitem[{{Hofmann} {et~al.}(1995){Hofmann}, {Deubner}, \&
  {Fleck}}]{Hofmann1995}
{Hofmann}, J., {Deubner}, F.-L., \& {Fleck}, B. 1995, in Astronomical Society
  of the Pacific Conference Series, Vol.~76, GONG 1994. Helio- and
  Astro-Seismology from the Earth and Space, ed. R.~K. {Ulrich}, E.~J.
  {Rhodes}, Jr., \& W.~{Dappen}, 342

\bibitem[{{Huang} {et~al.}(2014{\natexlab{a}}){Huang}, {Madjarska}, {Koleva},
  {Doyle}, {Duchlev}, {Dechev}, \& {Reardon}}]{2014A&A...566A.148H}
{Huang}, Z., {Madjarska}, M.~S., {Koleva}, K., {et~al.} 2014{\natexlab{a}},
  \aap, 566, A148

\bibitem[{{Huang} {et~al.}(2014{\natexlab{b}}){Huang}, {Madjarska}, {Xia},
  {Doyle}, {Galsgaard}, \& {Fu}}]{2014ApJ...797...88H}
{Huang}, Z., {Madjarska}, M.~S., {Xia}, L., {et~al.} 2014{\natexlab{b}}, \apj,
  797, 88

\bibitem[{{Hudson}(1972)}]{1972SoPh...24..414H}
{Hudson}, H.~S. 1972, \solphys, 24, 414

\bibitem[{{Hudson}(1991)}]{1991SoPh..133..357H}
{Hudson}, H.~S. 1991, \solphys, 133, 357

\bibitem[{{Ichimoto} {et~al.}(2008){Ichimoto}, {Lites}, {Elmore}, {Suematsu},
  {Tsuneta}, {Katsukawa}, {Shimizu}, {Shine}, {Tarbell}, {Title}, {Kiyohara},
  {Shinoda}, {Card}, {Lecinski}, {Streander}, {Nakagiri}, {Miyashita},
  {Noguchi}, {Hoffmann}, \& {Cruz}}]{2008SoPh..249..233I}
{Ichimoto}, K., {Lites}, B., {Elmore}, D., {et~al.} 2008, \solphys, 249, 233

\bibitem[{{Inglis} \& {Nakariakov}(2009)}]{2009A&A...493..259I}
{Inglis}, A.~R. \& {Nakariakov}, V.~M. 2009, \aap, 493, 259

\bibitem[{{Innes} {et~al.}(1997){Innes}, {Inhester}, {Axford}, \&
  {Wilhelm}}]{1997Natur.386..811I}
{Innes}, D.~E., {Inhester}, B., {Axford}, W.~I., \& {Wilhelm}, K. 1997, \nat,
  386, 811

\bibitem[{{Ionson}(1978)}]{1978ApJ...226..650I}
{Ionson}, J.~A. 1978, \apj, 226, 650

\bibitem[{{Irimajiri} {et~al.}(1995){Irimajiri}, {Takano}, {Nakajima},
  {Shibasaki}, {Hanaoka}, \& {Ichimoto}}]{1995SoPh..156..363I}
{Irimajiri}, Y., {Takano}, T., {Nakajima}, H., {et~al.} 1995, \solphys, 156,
  363

\bibitem[{{Jefferies} {et~al.}(2006){Jefferies}, {McIntosh}, {Armstrong},
  {Bogdan}, {Cacciani}, \& {Fleck}}]{2006ApJ...648L.151J}
{Jefferies}, S.~M., {McIntosh}, S.~W., {Armstrong}, J.~D., {et~al.} 2006,
  \apjl, 648, L151

\bibitem[{{Jess} {et~al.}(2008){Jess}, {Mathioudakis}, {Crockett}, \&
  {Keenan}}]{2008ApJ...688L.119J}
{Jess}, D.~B., {Mathioudakis}, M., {Crockett}, P.~J., \& {Keenan}, F.~P. 2008,
  \apjl, 688, L119

\bibitem[{{Jess} {et~al.}(2009){Jess}, {Mathioudakis}, {Erd{\'e}lyi},
  {Crockett}, {Keenan}, \& {Christian}}]{2009Sci...323.1582J}
{Jess}, D.~B., {Mathioudakis}, M., {Erd{\'e}lyi}, R., {et~al.} 2009, Science,
  323, 1582

\bibitem[{{Judge} {et~al.}(2003){Judge}, {Carlsson}, \& {Stein}}]{Judge2003}
{Judge}, P.~G., {Carlsson}, M., \& {Stein}, R.~F. 2003, \apj, 597, 1158

\bibitem[{{Judge} \& {Carpenter}(1998)}]{Judge1998}
{Judge}, P.~G. \& {Carpenter}, K.~G. 1998, \apj, 494, 828

\bibitem[{Junor {et~al.}(1999)Junor, Biretta, \& Livio}]{1999Natur.401..891J}
Junor, W., Biretta, J.~A., \& Livio, M. 1999, Nature, 401, 891

\bibitem[{{Jurdana-{\v S}epi{\'c}} {et~al.}(2009){Jurdana-{\v S}epi{\'c}},
  {Braj{\v s}a}, {{\v S}aina}, \& {W{\"o}hl}}]{2009CEAB...33..337J}
{Jurdana-{\v S}epi{\'c}}, R., {Braj{\v s}a}, R., {{\v S}aina}, B., \&
  {W{\"o}hl}, H. 2009, Central European Astrophysical Bulletin, 33, 337

\bibitem[{{Kakinuma} \& {Swarup}(1962)}]{1962ApJ...136..975K}
{Kakinuma}, T. \& {Swarup}, G. 1962, \apj, 136, 975

\bibitem[{{Kane} \& {Donnelly}(1971)}]{1971ApJ...164..151K}
{Kane}, S.~R. \& {Donnelly}, R.~F. 1971, \apj, 164, 151

\bibitem[{{Kane} {et~al.}(1998){Kane}, {Hurley}, {McTiernan}, {Boer}, {Niel},
  {Kosugi}, \& {Yoshimori}}]{1998ApJ...500.1003K}
{Kane}, S.~R., {Hurley}, K., {McTiernan}, J.~M., {et~al.} 1998, \apj, 500, 1003

\bibitem[{{Kane} {et~al.}(1983){Kane}, {Kai}, {Kosugi}, {Enome}, {Landecker},
  \& {McKenzie}}]{1983ApJ...271..376K}
{Kane}, S.~R., {Kai}, K., {Kosugi}, T., {et~al.} 1983, \apj, 271, 376

\bibitem[{{Karlick{\'y}} {et~al.}(2011){Karlick{\'y}}, {B{\'a}rta}, {D{\c
  a}browski}, \& {Heinzel}}]{2011SoPh..268..165K}
{Karlick{\'y}}, M., {B{\'a}rta}, M., {D{\c a}browski}, B.~P., \& {Heinzel}, P.
  2011, \solphys, 268, 165

\bibitem[{{Karlicky} \& {Henoux}(1992)}]{1992A&A...264..679K}
{Karlicky}, M. \& {Henoux}, J.-C. 1992, \aap, 264, 679

\bibitem[{{Karlick{\'y}} \& {H{\'e}noux}(2002)}]{2002A&A...383..713K}
{Karlick{\'y}}, M. \& {H{\'e}noux}, J.~C. 2002, \aap, 383, 713

\bibitem[{{Karzas} \& {Latter}(1961)}]{1961ApJS....6..167K}
{Karzas}, W.~J. \& {Latter}, R. 1961, \apjs, 6, 167

\bibitem[{{Katsukawa} {et~al.}(2007){Katsukawa}, {Berger}, {Ichimoto}, {Lites},
  {Nagata}, {Shimizu}, {Shine}, {Suematsu}, {Tarbell}, {Title}, \&
  {Tsuneta}}]{2007Sci...318.1594K}
{Katsukawa}, Y., {Berger}, T.~E., {Ichimoto}, K., {et~al.} 2007, Science, 318,
  1594

\bibitem[{{Kaufmann} {et~al.}(2009{\natexlab{a}}){Kaufmann}, {Gim{\'e}nez de
  Castro}, {Correia}, {Costa}, {Raulin}, \& {V{\'a}lio}}]{2009ApJ...697..420K}
{Kaufmann}, P., {Gim{\'e}nez de Castro}, C.~G., {Correia}, E., {et~al.}
  2009{\natexlab{a}}, \apj, 697, 420

\bibitem[{{Kaufmann} {et~al.}(2001){Kaufmann}, {Raulin}, {Correia}, {Costa},
  {de Castro}, {Silva}, {Levato}, {Rovira}, {Mandrini}, {Fern{\'a}ndez-Borda},
  \& {Bauer}}]{2001ApJ...548L..95K}
{Kaufmann}, P., {Raulin}, J.-P., {Correia}, E., {et~al.} 2001, \apjl, 548, L95

\bibitem[{{Kaufmann} {et~al.}(2004){Kaufmann}, {Raulin}, {de Castro}, {Levato},
  {Gary}, {Costa}, {Marun}, {Pereyra}, {Silva}, \&
  {Correia}}]{2004ApJ...603L.121K}
{Kaufmann}, P., {Raulin}, J.-P., {de Castro}, C.~G.~G., {et~al.} 2004, \apjl,
  603, L121

\bibitem[{{Kaufmann} {et~al.}(2009{\natexlab{b}}){Kaufmann}, {Trottet},
  {Gim{\'e}nez de Castro}, {Raulin}, {Krucker}, {Shih}, \&
  {Levato}}]{2009SoPh..255..131K}
{Kaufmann}, P., {Trottet}, G., {Gim{\'e}nez de Castro}, C.~G., {et~al.}
  2009{\natexlab{b}}, \solphys, 255, 131

\bibitem[{{Kaufmann} {et~al.}(2013){Kaufmann}, {White}, {Freeland}, {Marcon},
  {Fernandes}, {Kudaka}, {de Souza}, {Aballay}, {Fernandez}, {Godoy}, {Marun},
  {Valio}, {Raulin}, \& {Gim{\'e}nez de Castro}}]{2013ApJ...768..134K}
{Kaufmann}, P., {White}, S.~M., {Freeland}, S.~L., {et~al.} 2013, \apj, 768,
  134

\bibitem[{{Kaufmann} {et~al.}(2015){Kaufmann}, {White}, {Marcon}, {Kudaka},
  {Cabezas}, {Cassiano}, {Francile}, {Fernandes}, {Hidalgo Ramirez}, {Luoni},
  {Marun}, {Pereyra}, \& {Souza}}]{2015JGRA..120.4155K}
{Kaufmann}, P., {White}, S.~M., {Marcon}, R., {et~al.} 2015, Journal of
  Geophysical Research (Space Physics), 120, 4155

\bibitem[{{Kawaguchi}(1970)}]{1970PASJ...22..405K}
{Kawaguchi}, I. 1970, \pasj, 22, 405

\bibitem[{{Keil} {et~al.}(2011){Keil}, {Rimmele}, {Wagner}, {Elmore}, \& {ATST
  Team}}]{2011ASPC..437..319K}
{Keil}, S.~L., {Rimmele}, T.~R., {Wagner}, J., {Elmore}, D., \& {ATST Team}.
  2011, in Astronomical Society of the Pacific Conference Series, Vol. 437,
  Solar Polarization 6, ed. J.~R. {Kuhn}, D.~M. {Harrington}, H.~{Lin}, S.~V.
  {Berdyugina}, J.~{Trujillo-Bueno}, S.~L. {Keil}, \& T.~{Rimmele}, 319

\bibitem[{{Kerdraon} \& {Delouis}(1997)}]{1997LNP...483..192K}
{Kerdraon}, A. \& {Delouis}, J.-M. 1997, in Lecture Notes in Physics, Berlin
  Springer Verlag, Vol. 483, Coronal Physics from Radio and Space Observations,
  ed. G.~{Trottet}, 192

\bibitem[{{Khodachenko} {et~al.}(2004){Khodachenko}, {Arber}, {Rucker}, \&
  {Hanslmeier}}]{2004A&A...422.1073K}
{Khodachenko}, M.~L., {Arber}, T.~D., {Rucker}, H.~O., \& {Hanslmeier}, A.
  2004, \aap, 422, 1073

\bibitem[{{Kim} {et~al.}(2012){Kim}, {Nakariakov}, \&
  {Shibasaki}}]{2012ApJ...756L..36K}
{Kim}, S., {Nakariakov}, V.~M., \& {Shibasaki}, K. 2012, \apjl, 756, L36

\bibitem[{{Kitiashvili} {et~al.}(2011){Kitiashvili}, {Kosovichev}, {Mansour},
  \& {Wray}}]{2011ApJ...727L..50K}
{Kitiashvili}, I.~N., {Kosovichev}, A.~G., {Mansour}, N.~N., \& {Wray}, A.~A.
  2011, \apjl, 727, L50

\bibitem[{{Kjeldseth-Moe} \& {Brekke}(1998)}]{1998SoPh..182...73K}
{Kjeldseth-Moe}, O. \& {Brekke}, P. 1998, \solphys, 182, 73

\bibitem[{{Kliem} {et~al.}(2000){Kliem}, {Karlick{\'y}}, \&
  {Benz}}]{2000A&A...360..715K}
{Kliem}, B., {Karlick{\'y}}, M., \& {Benz}, A.~O. 2000, \aap, 360, 715

\bibitem[{{Klimchuk}(2006)}]{2006SoPh..234...41K}
{Klimchuk}, J.~A. 2006, \solphys, 234, 41

\bibitem[{{Klopf} {et~al.}(2014){Klopf}, {Kaufmann}, {Raulin}, \&
  {Szpigel}}]{2014ApJ...791...31K}
{Klopf}, J.~M., {Kaufmann}, P., {Raulin}, J.-P., \& {Szpigel}, S. 2014, \apj,
  791, 31

\bibitem[{{Kobayashi} {et~al.}(2014){Kobayashi}, {Cirtain}, {Winebarger},
  {Korreck}, {Golub}, {Walsh}, {De Pontieu}, {DeForest}, {Title}, {Kuzin},
  {Savage}, {Beabout}, {Beabout}, {Podgorski}, {Caldwell}, {McCracken},
  {Ordway}, {Bergner}, {Gates}, {McKillop}, {Cheimets}, {Platt}, {Mitchell}, \&
  {Windt}}]{2014SoPh..289.4393K}
{Kobayashi}, K., {Cirtain}, J., {Winebarger}, A.~R., {et~al.} 2014, \solphys,
  289, 4393

\bibitem[{{Kochanov} {et~al.}(2013){Kochanov}, {Anfinogentov}, {Prosovetsky},
  {Rudenko}, \& {Grechnev}}]{2013PASJ...65S..19K}
{Kochanov}, A.~A., {Anfinogentov}, S.~A., {Prosovetsky}, D.~V., {Rudenko},
  G.~V., \& {Grechnev}, V.~V. 2013, \pasj, 65, 19

\bibitem[{{Kohl} {et~al.}(1995){Kohl}, {Esser}, {Gardner}, {Habbal},
  {Daigneau}, {Dennis}, {Nystrom}, {Panasyuk}, {Raymond}, {Smith}, {Strachan},
  {van Ballegooijen}, {Noci}, {Fineschi}, {Romoli}, {Ciaravella}, {Modigliani},
  {Huber}, {Antonucci}, {Benna}, {Giordano}, {Tondello}, {Nicolosi}, {Naletto},
  {Pernechele}, {Spadaro}, {Poletto}, {Livi}, {von der L{\"u}he}, {Geiss},
  {Timothy}, {Gloeckler}, {Allegra}, {Basile}, {Brusa}, {Wood}, {Siegmund},
  {Fowler}, {Fisher}, \& {Jhabvala}}]{1995SoPh..162..313K}
{Kohl}, J.~L., {Esser}, R., {Gardner}, L.~D., {et~al.} 1995, \solphys, 162, 313

\bibitem[{{Kontar} \& {Brown}(2006)}]{2006ApJ...653L.149K}
{Kontar}, E.~P. \& {Brown}, J.~C. 2006, \apjl, 653, L149

\bibitem[{{Kontar} {et~al.}(2011){Kontar}, {Brown}, {Emslie}, {Hajdas},
  {Holman}, {Hurford}, {Ka{\v s}parov{\'a}}, {Mallik}, {Massone}, {McConnell},
  {Piana}, {Prato}, {Schmahl}, \& {Suarez-Garcia}}]{2011SSRv..159..301K}
{Kontar}, E.~P., {Brown}, J.~C., {Emslie}, A.~G., {et~al.} 2011, \ssr, 159, 301

\bibitem[{{Kontar} {et~al.}(2010){Kontar}, {Hannah}, {Jeffrey}, \&
  {Battaglia}}]{2010ApJ...717..250K}
{Kontar}, E.~P., {Hannah}, I.~G., {Jeffrey}, N.~L.~S., \& {Battaglia}, M. 2010,
  \apj, 717, 250

\bibitem[{{Kontar} {et~al.}(2008){Kontar}, {Hannah}, \&
  {MacKinnon}}]{2008A&A...489L..57K}
{Kontar}, E.~P., {Hannah}, I.~G., \& {MacKinnon}, A.~L. 2008, \aap, 489, L57

\bibitem[{{Kosugi} {et~al.}(1986){Kosugi}, {Ishiguro}, \&
  {Shibasaki}}]{1986PASJ...38....1K}
{Kosugi}, T., {Ishiguro}, M., \& {Shibasaki}, K. 1986, \pasj, 38, 1

\bibitem[{{Kosugi} {et~al.}(2007){Kosugi}, {Matsuzaki}, {Sakao}, {Shimizu},
  {Sone}, {Tachikawa}, {Hashimoto}, {Minesugi}, {Ohnishi}, {Yamada}, {Tsuneta},
  {Hara}, {Ichimoto}, {Suematsu}, {Shimojo}, {Watanabe}, {Shimada}, {Davis},
  {Hill}, {Owens}, {Title}, {Culhane}, {Harra}, {Doschek}, \&
  {Golub}}]{2007SoPh..243....3K}
{Kosugi}, T., {Matsuzaki}, K., {Sakao}, T., {et~al.} 2007, \solphys, 243, 3

\bibitem[{{Krasnoselskikh} {et~al.}(2010){Krasnoselskikh}, {Vekstein},
  {Hudson}, {Bale}, \& {Abbett}}]{2010ApJ...724.1542K}
{Krasnoselskikh}, V., {Vekstein}, G., {Hudson}, H.~S., {Bale}, S.~D., \&
  {Abbett}, W.~P. 2010, \apj, 724, 1542

\bibitem[{{Krucker} \& {Benz}(2000)}]{2000SoPh..191..341K}
{Krucker}, S. \& {Benz}, A.~O. 2000, \solphys, 191, 341

\bibitem[{{Krucker} {et~al.}(2002){Krucker}, {Christe}, {Lin}, {Hurford}, \&
  {Schwartz}}]{2002SoPh..210..445K}
{Krucker}, S., {Christe}, S., {Lin}, R.~P., {Hurford}, G.~J., \& {Schwartz},
  R.~A. 2002, \solphys, 210, 445

\bibitem[{{Krucker} {et~al.}(2013){Krucker}, {Gim{\'e}nez de Castro}, {Hudson},
  {Trottet}, {Bastian}, {Hales}, {Ka{\v s}parov{\'a}}, {Klein}, {Kretzschmar},
  {L{\"u}thi}, {Mackinnon}, {Pohjolainen}, \& {White}}]{2013A&ARv..21...58K}
{Krucker}, S., {Gim{\'e}nez de Castro}, C.~G., {Hudson}, H.~S., {et~al.} 2013,
  \aapr, 21, 58

\bibitem[{{Krucker} {et~al.}(2011){Krucker}, {Hudson}, {Jeffrey}, {Battaglia},
  {Kontar}, {Benz}, {Csillaghy}, \& {Lin}}]{2011ApJ...739...96K}
{Krucker}, S., {Hudson}, H.~S., {Jeffrey}, N.~L.~S., {et~al.} 2011, \apj, 739,
  96

\bibitem[{{Krucker} {et~al.}(2015){Krucker}, {Saint-Hilaire}, {Hudson},
  {Haberreiter}, {Martinez-Oliveros}, {Fivian}, {Hurford}, {Kleint},
  {Battaglia}, {Kuhar}, \& {Arnold}}]{2015ApJ...802...19K}
{Krucker}, S., {Saint-Hilaire}, P., {Hudson}, H.~S., {et~al.} 2015, \apj, 802,
  19

\bibitem[{{Kundu}(1959)}]{1959AnAp...22....1K}
{Kundu}, M.~R. 1959, Annales d'Astrophysique, 22, 1

\bibitem[{{Kundu}(1965)}]{1965sra..book.....K}
{Kundu}, M.~R. 1965, {Solar radio astronomy} (Interscience Publication, New
  York)

\bibitem[{{Kundu} \& {Alissandrakis}(1975)}]{1975Natur.257..465K}
{Kundu}, M.~R. \& {Alissandrakis}, C.~E. 1975, \nat, 257, 465

\bibitem[{{Kundu} {et~al.}(1979){Kundu}, {Rao}, {Erskine}, \&
  {Bregman}}]{1979ApJ...234.1122K}
{Kundu}, M.~R., {Rao}, A.~P., {Erskine}, F.~T., \& {Bregman}, J.~D. 1979, \apj,
  234, 1122

\bibitem[{{Kundu} {et~al.}(2006){Kundu}, {Schmahl}, {Grigis}, {Garaimov}, \&
  {Shibasaki}}]{2006A&A...451..691K}
{Kundu}, M.~R., {Schmahl}, E.~J., {Grigis}, P.~C., {Garaimov}, V.~I., \&
  {Shibasaki}, K. 2006, \aap, 451, 691

\bibitem[{{Kupriyanova} {et~al.}(2010){Kupriyanova}, {Melnikov}, {Nakariakov},
  \& {Shibasaki}}]{2010SoPh..267..329K}
{Kupriyanova}, E.~G., {Melnikov}, V.~F., {Nakariakov}, V.~M., \& {Shibasaki},
  K. 2010, Solar Physics, 267, 329

\bibitem[{{Kuznetsov} {et~al.}(2011){Kuznetsov}, {Nita}, \&
  {Fleishman}}]{2011ApJ...742...87K}
{Kuznetsov}, A.~A., {Nita}, G.~M., \& {Fleishman}, G.~D. 2011, \apj, 742, 87

\bibitem[{{Labrosse} {et~al.}(2010){Labrosse}, {Heinzel}, {Vial}, {Kucera},
  {Parenti}, {Gun{\'a}r}, {Schmieder}, \& {Kilper}}]{2010SSRv..151..243L}
{Labrosse}, N., {Heinzel}, P., {Vial}, J., {et~al.} 2010, Space Science
  Reviews, 151, 243

\bibitem[{{Lagg} {et~al.}(2014){Lagg}, {Solanki}, {van Noort}, \&
  {Danilovic}}]{2014A&A...568A..60L}
{Lagg}, A., {Solanki}, S.~K., {van Noort}, M., \& {Danilovic}, S. 2014, \aap,
  568, A60

\bibitem[{{Lantos} \& {Kundu}(1972)}]{1972A&A....21..119L}
{Lantos}, P. \& {Kundu}, M.~R. 1972, \aap, 21, 119

\bibitem[{{Leake} {et~al.}(2005){Leake}, {Arber}, \&
  {Khodachenko}}]{2005A&A...442.1091L}
{Leake}, J.~E., {Arber}, T.~D., \& {Khodachenko}, M.~L. 2005, \aap, 442, 1091

\bibitem[{{Lee}(2007)}]{2007SSRv..133...73L}
{Lee}, J. 2007, \ssr, 133, 73

\bibitem[{{Leenaarts} {et~al.}(2013){Leenaarts}, {Pereira}, {Carlsson},
  {Uitenbroek}, \& {De Pontieu}}]{2013ApJ...772...89L}
{Leenaarts}, J., {Pereira}, T.~M.~D., {Carlsson}, M., {Uitenbroek}, H., \& {De
  Pontieu}, B. 2013, \apj, 772, 89

\bibitem[{{Leenaarts} \& {Wedemeyer-B{\"o}hm}(2006)}]{2006ASPC..354..306L}
{Leenaarts}, J. \& {Wedemeyer-B{\"o}hm}, S. 2006, in Astronomical Society of
  the Pacific Conference Series, Vol. 354, Solar MHD Theory and Observations: A
  High Spatial Resolution Perspective, ed. J.~{Leibacher}, R.~F. {Stein}, \&
  H.~{Uitenbroek}, 306

\bibitem[{{Lemen} {et~al.}(2012){Lemen}, {Title}, {Akin}, {Boerner}, {Chou},
  {Drake}, {Duncan}, {Edwards}, {Friedlaender}, {Heyman}, {Hurlburt}, {Katz},
  {Kushner}, {Levay}, {Lindgren}, {Mathur}, {McFeaters}, {Mitchell}, {Rehse},
  {Schrijver}, {Springer}, {Stern}, {Tarbell}, {Wuelser}, {Wolfson}, {Yanari},
  {Bookbinder}, {Cheimets}, {Caldwell}, {Deluca}, {Gates}, {Golub}, {Park},
  {Podgorski}, {Bush}, {Scherrer}, {Gummin}, {Smith}, {Auker}, {Jerram},
  {Pool}, {Soufli}, {Windt}, {Beardsley}, {Clapp}, {Lang}, \&
  {Waltham}}]{2012SoPh..275...17L}
{Lemen}, J.~R., {Title}, A.~M., {Akin}, D.~J., {et~al.} 2012, \solphys, 275, 17

\bibitem[{{Leroy}(1972)}]{Leroy_1972SoPh...25..413L}
{Leroy}, J. 1972, \solphys, 25, 413

\bibitem[{{Levens} {et~al.}(2015){Levens}, {Labrosse}, {Fletcher}, \&
  {Schmieder}}]{Levens}
{Levens}, P., {Labrosse}, N., {Fletcher}, L., \& {Schmieder}, B. 2015, \aap, 582, id.A27, 11 

\bibitem[{{Li} {et~al.}(2012){Li}, {Morgan}, {Leonard}, \&
  {Jeska}}]{2012ApJ...752L..22L}
{Li}, X., {Morgan}, H., {Leonard}, D., \& {Jeska}, L. 2012, \apjl, 752, L22

\bibitem[{{Lin} {et~al.}(2002){Lin}, {Dennis}, {Hurford}, {Smith}, {Zehnder},
  {Harvey}, {Curtis}, {Pankow}, {Turin}, {Bester}, {Csillaghy}, {Lewis},
  {Madden}, {van Beek}, {Appleby}, {Raudorf}, {McTiernan}, {Ramaty}, {Schmahl},
  {Schwartz}, {Krucker}, {Abiad}, {Quinn}, {Berg}, {Hashii}, {Sterling},
  {Jackson}, {Pratt}, {Campbell}, {Malone}, {Landis}, {Barrington-Leigh},
  {Slassi-Sennou}, {Cork}, {Clark}, {Amato}, {Orwig}, {Boyle}, {Banks},
  {Shirey}, {Tolbert}, {Zarro}, {Snow}, {Thomsen}, {Henneck}, {McHedlishvili},
  {Ming}, {Fivian}, {Jordan}, {Wanner}, {Crubb}, {Preble}, {Matranga}, {Benz},
  {Hudson}, {Canfield}, {Holman}, {Crannell}, {Kosugi}, {Emslie}, {Vilmer},
  {Brown}, {Johns-Krull}, {Aschwanden}, {Metcalf}, \&
  {Conway}}]{2002SoPh..210....3L}
{Lin}, R.~P., {Dennis}, B.~R., {Hurford}, G.~J., {et~al.} 2002, \solphys, 210,
  3

\bibitem[{{Lin}(2011)}]{Lin_2011SSRv..158..237L}
{Lin}, Y. 2011, \ssr, 158, 237

\bibitem[{{Lin} {et~al.}(2009){Lin}, {Soler}, {Engvold}, {Ballester},
  {Langangen}, {Oliver}, \& {Rouppe van der Voort}}]{2009ApJ...704..870L}
{Lin}, Y., {Soler}, R., {Engvold}, O., {et~al.} 2009, \apj, 704, 870

\bibitem[{{Lindsey} \& {Hudson}(1976)}]{1976ApJ...203..753L}
{Lindsey}, C. \& {Hudson}, H.~S. 1976, \apj, 203, 753

\bibitem[{{Liseau} {et~al.}(2015){Liseau}, {Vlemmings}, {Bayo}, {Bertone},
  {Black}, {del Burgo}, {Chavez}, {Danchi}, {De la Luz}, {Eiroa}, {Ertel},
  {Fridlund}, {Justtanont}, {Krivov}, {Marshall}, {Mora}, {Montesinos},
  {Nyman}, {Olofsson}, {Sanz-Forcada}, {Th{\'e}bault}, \&
  {White}}]{2015A&A...573L...4L}
{Liseau}, R., {Vlemmings}, W., {Bayo}, A., {et~al.} 2015, \aap, 573, L4

\bibitem[{{Lites}(2011)}]{2011ApJ...737...52L}
{Lites}, B.~W. 2011, \apj, 737, 52

\bibitem[{{Lites} {et~al.}(1982){Lites}, {Chipman}, \& {White}}]{Lites1982}
{Lites}, B.~W., {Chipman}, E.~G., \& {White}, O.~R. 1982, \apj, 253, 367

\bibitem[{{Liu} {et~al.}(2012){Liu}, {Berger}, \& {Low}}]{2012ApJ...745L..21L}
{Liu}, W., {Berger}, T.~E., \& {Low}, B.~C. 2012, \apjl, 745, L21

\bibitem[{Lobanov \& Zensus(2001)}]{2001Sci...294..128L}
Lobanov, A.~P. \& Zensus, J.~A. 2001, Science, 294, 128

\bibitem[{{Loukitcheva} {et~al.}(2004){Loukitcheva}, {Solanki}, {Carlsson}, \&
  {Stein}}]{2004A&A...419..747L}
{Loukitcheva}, M., {Solanki}, S.~K., {Carlsson}, M., \& {Stein}, R.~F. 2004,
  \aap, 419, 747

\bibitem[{{Loukitcheva} {et~al.}(2015{\natexlab{a}}){Loukitcheva}, {Solanki},
  {Carlsson}, \& {White}}]{2015A&A...575A..15L}
{Loukitcheva}, M., {Solanki}, S.~K., {Carlsson}, M., \& {White}, S.~M.
  2015{\natexlab{a}}, \aap, 575, A15

\bibitem[{{Loukitcheva} {et~al.}(2006){Loukitcheva}, {Solanki}, \&
  {White}}]{2006A&A...456..713L}
{Loukitcheva}, M., {Solanki}, S.~K., \& {White}, S. 2006, \aap, 456, 713

\bibitem[{{Loukitcheva} {et~al.}(2009){Loukitcheva}, {Solanki}, \&
  {White}}]{2009A&A...497..273L}
{Loukitcheva}, M., {Solanki}, S.~K., \& {White}, S.~M. 2009, \aap, 497, 273

\bibitem[{{Loukitcheva} {et~al.}(2014){Loukitcheva}, {Solanki}, \&
  {White}}]{2014A&A...561A.133L}
{Loukitcheva}, M., {Solanki}, S.~K., \& {White}, S.~M. 2014, \aap, 561, A133

\bibitem[{{Loukitcheva} {et~al.}(2015{\natexlab{b}}){Loukitcheva}, {Solanki},
  {White}, \& {Carlsson}}]{2015arXiv150805686L}
{Loukitcheva}, M., {Solanki}, S.~K., {White}, S.~M., \& {Carlsson}, M.
  2015{\natexlab{b}}, ArXiv e-prints

\bibitem[{{Loukitcheva} {et~al.}(2008){Loukitcheva}, {Solanki}, \&
  {White}}]{2008Ap&SS.313..197L}
{Loukitcheva}, M.~A., {Solanki}, S.~K., \& {White}, S. 2008, \apss, 313, 197

\bibitem[{{L{\"u}thi} {et~al.}(2004){L{\"u}thi}, {L{\"u}di}, \&
  {Magun}}]{2004A&A...420..361L}
{L{\"u}thi}, T., {L{\"u}di}, A., \& {Magun}, A. 2004, \aap, 420, 361

\bibitem[{{Mackay} {et~al.}(2010){Mackay}, {Karpen}, {Ballester}, {Schmieder},
  \& {Aulanier}}]{2010SSRv..151..333M}
{Mackay}, D.~H., {Karpen}, J.~T., {Ballester}, J.~L., {Schmieder}, B., \&
  {Aulanier}, G. 2010, \ssr, 151, 333

\bibitem[{{MacNeice} {et~al.}(1984){MacNeice}, {Burgess}, {McWhirter}, \&
  {Spicer}}]{1984SoPh...90..357M}
{MacNeice}, P., {Burgess}, A., {McWhirter}, R.~W.~P., \& {Spicer}, D.~S. 1984,
  \solphys, 90, 357

\bibitem[{{Madsen} {et~al.}(2015){Madsen}, {Tian}, \&
  {DeLuca}}]{2015ApJ...800..129M}
{Madsen}, C.~A., {Tian}, H., \& {DeLuca}, E.~E. 2015, \apj, 800, 129

\bibitem[{{Magara}(2010)}]{2010ApJ...715L..40M}
{Magara}, T. 2010, \apjl, 715, L40

\bibitem[{Mahmud {et~al.}(2013)Mahmud, Coughlan, Murphy, Gabuzda, \&
  Hallahan}]{2013MNRAS.431..695M}
Mahmud, M., Coughlan, C.~P., Murphy, E., Gabuzda, D.~C., \& Hallahan, D.~R.
  2013, Monthly Notices of the Royal Astronomical Society, 431, 695

\bibitem[{{Makhmutov} {et~al.}(2003){Makhmutov}, {Raulin}, {Gim{\'e}nez de
  Castro}, {Kaufmann}, \& {Correia}}]{2003SoPh..218..211M}
{Makhmutov}, V.~S., {Raulin}, J.-P., {Gim{\'e}nez de Castro}, C.~G.,
  {Kaufmann}, P., \& {Correia}, E. 2003, \solphys, 218, 211

\bibitem[{{Mart{\'{\i}}nez Gonz{\'a}lez} {et~al.}(2012){Mart{\'{\i}}nez
  Gonz{\'a}lez}, {Asensio Ramos}, {Manso Sainz}, {Beck}, \&
  {Belluzzi}}]{2012ApJ...759...16M}
{Mart{\'{\i}}nez Gonz{\'a}lez}, M.~J., {Asensio Ramos}, A., {Manso Sainz}, R.,
  {Beck}, C., \& {Belluzzi}, L. 2012, \apj, 759, 16

\bibitem[{{Mart{\'{\i}}nez Oliveros} {et~al.}(2012){Mart{\'{\i}}nez Oliveros},
  {Hudson}, {Hurford}, {Krucker}, {Lin}, {Lindsey}, {Couvidat}, {Schou}, \&
  {Thompson}}]{2012ApJ...753L..26M}
{Mart{\'{\i}}nez Oliveros}, J.-C., {Hudson}, H.~S., {Hurford}, G.~J., {et~al.}
  2012, \apjl, 753, L26

\bibitem[{{Mart{\'{\i}}nez-Sykora} {et~al.}(2012){Mart{\'{\i}}nez-Sykora}, {De
  Pontieu}, \& {Hansteen}}]{2012ApJ...753..161M}
{Mart{\'{\i}}nez-Sykora}, J., {De Pontieu}, B., \& {Hansteen}, V. 2012, \apj,
  753, 161

\bibitem[{{Masuda} {et~al.}(2013){Masuda}, {Shimojo}, {Kawate}, {Ishikawa}, \&
  {Ohno}}]{2013PASJ...65S...1M}
{Masuda}, S., {Shimojo}, M., {Kawate}, T., {Ishikawa}, S.-n., \& {Ohno}, M.
  2013, \pasj, 65, 1

\bibitem[{{McIntosh} {et~al.}(2011){McIntosh}, {de Pontieu}, {Carlsson},
  {Hansteen}, {Boerner}, \& {Goossens}}]{2011Natur.475..477M}
{McIntosh}, S.~W., {de Pontieu}, B., {Carlsson}, M., {et~al.} 2011, \nat, 475,
  477

\bibitem[{{Mein} \& {Mein}(1976)}]{Mein1976}
{Mein}, N. \& {Mein}, P. 1976, \solphys, 49, 231

\bibitem[{{Mein}(1971)}]{Mein1971}
{Mein}, P. 1971, \solphys, 20, 3

\bibitem[{{Melrose}(1968)}]{1968Ap&SS...2..171M}
{Melrose}, D.~B. 1968, \apss, 2, 171

\bibitem[{{Mendoza-Brice{\~n}o} {et~al.}(2005){Mendoza-Brice{\~n}o},
  {Sigalotti}, \& {Erd{\'e}lyi}}]{2005ApJ...624.1080M}
{Mendoza-Brice{\~n}o}, C.~A., {Sigalotti}, L.~D.~G., \& {Erd{\'e}lyi}, R. 2005,
  \apj, 624, 1080

\bibitem[{{Miki{\'c}} {et~al.}(2013){Miki{\'c}}, {Lionello}, {Mok}, {Linker},
  \& {Winebarger}}]{2013ApJ...773...94M}
{Miki{\'c}}, Z., {Lionello}, R., {Mok}, Y., {Linker}, J.~A., \& {Winebarger},
  A.~R. 2013, \apj, 773, 94

\bibitem[{Mirabel \& Rodr{\'\i}guez(1998)}]{1998Natur.392..673M}
Mirabel, I.~F. \& Rodr{\'\i}guez, L.~F. 1998, Nature, 392, 673

\bibitem[{{Moll} {et~al.}(2011){Moll}, {Cameron}, \&
  {Sch{\"u}ssler}}]{2011A&A...533A.126M}
{Moll}, R., {Cameron}, R.~H., \& {Sch{\"u}ssler}, M. 2011, \aap, 533, A126+

\bibitem[{{Moll} {et~al.}(2012){Moll}, {Cameron}, \&
  {Sch{\"u}ssler}}]{2012A&A...541A..68M}
{Moll}, R., {Cameron}, R.~H., \& {Sch{\"u}ssler}, M. 2012, \aap, 541, A68

\bibitem[{{Mossessian} \& {Fleishman}(2012)}]{2012ApJ...748..140M}
{Mossessian}, G. \& {Fleishman}, G.~D. 2012, \apj, 748, 140

\bibitem[{{M{\"u}ller} {et~al.}(2005{\natexlab{a}}){M{\"u}ller}, {de Groof},
  {de Pontieu}, \& {Hansteen}}]{2005ESASP.600E..30M}
{M{\"u}ller}, D.~A.~N., {de Groof}, A., {de Pontieu}, B., \& {Hansteen}, V.~H.
  2005{\natexlab{a}}, in ESA Special Publication, Vol. 600, The Dynamic Sun:
  Challenges for Theory and Observations, 30

\bibitem[{{M{\"u}ller} {et~al.}(2005{\natexlab{b}}){M{\"u}ller}, {De Groof},
  {Hansteen}, \& {Peter}}]{2005A&A...436.1067M}
{M{\"u}ller}, D.~A.~N., {De Groof}, A., {Hansteen}, V.~H., \& {Peter}, H.
  2005{\natexlab{b}}, \aap, 436, 1067

\bibitem[{{M{\"u}ller} {et~al.}(2003){M{\"u}ller}, {Hansteen}, \&
  {Peter}}]{2003A&A...411..605M}
{M{\"u}ller}, D.~A.~N., {Hansteen}, V.~H., \& {Peter}, H. 2003, \aap, 411, 605

\bibitem[{{M{\"u}ller} {et~al.}(2004){M{\"u}ller}, {Peter}, \&
  {Hansteen}}]{2004A&A...424..289M}
{M{\"u}ller}, D.~A.~N., {Peter}, H., \& {Hansteen}, V.~H. 2004, \aap, 424, 289

\bibitem[{{Murcray} {et~al.}(1981){Murcray}, {Goldman}, {Murcray}, {Bradford},
  {Murcray}, {Coffey}, \& {Mankin}}]{murcray1981}
{Murcray}, F.~J., {Goldman}, A., {Murcray}, F.~H., {et~al.} 1981, \apjl, 247,
  L97

\bibitem[{{Nagai} \& {Emslie}(1984)}]{1984ApJ...279..896N}
{Nagai}, F. \& {Emslie}, A.~G. 1984, \apj, 279, 896

\bibitem[{{Nakajima} {et~al.}(1983){Nakajima}, {Kosugi}, {Kai}, \&
  {Enome}}]{1983Natur.305..292N}
{Nakajima}, H., {Kosugi}, T., {Kai}, K., \& {Enome}, S. 1983, \nat, 305, 292

\bibitem[{{Nakajima} {et~al.}(1994){Nakajima}, {Nishio}, {Enome}, {Shibasaki},
  {Takano}, {Hanaoka}, {Torii}, {Sekiguchi}, {Bushimata}, {Kawashima},
  {Shinohara}, {Irimajiri}, {Koshiishi}, {Kosugi}, {Shiomi}, {Sawa}, \&
  {Kai}}]{1994IEEEP..82..705N}
{Nakajima}, H., {Nishio}, M., {Enome}, S., {et~al.} 1994, IEEE Proceedings, 82,
  705

\bibitem[{{Nakariakov} {et~al.}(2010){Nakariakov}, {Foullon}, {Myagkova}, \&
  {Inglis}}]{2010ApJ...708L..47N}
{Nakariakov}, V.~M., {Foullon}, C., {Myagkova}, I.~N., \& {Inglis}, A.~R. 2010,
  \apjl, 708, L47

\bibitem[{{Nakariakov} {et~al.}(2006){Nakariakov}, {Foullon}, {Verwichte}, \&
  {Young}}]{2006A&A...452..343N}
{Nakariakov}, V.~M., {Foullon}, C., {Verwichte}, E., \& {Young}, N.~P. 2006,
  \aap, 452, 343

\bibitem[{{Nakariakov} \& {Melnikov}(2006)}]{2006A&A...446.1151N}
{Nakariakov}, V.~M. \& {Melnikov}, V.~F. 2006, \aap, 446, 1151

\bibitem[{{Nakariakov} \& {Melnikov}(2009)}]{2009SSRv..149..119N}
{Nakariakov}, V.~M. \& {Melnikov}, V.~F. 2009, \ssr, 149, 119

\bibitem[{{Nelson} {et~al.}(2015){Nelson}, {Scullion}, {Doyle}, {Freij}, \&
  {Erd{\'e}lyi}}]{2015ApJ...798...19N}
{Nelson}, C.~J., {Scullion}, E.~M., {Doyle}, J.~G., {Freij}, N., \&
  {Erd{\'e}lyi}, R. 2015, \apj, 798, 19

\bibitem[{{Nindos} {et~al.}(2002){Nindos}, {Alissandrakis}, {Gelfreikh},
  {Bogod}, \& {Gontikakis}}]{2002A&A...386..658N}
{Nindos}, A., {Alissandrakis}, C.~E., {Gelfreikh}, G.~B., {Bogod}, V.~M., \&
  {Gontikakis}, C. 2002, \aap, 386, 658

\bibitem[{{Nindos} {et~al.}(1999{\natexlab{a}}){Nindos}, {Kundu}, \&
  {White}}]{1999ApJ...513..983N}
{Nindos}, A., {Kundu}, M.~R., \& {White}, S.~M. 1999{\natexlab{a}}, \apj, 513,
  983

\bibitem[{{Nindos} {et~al.}(1999{\natexlab{b}}){Nindos}, {Kundu}, {White},
  {Gary}, {Shibasaki}, \& {Dere}}]{1999ApJ...527..415N}
{Nindos}, A., {Kundu}, M.~R., {White}, S.~M., {et~al.} 1999{\natexlab{b}},
  \apj, 527, 415

\bibitem[{{Nita} {et~al.}(2015){Nita}, {Fleishman}, {Kuznetsov}, {Kontar}, \&
  {Gary}}]{2015ApJ...799..236N}
{Nita}, G.~M., {Fleishman}, G.~D., {Kuznetsov}, A.~A., {Kontar}, E.~P., \&
  {Gary}, D.~E. 2015, \apj, 799, 236

\bibitem[{{Nita} {et~al.}(2004){Nita}, {Gary}, \& {Lee}}]{2004ApJ...605..528N}
{Nita}, G.~M., {Gary}, D.~E., \& {Lee}, J. 2004, \apj, 605, 528

\bibitem[{{Nitta} {et~al.}(2014){Nitta}, {Sun}, {Hoeksema}, \&
  {DeRosa}}]{2014ApJ...780L..23N}
{Nitta}, N.~V., {Sun}, X., {Hoeksema}, J.~T., \& {DeRosa}, M.~L. 2014, \apjl,
  780, L23

\bibitem[{{Noyes} \& {Hall}(1972)}]{noyes72b}
{Noyes}, R.~W. \& {Hall}, D.~N.~B. 1972, \apjl, 176, L89+

\bibitem[{{Ofman} {et~al.}(1994){Ofman}, {Davila}, \&
  {Steinolfson}}]{Ofman_1994GeoRL..21.2259O}
{Ofman}, L., {Davila}, J.~M., \& {Steinolfson}, R.~S. 1994, \grl, 21, 2259

\bibitem[{{Ohki} \& {Hudson}(1975)}]{1975SoPh...43..405O}
{Ohki}, K. \& {Hudson}, H.~S. 1975, \solphys, 43, 405

\bibitem[{{Oka} {et~al.}(2015){Oka}, {Krucker}, {Hudson}, \&
  {Saint-Hilaire}}]{2015ApJ...799..129O}
{Oka}, M., {Krucker}, S., {Hudson}, H.~S., \& {Saint-Hilaire}, P. 2015, \apj,
  799, 129

\bibitem[{{Okamoto} {et~al.}(2015{\natexlab{b}}){Okamoto}, {Antolin}, {De
  Pontieu}, {Uitenbroek}, {Van Doorsselaere}, \&
  {Yokoyama}}]{2015ApJ...809...71O}
{Okamoto}, T.~J., {Antolin}, P., {De Pontieu}, B., {et~al.} 2015{\natexlab{b}},
  \apj, 809, 71

\bibitem[{{Okamoto} {et~al.}(2007){Okamoto}, {Tsuneta}, {Berger}, {Ichimoto},
  {Katsukawa}, {Lites}, {Nagata}, {Shibata}, {Shimizu}, {Shine}, {Suematsu},
  {Tarbell}, \& {Title}}]{2007Sci...318.1577O}
{Okamoto}, T.~J., {Tsuneta}, S., {Berger}, T.~E., {et~al.} 2007, Science, 318,
  1577

\bibitem[{{Oliver} \& {Ballester}(2002)}]{2002SoPh..206...45O}
{Oliver}, R. \& {Ballester}, J.~L. 2002, \solphys, 206, 45

\bibitem[{{Orozco Su{\'a}rez} {et~al.}(2012){Orozco Su{\'a}rez}, {Asensio
  Ramos}, \& {Trujillo Bueno}}]{2012ApJ...761L..25O}
{Orozco Su{\'a}rez}, D., {Asensio Ramos}, A., \& {Trujillo Bueno}, J. 2012,
  \apjl, 761, L25

\bibitem[{{O'Shea} {et~al.}(2007){O'Shea}, {Banerjee}, \&
  {Doyle}}]{2007A&A...475L..25O}
{O'Shea}, E., {Banerjee}, D., \& {Doyle}, J.~G. 2007, \aap, 475, L25

\bibitem[{{Oster}(1961{\natexlab{a}})}]{1961RvMP...33..525O}
{Oster}, L. 1961{\natexlab{a}}, Reviews of Modern Physics, 33, 525

\bibitem[{{Oster}(1961{\natexlab{b}})}]{1961ApJ...134.1010O}
{Oster}, L. 1961{\natexlab{b}}, \apj, 134, 1010

\bibitem[{{Oster}(1970)}]{1970A&A.....9..318O}
{Oster}, L. 1970, \aap, 9, 318

\bibitem[{{Panesar} {et~al.}(2013){Panesar}, {Innes}, {Tiwari}, \&
  {Low}}]{2013A&A...549A.105P}
{Panesar}, N.~K., {Innes}, D.~E., {Tiwari}, S.~K., \& {Low}, B.~C. 2013, \aap,
  549, A105

\bibitem[{{Parenti}(2014)}]{2014LRSP...11....1P}
{Parenti}, S. 2014, Living Reviews in Solar Physics, 11, 1

\bibitem[{{Parker}(1953)}]{1953ApJ...117..431P}
{Parker}, E.~N. 1953, \apj, 117, 431

\bibitem[{{Parker}(1988)}]{1988ApJ...330..474P}
{Parker}, E.~N. 1988, \apj, 330, 474

\bibitem[{{Parks} \& {Winckler}(1969)}]{1969ApJ...155L.117P}
{Parks}, G.~K. \& {Winckler}, J.~R. 1969, \apjl, 155, L117

\bibitem[{{Parnell} \& {De Moortel}(2012)}]{2012RSPTA.370.3217P}
{Parnell}, C.~E. \& {De Moortel}, I. 2012, Royal Society of London
  Philosophical Transactions Series A, 370, 3217

\bibitem[{{Parnell} \& {Jupp}(2000)}]{2000ApJ...529..554P}
{Parnell}, C.~E. \& {Jupp}, P.~E. 2000, \apj, 529, 554

\bibitem[{{Peach}(2014)}]{peach2014}
{Peach}, G. 2014, Advances in Space Research, 54, 1180

\bibitem[{{Pereira} {et~al.}(2013){Pereira}, {Leenaarts}, {De Pontieu},
  {Carlsson}, \& {Uitenbroek}}]{2013ApJ...778..143P}
{Pereira}, T.~M.~D., {Leenaarts}, J., {De Pontieu}, B., {Carlsson}, M., \&
  {Uitenbroek}, H. 2013, \apj, 778, 143

\bibitem[{{P{\'e}rez} {et~al.}(1999){P{\'e}rez}, {Doyle}, {O'Shea}, \&
  {Keenan}}]{1999A&A...351.1139P}
{P{\'e}rez}, M.~E., {Doyle}, J.~G., {O'Shea}, E., \& {Keenan}, F.~P. 1999,
  \aap, 351, 1139

\bibitem[{{P{\'e}rez-Le{\'o}n} {et~al.}(2013){P{\'e}rez-Le{\'o}n}, {Hiriart},
  \& {Mendoza-Torres}}]{2013RMxAA..49....3P}
{P{\'e}rez-Le{\'o}n}, J.~E., {Hiriart}, D., \& {Mendoza-Torres}, J.~E. 2013,
  \rmxaa, 49, 3

\bibitem[{{P{\'e}rez Mart{\'{\i}}nez} {et~al.}(2011){P{\'e}rez
  Mart{\'{\i}}nez}, {Schr{\"o}der}, \& {Cuntz}}]{Perez2011}
{P{\'e}rez Mart{\'{\i}}nez}, M.~I., {Schr{\"o}der}, K.-P., \& {Cuntz}, M. 2011,
  \mnras, 414, 418

\bibitem[{{Peter} {et~al.}(2013){Peter}, {Bingert}, {Klimchuk}, {de Forest},
  {Cirtain}, {Golub}, {Winebarger}, {Kobayashi}, \&
  {Korreck}}]{2013A&A...556A.104P}
{Peter}, H., {Bingert}, S., {Klimchuk}, J.~A., {et~al.} 2013, \aap, 556, A104

\bibitem[{{Peter} {et~al.}(2014){Peter}, {Tian}, {Curdt}, {Schmit}, {Innes},
  {De Pontieu}, {Lemen}, {Title}, {Boerner}, {Hurlburt}, {Tarbell}, {Wuelser},
  {Mart{\'{\i}}nez-Sykora}, {Kleint}, {Golub}, {McKillop}, {Reeves}, {Saar},
  {Testa}, {Kankelborg}, {Jaeggli}, {Carlsson}, \&
  {Hansteen}}]{2014Sci...346C.315P}
{Peter}, H., {Tian}, H., {Curdt}, W., {et~al.} 2014, Science, 346, C315

\bibitem[{{Phillips} {et~al.}(2015){Phillips}, {Hills}, {Bastian}, {Hudson},
  {Marson}, \& {Wedemeyer}}]{Phillips2015}
{Phillips}, N., {Hills}, R., {Bastian}, T.~S., {et~al.} 2015, ASPCS Vol. 499, 347

\bibitem[{{Pohjolainen} {et~al.}(2000){Pohjolainen}, {Portier-Fozzani}, \&
  {Ragaigne}}]{2000A&AS..143..227P}
{Pohjolainen}, S., {Portier-Fozzani}, F., \& {Ragaigne}, D. 2000, \aaps, 143,
  227

\bibitem[{Pudritz {et~al.}(2012)Pudritz, Hardcastle, \&
  Gabuzda}]{2012SSRv..169...27P}
Pudritz, R.~E., Hardcastle, M.~J., \& Gabuzda, D.~C. 2012, Space Sci Rev, 169,
  27

\bibitem[{{Puschmann} {et~al.}(2012){Puschmann}, {Balthasar}, {Bauer}, {Hahn},
  {Popow}, {Seelemann}, {Volkmer}, {Woche}, \& {Denker}}]{2012ASPC..463..423P}
{Puschmann}, K.~G., {Balthasar}, H., {Bauer}, S.-M., {et~al.} 2012, in
  Astronomical Society of the Pacific Conference Series, Vol. 463, Second
  ATST-EAST Meeting: Magnetic Fields from the Photosphere to the Corona., ed.
  T.~R. {Rimmele}, A.~{Tritschler}, F.~{W{\"o}ger}, M.~{Collados Vera},
  H.~{Socas-Navarro}, R.~{Schlichenmaier}, M.~{Carlsson}, T.~{Berger},
  A.~{Cadavid}, P.~R. {Gilbert}, P.~R. {Goode}, \& M.~{Kn{\"o}lker}, 423

\bibitem[{{Qiu} {et~al.}(2004){Qiu}, {Liu}, {Gary}, {Nita}, \&
  {Wang}}]{2004ApJ...612..530Q}
{Qiu}, J., {Liu}, C., {Gary}, D.~E., {Nita}, G.~M., \& {Wang}, H. 2004, \apj,
  612, 530

\bibitem[{{Ramaty}(1969)}]{1969ApJ...158..753R}
{Ramaty}, R. 1969, \apj, 158, 753

\bibitem[{{Raoult} {et~al.}(1979){Raoult}, {Lantos}, \&
  {Fuerst}}]{1979SoPh...61..335R}
{Raoult}, A., {Lantos}, P., \& {Fuerst}, E. 1979, \solphys, 61, 335

\bibitem[{{Raulin} {et~al.}(2004){Raulin}, {Makhmutov}, {Kaufmann}, {Pacini},
  {L{\"u}thi}, {Hudson}, \& {Gary}}]{2004SoPh..223..181R}
{Raulin}, J.~P., {Makhmutov}, V.~S., {Kaufmann}, P., {et~al.} 2004, \solphys,
  223, 181

\bibitem[{{Raulin} {et~al.}(1999){Raulin}, {White}, {Kundu}, {Silva}, \&
  {Shibasaki}}]{1999ApJ...522..547R}
{Raulin}, J.-P., {White}, S.~M., {Kundu}, M.~R., {Silva}, A.~V.~R., \&
  {Shibasaki}, K. 1999, \apj, 522, 547

\bibitem[{{Reale}(2010)}]{2010LRSP....7....5R}
{Reale}, F. 2010, Living Reviews in Solar Physics, 7, 5

\bibitem[{{Reznikova} {et~al.}(2014){Reznikova}, {Antolin}, \& {Van
  Doorsselaere}}]{2014ApJ...785...86R}
{Reznikova}, V.~E., {Antolin}, P., \& {Van Doorsselaere}, T. 2014, \apj, 785,
  86

\bibitem[{{Reznikova} {et~al.}(2012){Reznikova}, {Shibasaki}, {Sych}, \&
  {Nakariakov}}]{2012ApJ...746..119R}
{Reznikova}, V.~E., {Shibasaki}, K., {Sych}, R.~A., \& {Nakariakov}, V.~M.
  2012, \apj, 746, 119

\bibitem[{{Reznikova} {et~al.}(2015){Reznikova}, {Van Doorsselaere}, \&
  {Kuznetsov}}]{2015REZNIKOVA}
{Reznikova}, V.~E., {Van Doorsselaere}, T., \& {Kuznetsov}, A.~A. 2015, {\aap}

\bibitem[{{Riethm{\"u}ller} {et~al.}(2013){Riethm{\"u}ller}, {Solanki}, {van
  Noort}, \& {Tiwari}}]{2013A&A...554A..53R}
{Riethm{\"u}ller}, T.~L., {Solanki}, S.~K., {van Noort}, M., \& {Tiwari}, S.~K.
  2013, \aap, 554, A53

\bibitem[{{Rimmele}(1997)}]{1997ApJ...490..458R}
{Rimmele}, T.~R. 1997, \apj, 490, 458

\bibitem[{{Rom{\v s}tajn} {et~al.}(2009){Rom{\v s}tajn}, {Braj{\v s}a},
  {W{\"o}hl}, {Benz}, {Temmer}, {Ro{\v s}a}, \& {Ru{\v
  z}djak}}]{2009CEAB...33...79R}
{Rom{\v s}tajn}, I., {Braj{\v s}a}, R., {W{\"o}hl}, H., {et~al.} 2009, Central
  European Astrophysical Bulletin, 33, 79

\bibitem[{{Rouppe van der Voort} \& {de la Cruz
  Rodr{\'{\i}}guez}(2013)}]{2013ApJ...776...56R}
{Rouppe van der Voort}, L. \& {de la Cruz Rodr{\'{\i}}guez}, J. 2013, \apj,
  776, 56

\bibitem[{{Russell} \& {Fletcher}(2013)}]{2013ApJ...765...81R}
{Russell}, A.~J.~B. \& {Fletcher}, L. 2013, \apj, 765, 81

\bibitem[{{Rutten}(2006)}]{2006ASPC..354..276R}
{Rutten}, R.~J. 2006, in Astronomical Society of the Pacific Conference Series,
  Vol. 354, Solar MHD Theory and Observations: A High Spatial Resolution
  Perspective, ed. J.~{Leibacher}, R.~F. {Stein}, \& H.~{Uitenbroek}, 276--283

\bibitem[{{Rutten} {et~al.}(2013){Rutten}, {Vissers}, {Rouppe van der Voort},
  {S{\"u}tterlin}, \& {Vitas}}]{2013JPhCS.440a2007R}
{Rutten}, R.~J., {Vissers}, G.~J.~M., {Rouppe van der Voort}, L.~H.~M.,
  {S{\"u}tterlin}, P., \& {Vitas}, N. 2013, Journal of Physics Conference
  Series, 440, 012007

\bibitem[{{Saint-Hilaire} {et~al.}(2010){Saint-Hilaire}, {Krucker}, \&
  {Lin}}]{2010ApJ...721.1933S}
{Saint-Hilaire}, P., {Krucker}, S., \& {Lin}, R.~P. 2010, \apj, 721, 1933

\bibitem[{{Sakurai} {et~al.}(1991){Sakurai}, {Goossens}, \&
  {Hollweg}}]{1991SoPh..133..227S}
{Sakurai}, T., {Goossens}, M., \& {Hollweg}, J.~V. 1991, \solphys, 133, 227

\bibitem[{{Schad} {et~al.}(2013){Schad}, {Penn}, \&
  {Lin}}]{2013ApJ...768..111S}
{Schad}, T.~A., {Penn}, M.~J., \& {Lin}, H. 2013, \apj, 768, 111

\bibitem[{{Schaffenberger} {et~al.}(2006){Schaffenberger},
  {Wedemeyer-B{\"o}hm}, {Steiner}, \& {Freytag}}]{2006ASPC..354..345S}
{Schaffenberger}, W., {Wedemeyer-B{\"o}hm}, S., {Steiner}, O., \& {Freytag}, B.
  2006, in Astronomical Society of the Pacific Conference Series, Vol. 354,
  Solar MHD Theory and Observations: A High Spatial Resolution Perspective, ed.
  J.~{Leibacher}, R.~F. {Stein}, \& H.~{Uitenbroek}, 345--+

\bibitem[{{Scherrer} {et~al.}(2012){Scherrer}, {Schou}, {Bush}, {Kosovichev},
  {Bogart}, {Hoeksema}, {Liu}, {Duvall}, {Zhao}, {Title}, {Schrijver},
  {Tarbell}, \& {Tomczyk}}]{2012SoPh..275..207S}
{Scherrer}, P.~H., {Schou}, J., {Bush}, R.~I., {et~al.} 2012, \solphys, 275,
  207

\bibitem[{{Schmelz} {et~al.}(2011){Schmelz}, {Worley}, {Anderson}, {Pathak},
  {Kimble}, {Jenkins}, \& {Saar}}]{2011ApJ...739...33S}
{Schmelz}, J.~T., {Worley}, B.~T., {Anderson}, D.~J., {et~al.} 2011, \apj, 739,
  33

\bibitem[{{Schmidt} {et~al.}(2012){Schmidt}, {von der L{\"u}he}, {Volkmer},
  {Denker}, {Solanki}, {Balthasar}, {Bello Gonzalez}, {Berkefeld}, {Collados},
  {Fischer}, {Halbgewachs}, {Heidecke}, {Hofmann}, {Kneer}, {Lagg}, {Nicklas},
  {Popow}, {Puschmann}, {Schmidt}, {Sigwarth}, {Sobotka}, {Soltau}, {Staude},
  {Strassmeier}, \& {Waldmann }}]{2012AN....333..796S}
{Schmidt}, W., {von der L{\"u}he}, O., {Volkmer}, R., {et~al.} 2012,
  Astronomische Nachrichten, 333, 796

\bibitem[{{Schmit} {et~al.}(2014){Schmit}, {Innes}, {Ayres}, {Peter}, {Curdt},
  \& {Jaeggli}}]{Schmitt2014}
{Schmit}, D.~J., {Innes}, D., {Ayres}, T., {et~al.} 2014, \aap, 569, L7

\bibitem[{{Schrijver}(1987)}]{Schrijver1987}
{Schrijver}, C.~J. 1987, \aap, 172, 111

\bibitem[{{Schrijver}(1995)}]{Schrijver1995}
{Schrijver}, C.~J. 1995, \aapr, 6, 181

\bibitem[{{Schrijver}(2001)}]{Schrijver_2001SoPh..198..325S}
{Schrijver}, C.~J. 2001, \solphys, 198, 325

\bibitem[{{Schr{\"o}der} {et~al.}(2012){Schr{\"o}der}, {Mittag}, {P{\'e}rez
  Mart{\'{\i}}nez}, {Cuntz}, \& {Schmitt}}]{Schroeder2012}
{Schr{\"o}der}, K.-P., {Mittag}, M., {P{\'e}rez Mart{\'{\i}}nez}, M.~I.,
  {Cuntz}, M., \& {Schmitt}, J.~H.~M.~M. 2012, \aap, 540, A130

\bibitem[{{Scullion} {et~al.}(2014){Scullion}, {Rouppe van der Voort},
  {Wedemeyer}, \& {Antolin}}]{2014ApJ...797...36S}
{Scullion}, E., {Rouppe van der Voort}, L., {Wedemeyer}, S., \& {Antolin}, P.
  2014, \apj, 797, 36

\bibitem[{{Selhorst} {et~al.}(2011){Selhorst}, {Gim{\'e}nez de Castro},
  {V{\'a}lio}, {Costa}, \& {Shibasaki}}]{2011ApJ...734...64S}
{Selhorst}, C.~L., {Gim{\'e}nez de Castro}, C.~G., {V{\'a}lio}, A., {Costa},
  J.~E.~R., \& {Shibasaki}, K. 2011, \apj, 734, 64

\bibitem[{{Selhorst} {et~al.}(2010){Selhorst}, {Gim{\'e}nez de Castro}, {Varela
  Saraiva}, \& {Costa}}]{2010A&A...509A..51S}
{Selhorst}, C.~L., {Gim{\'e}nez de Castro}, C.~G., {Varela Saraiva}, A.~C., \&
  {Costa}, J.~E.~R. 2010, \aap, 509, A51

\bibitem[{{Selhorst} {et~al.}(2005{\natexlab{a}}){Selhorst}, {Silva}, \&
  {Costa}}]{2005A&A...433..365S}
{Selhorst}, C.~L., {Silva}, A.~V.~R., \& {Costa}, J.~E.~R. 2005{\natexlab{a}},
  \aap, 433, 365

\bibitem[{{Selhorst} {et~al.}(2005{\natexlab{b}}){Selhorst}, {Silva}, \&
  {Costa}}]{2005A&A...440..367S}
{Selhorst}, C.~L., {Silva}, A.~V.~R., \& {Costa}, J.~E.~R. 2005{\natexlab{b}},
  \aap, 440, 367

\bibitem[{{Selhorst} {et~al.}(2003){Selhorst}, {Silva}, {Costa}, \&
  {Shibasaki}}]{2003A&A...401.1143S}
{Selhorst}, C.~L., {Silva}, A.~V.~R., {Costa}, J.~E.~R., \& {Shibasaki}, K.
  2003, \aap, 401, 1143

\bibitem[{{Shibasaki}(1998)}]{1998ASPC..140..373S}
{Shibasaki}, K. 1998, in Astronomical Society of the Pacific Conference Series,
  Vol. 140, Synoptic Solar Physics, ed. K.~S. {Balasubramaniam}, J.~{Harvey},
  \& D.~{Rabin}, 373

\bibitem[{{Shibasaki} {et~al.}(2011){Shibasaki}, {Alissandrakis}, \&
  {Pohjolainen}}]{2011SoPh..273..309S}
{Shibasaki}, K., {Alissandrakis}, C.~E., \& {Pohjolainen}, S. 2011, \solphys,
  273, 309

\bibitem[{{Shibayama} {et~al.}(2013){Shibayama}, {Maehara}, {Notsu}, {Notsu},
  {Nagao}, {Honda}, {Ishii}, {Nogami}, \& {Shibata}}]{shibayama2013}
{Shibayama}, T., {Maehara}, H., {Notsu}, S., {et~al.} 2013, \apjs, 209, 5

\bibitem[{{Shimizu}(2011)}]{2011ApJ...738...83S}
{Shimizu}, T. 2011, \apj, 738, 83

\bibitem[{{Silva} {et~al.}(1996){Silva}, {White}, {Lin}, {de Pater},
  {Shibasaki}, {Hudson}, \& {Kundu}}]{1996ApJ...458L..49S}
{Silva}, A.~V.~R., {White}, S.~M., {Lin}, R.~P., {et~al.} 1996, \apjl, 458, L49

\bibitem[{{Sim{\~o}es} {et~al.}(2015){Sim{\~o}es}, {Hudson}, \&
  {Fletcher}}]{2015SoPh..tmp...50S}
{Sim{\~o}es}, P.~J.~A., {Hudson}, H.~S., \& {Fletcher}, L. 2015, \solphys

\bibitem[{{Skartlien} {et~al.}(1994){Skartlien}, {Carlsson}, \&
  {Stein}}]{Skartlien1994}
{Skartlien}, R., {Carlsson}, M., \& {Stein}, R.~F. 1994, in Chromospheric
  Dynamics, ed. M.~{Carlsson}, 79

\bibitem[{{Skartlien} {et~al.}(2000){Skartlien}, {Stein}, \&
  {Nordlund}}]{2000ApJ...541..468S}
{Skartlien}, R., {Stein}, R.~F., \& {Nordlund}, {\AA}. 2000, \apj, 541, 468

\bibitem[{Sofue(2007)}]{2007PASJ...59..189S}
Sofue, Y. 2007, Publications of the Astronomical Society of Japan, 59, 189

\bibitem[{{Solanki}(2003)}]{2003A&ARv..11..153S}
{Solanki}, S.~K. 2003, \aapr, 11, 153

\bibitem[{{Solanki} {et~al.}(2003){Solanki}, {Lagg}, {Woch}, {Krupp}, \&
  {Collados}}]{2003Natur.425..692S}
{Solanki}, S.~K., {Lagg}, A., {Woch}, J., {Krupp}, N., \& {Collados}, M. 2003,
  \nat, 425, 692

\bibitem[{{Solanki} {et~al.}(1994){Solanki}, {Livingston}, \&
  {Ayres}}]{1994Sci...263...64S}
{Solanki}, S.~K., {Livingston}, W., \& {Ayres}, T. 1994, Science, 263, 64

\bibitem[{{Soler} {et~al.}(2015){Soler}, {Ballester}, \&
  {Zaqarashvili}}]{2015A&A...573A..79S}
{Soler}, R., {Ballester}, J.~L., \& {Zaqarashvili}, T.~V. 2015, \aap, 573, A79

\bibitem[{{Soler} {et~al.}(2013{\natexlab{a}}){Soler}, {Carbonell}, \&
  {Ballester}}]{2013ApJS..209...16S}
{Soler}, R., {Carbonell}, M., \& {Ballester}, J.~L. 2013{\natexlab{a}}, \apjs,
  209, 16

\bibitem[{{Soler} {et~al.}(2013{\natexlab{b}}){Soler}, {Carbonell},
  {Ballester}, \& {Terradas}}]{2013ApJ...767..171S}
{Soler}, R., {Carbonell}, M., {Ballester}, J.~L., \& {Terradas}, J.
  2013{\natexlab{b}}, \apj, 767, 171

\bibitem[{{Soler} {et~al.}(2010){Soler}, {Terradas}, {Oliver}, {Ballester}, \&
  {Goossens}}]{Soler_2010ApJ...712..875S}
{Soler}, R., {Terradas}, J., {Oliver}, R., {Ballester}, J.~L., \& {Goossens},
  M. 2010, \apj, 712, 875

\bibitem[{{Song} \& {Vasyli{\=u}nas}(2011)}]{2011JGRA..116.9104S}
{Song}, P. \& {Vasyli{\=u}nas}, V.~M. 2011, Journal of Geophysical Research
  (Space Physics), 116, 9104

\bibitem[{{Song} \& {Vasyli{\=u}nas}(2014)}]{2014ApJ...796L..23S}
{Song}, P. \& {Vasyli{\=u}nas}, V.~M. 2014, \apjl, 796, L23

\bibitem[{{Staiger} {et~al.}(1984){Staiger}, {Mattig}, {Schmieder}, \&
  {Deubner}}]{Staiger1984}
{Staiger}, J., {Mattig}, W., {Schmieder}, B., \& {Deubner}, F.-L. 1984,
  \memsai, 55, 147

\bibitem[{{Stallcop}(1974)}]{1974ApJ...187..179S}
{Stallcop}, J.~R. 1974, \apj, 187, 179

\bibitem[{{Straus} {et~al.}(2008){Straus}, {Fleck}, {Jefferies}, G.,
  {McIntosh}, {Reardon}, {Severino}, \& {Steffen}}]{2008ApJ...681L.125S}
{Straus}, T., {Fleck}, B., {Jefferies}, S., {et~al.} 2008, \apjl, 681

\bibitem[{{Su} {et~al.}(2012){Su}, {Wang}, {Veronig}, {Temmer}, \&
  {Gan}}]{2012ApJ...756L..41S}
{Su}, Y., {Wang}, T., {Veronig}, A., {Temmer}, M., \& {Gan}, W. 2012, \apjl,
  756, L41

\bibitem[{{Susino} {et~al.}(2010){Susino}, {Lanzafame}, {Lanza}, \&
  {Spadaro}}]{2010ApJ...709..499S}
{Susino}, R., {Lanzafame}, A.~C., {Lanza}, A.~F., \& {Spadaro}, D. 2010, \apj,
  709, 499

\bibitem[{{Sych} \& {Nakariakov}(2014)}]{2014A&A...569A..72S}
{Sych}, R. \& {Nakariakov}, V.~M. 2014, \aap, 569, A72

\bibitem[{{Tajima} {et~al.}(1987){Tajima}, {Sakai}, {Nakajima}, {Kosugi},
  {Brunel}, \& {Kundu}}]{1987ApJ...321.1031T}
{Tajima}, T., {Sakai}, J., {Nakajima}, H., {et~al.} 1987, \apj, 321, 1031

\bibitem[{{Tandberg-Hanssen}(1995)}]{Tandberg-Hansen_1995ASSL..199.....T}
{Tandberg-Hanssen}, E., ed. 1995, Astrophysics and Space Science Library, Vol.
  199, {The nature of solar prominences} (Dordrecht: Kluwer Academic
  Publishers)

\bibitem[{{Taylor} {et~al.}(1999){Taylor}, {Carilli}, \&
  {Perley}}]{1999ASPC..180.....T}
{Taylor}, G.~B., {Carilli}, C.~L., \& {Perley}, R.~A., eds. 1999, Astronomical
  Society of the Pacific Conference Series, Vol. 180, {Synthesis Imaging in
  Radio Astronomy II}

\bibitem[{{Terradas} {et~al.}(2008{\natexlab{a}}){Terradas}, {Andries},
  {Goossens}, {Arregui}, {Oliver}, \&
  {Ballester}}]{Terradas_2008ApJ...687L.115T}
{Terradas}, J., {Andries}, J., {Goossens}, M., {et~al.} 2008{\natexlab{a}},
  \apjl, 687, L115

\bibitem[{{Terradas} {et~al.}(2008{\natexlab{b}}){Terradas}, {Arregui},
  {Oliver}, {Ballester}, {Andries}, \& {Goossens}}]{2008ApJ...679.1611T}
{Terradas}, J., {Arregui}, I., {Oliver}, R., {et~al.} 2008{\natexlab{b}}, \apj,
  679, 1611

\bibitem[{{Testa} {et~al.}(2014{\natexlab{a}}){Testa}, {De Pontieu}, {Allred},
  {Carlsson}, {Reale}, {Daw}, {Hansteen}, {Martinez-Sykora}, {Liu}, {DeLuca},
  {Golub}, {McKillop}, {Reeves}, {Saar}, {Tian}, {Lemen}, {Title}, {Boerner},
  {Hurlburt}, {Tarbell}, {Wuelser}, {Kleint}, {Kankelborg}, \&
  {Jaeggli}}]{Testa2014}
{Testa}, P., {De Pontieu}, B., {Allred}, J., {et~al.} 2014{\natexlab{a}},
  Science, 346, B315

\bibitem[{{Testa} {et~al.}(2014{\natexlab{b}}){Testa}, {De Pontieu}, {Allred},
  {Carlsson}, {Reale}, {Daw}, {Hansteen}, {Martinez-Sykora}, {Liu}, {DeLuca},
  {Golub}, {McKillop}, {Reeves}, {Saar}, {Tian}, {Lemen}, {Title}, {Boerner},
  {Hurlburt}, {Tarbell}, {Wuelser}, {Kleint}, {Kankelborg}, \&
  {Jaeggli}}]{2014Sci...346B.315T}
{Testa}, P., {De Pontieu}, B., {Allred}, J., {et~al.} 2014{\natexlab{b}},
  Science, 346, B315

\bibitem[{{Tian} {et~al.}(2014{\natexlab{a}}){Tian}, {DeLuca}, {Cranmer}, {De
  Pontieu}, {Peter}, {Mart{\'{\i}}nez-Sykora}, {Golub}, {McKillop}, {Reeves},
  {Miralles}, {McCauley}, {Saar}, {Testa}, {Weber}, {Murphy}, {Lemen}, {Title},
  {Boerner}, {Hurlburt}, {Tarbell}, {Wuelser}, {Kleint}, {Kankelborg},
  {Jaeggli}, {Carlsson}, {Hansteen}, \& {McIntosh}}]{Tian2014}
{Tian}, H., {DeLuca}, E.~E., {Cranmer}, S.~R., {et~al.} 2014{\natexlab{a}},
  Science, 346, A315

\bibitem[{{Tian} {et~al.}(2014{\natexlab{b}}){Tian}, {Kleint}, {Peter},
  {Weber}, {Testa}, {DeLuca}, {Golub}, \& {Schanche}}]{2014ApJ...790L..29T}
{Tian}, H., {Kleint}, L., {Peter}, H., {et~al.} 2014{\natexlab{b}}, \apjl, 790,
  L29

\bibitem[{{Tiwari} {et~al.}(2014){Tiwari}, {Alexander}, {Winebarger}, \&
  {Moore}}]{2014ApJ...795L..24T}
{Tiwari}, S.~K., {Alexander}, C.~E., {Winebarger}, A.~R., \& {Moore}, R.~L.
  2014, \apjl, 795, L24

\bibitem[{{Tiwari} {et~al.}(2015){Tiwari}, {Moore}, {Winebarger}, \&
  {Alpert}}]{2015arXiv151107900T}
{Tiwari}, S.~K., {Moore}, R.~L., {Winebarger}, A.~R., \& {Alpert}, S.~E. 2015,
  ArXiv e-prints

\bibitem[{{Tiwari} {et~al.}(2013){Tiwari}, {van Noort}, {Lagg}, \&
  {Solanki}}]{2013A&A...557A..25T}
{Tiwari}, S.~K., {van Noort}, M., {Lagg}, A., \& {Solanki}, S.~K. 2013, \aap,
  557, A25

\bibitem[{{Tomczyk} {et~al.}(2007){Tomczyk}, {McIntosh}, {Keil}, {Judge},
  {Schad}, {Seeley}, \& {Edmondson}}]{2007Sci...317.1192T}
{Tomczyk}, S., {McIntosh}, S.~W., {Keil}, S.~L., {et~al.} 2007, Science, 317,
  1192

\bibitem[{{Tripathi} {et~al.}(2009){Tripathi}, {Mason}, {Dwivedi}, {del Zanna},
  \& {Young}}]{2009ApJ...694.1256T}
{Tripathi}, D., {Mason}, H.~E., {Dwivedi}, B.~N., {del Zanna}, G., \& {Young},
  P.~R. 2009, \apj, 694, 1256

\bibitem[{{Trottet} {et~al.}(2011){Trottet}, {Raulin}, {Gim{\'e}nez de Castro},
  {L{\"u}thi}, {Caspi}, {Mandrini}, {Luoni}, \&
  {Kaufmann}}]{2011SoPh..273..339T}
{Trottet}, G., {Raulin}, J.-P., {Gim{\'e}nez de Castro}, G., {et~al.} 2011,
  \solphys, 273, 339

\bibitem[{{Trottet} {et~al.}(2002){Trottet}, {Raulin}, {Kaufmann},
  {Siarkowski}, {Klein}, \& {Gary}}]{2002A&A...381..694T}
{Trottet}, G., {Raulin}, J.-P., {Kaufmann}, P., {et~al.} 2002, \aap, 381, 694

\bibitem[{{Trottet} {et~al.}(2015){Trottet}, {Raulin}, {Mackinnon},
  {Gim{\'e}nez de Castro}, {Sim{\~o}es}, {Cabezas}, {de La Luz}, {Luoni}, \&
  {Kaufmann}}]{2015SoPh..290.2809T}
{Trottet}, G., {Raulin}, J.-P., {Mackinnon}, A., {et~al.} 2015, \solphys, 290,
  2809

\bibitem[{{Tsiropoula} {et~al.}(2000){Tsiropoula}, {Alissandrakis}, \&
  {Mein}}]{2000A&A...355..375T}
{Tsiropoula}, G., {Alissandrakis}, C.~E., \& {Mein}, P. 2000, \aap, 355, 375

\bibitem[{{Tsiropoula} {et~al.}(2012){Tsiropoula}, {Tziotziou}, {Kontogiannis},
  {Madjarska}, {Doyle}, \& {Suematsu}}]{2012SSRv..169..181T}
{Tsiropoula}, G., {Tziotziou}, K., {Kontogiannis}, I., {et~al.} 2012, \ssr,
  169, 181

\bibitem[{{Tsuneta} {et~al.}(2008){Tsuneta}, {Ichimoto}, {Katsukawa}, {Nagata},
  {Otsubo}, {Shimizu}, {Suematsu}, {Nakagiri}, {Noguchi}, {Tarbell}, {Title},
  {Shine}, {Rosenberg}, {Hoffmann}, {Jurcevich}, {Kushner}, {Levay}, {Lites},
  {Elmore}, {Matsushita}, {Kawaguchi}, {Saito}, {Mikami}, {Hill}, \&
  {Owens}}]{2008SoPh..249..167T}
{Tsuneta}, S., {Ichimoto}, K., {Katsukawa}, Y., {et~al.} 2008, \solphys, 249,
  167

\bibitem[{{Tu} \& {Song}(2013)}]{2013ApJ...777...53T}
{Tu}, J. \& {Song}, P. 2013, \apj, 777, 53

\bibitem[{{Tun} {et~al.}(2011){Tun}, {Gary}, \&
  {Georgoulis}}]{2011ApJ...728....1T}
{Tun}, S.~D., {Gary}, D.~E., \& {Georgoulis}, M.~K. 2011, \apj, 728, 1

\bibitem[{{Ugarte-Urra} {et~al.}(2006){Ugarte-Urra}, {Winebarger}, \&
  {Warren}}]{2006ApJ...643.1245U}
{Ugarte-Urra}, I., {Winebarger}, A.~R., \& {Warren}, H.~P. 2006, \apj, 643,
  1245

\bibitem[{{Uitenbroek}(2000)}]{uitenbroek00b}
{Uitenbroek}, H. 2000, \apj, 536, 481

\bibitem[{{Ulmschneider} {et~al.}(1991){Ulmschneider}, {Priest}, \&
  {Rosner}}]{Ulmschneider1991}
{Ulmschneider}, P., {Priest}, E.~R., \& {Rosner}, R., eds. 1991, {Mechanisms of
  Chromospheric and Coronal Heating} (Springer-Verlag)

\bibitem[{{van Ballegooijen} \& {Cranmer}(2010)}]{2010ApJ...711..164V}
{van Ballegooijen}, A.~A. \& {Cranmer}, S.~R. 2010, \apj, 711, 164

\bibitem[{{van den Oord}(1990)}]{1990A&amp;A...234..496V}
{van den Oord}, G.~H.~J. 1990, \aap, 234, 496

\bibitem[{{Van Doorsselaere} {et~al.}(2011){Van Doorsselaere}, {De Groof},
  {Zender}, {Berghmans}, \& {Goossens}}]{2011ApJ...740...90V}
{Van Doorsselaere}, T., {De Groof}, A., {Zender}, J., {Berghmans}, D., \&
  {Goossens}, M. 2011, \apj, 740, 90

\bibitem[{{Van Doorsselaere} {et~al.}(2014){Van Doorsselaere}, {Gijsen},
  {Andries}, \& {Verth}}]{2014ApJ...795...18V}
{Van Doorsselaere}, T., {Gijsen}, S.~E., {Andries}, J., \& {Verth}, G. 2014,
  \apj, 795, 18

\bibitem[{{Van Doorsselaere} {et~al.}(2008){Van Doorsselaere}, {Nakariakov}, \&
  {Verwichte}}]{2008ApJ...676L..73V}
{Van Doorsselaere}, T., {Nakariakov}, V.~M., \& {Verwichte}, E. 2008, \apjl,
  676, L73

\bibitem[{{van Noort}(2012)}]{2012A&A...548A...5V}
{van Noort}, M. 2012, \aap, 548, A5

\bibitem[{{van Noort} {et~al.}(2013){van Noort}, {Lagg}, {Tiwari}, \&
  {Solanki}}]{2013A&A...557A..24V}
{van Noort}, M., {Lagg}, A., {Tiwari}, S.~K., \& {Solanki}, S.~K. 2013, \aap,
  557, A24

\bibitem[{{Vernazza} {et~al.}(1981){Vernazza}, {Avrett}, \&
  {Loeser}}]{1981ApJS...45..635V}
{Vernazza}, J.~E., {Avrett}, E.~H., \& {Loeser}, R. 1981, \apjs, 45, 635

\bibitem[{{Verth} {et~al.}(2011){Verth}, {Goossens}, \&
  {He}}]{2011ApJ...733L..15V}
{Verth}, G., {Goossens}, M., \& {He}, J.-S. 2011, \apjl, 733, L15

\bibitem[{{Verth} {et~al.}(2010){Verth}, {Terradas}, \&
  {Goossens}}]{2010ApJ...718L.102V}
{Verth}, G., {Terradas}, J., \& {Goossens}, M. 2010, \apjl, 718, L102

\bibitem[{{Vial} \& {Engvold}(2015)}]{2015ASSL..415.....V}
{Vial}, J.-C. \& {Engvold}, O., eds. 2015, Astrophysics and Space Science
  Library, Vol. 415, {Solar Prominences}

\bibitem[{{Vissers} {et~al.}(2013){Vissers}, {Rouppe van der Voort}, \&
  {Rutten}}]{2013ApJ...774...32V}
{Vissers}, G.~J.~M., {Rouppe van der Voort}, L.~H.~M., \& {Rutten}, R.~J. 2013,
  \apj, 774, 32

\bibitem[{{V{\"o}gler} \& {Sch{\"u}ssler}(2007)}]{2007A&A...465L..43V}
{V{\"o}gler}, A. \& {Sch{\"u}ssler}, M. 2007, \aap, 465, L43

\bibitem[{{Vr\v snak} {et~al.}(1992){Vr\v snak}, {Pohjolainen}, {Urpo},
  {Terasranta}, {Brajsa}, {Ruzdjak}, {Mouradian}, \&
  {Jurac}}]{1992SoPh..137...67V}
{Vr\v snak}, B., {Pohjolainen}, S., {Urpo}, S., {et~al.} 1992, \solphys, 137,
  67

\bibitem[{{Walkowicz} {et~al.}(2011){Walkowicz}, {Basri}, {Batalha},
  {Gilliland}, {Jenkins}, {Borucki}, {Koch}, {Caldwell}, {Dupree}, {Latham},
  {Meibom}, {Howell}, {Brown}, \& {Bryson}}]{walkowicz2011}
{Walkowicz}, L.~M., {Basri}, G., {Batalha}, N., {et~al.} 2011, \aj, 141, 50

\bibitem[{{Wang} {et~al.}(2015){Wang}, {Gary}, {Fleishman}, \&
  {White}}]{2015ApJ...805...93W}
{Wang}, Z., {Gary}, D.~E., {Fleishman}, G.~D., \& {White}, S.~M. 2015, \apj,
  805, 93

\bibitem[{{Warren} {et~al.}(2001){Warren}, {Mariska}, \& {Lean}}]{Warren2001}
{Warren}, H.~P., {Mariska}, J.~T., \& {Lean}, J. 2001, \jgr, 106, 15745

\bibitem[{{Watanabe} {et~al.}(2008){Watanabe}, {Kitai}, {Okamoto}, {Nishida},
  {Kiyohara}, {Ueno}, {Hagino}, {Ishii}, \& {Shibata}}]{2008ApJ...684..736W}
{Watanabe}, H., {Kitai}, R., {Okamoto}, K., {et~al.} 2008, \apj, 684, 736

\bibitem[{{Wedemeyer} {et~al.}(2015){Wedemeyer}, {Bastian}, {Brajsa}, {Barta},
  {Hudson}, {Fleishman}, {Loukitcheva}, {Fleck}, {Kontar}, {De Pontieu},
  {Tiwari}, {Kato}, {Soler}, {Yagoubov}, {Black}, {Antolin}, {Gunar},
  {Labrosse}, {Benz}, {Nindos}, {Steffen}, {Scullion}, {Doyle}, {Zaqarashvili},
  {Hanslmeier}, {Nakariakov}, {Heinzel}, {Ayres}, {Karlicky}, \& {the SSALMON
  Group}}]{ssalmon_espm15}
{Wedemeyer}, S., {Bastian}, T., {Brajsa}, R., {et~al.} 2015, ArXiv e-prints

\bibitem[{{Wedemeyer} {et~al.}(2003{\natexlab{a}}){Wedemeyer}, {Freytag},
  {Steffen}, {Ludwig}, \& {Holweger}}]{wedemeyer03a}
{Wedemeyer}, S., {Freytag}, B., {Steffen}, M., {Ludwig}, H.-G., \& {Holweger},
  H. 2003{\natexlab{a}}, Astron.~Nachr., 324, No.~4, 410

\bibitem[{{Wedemeyer} {et~al.}(2003{\natexlab{b}}){Wedemeyer}, {Freytag},
  {Steffen}, {Ludwig}, \& {Holweger}}]{wedemeyer03c}
{Wedemeyer}, S., {Freytag}, B., {Steffen}, M., {Ludwig}, H.-G., \& {Holweger},
  H. 2003{\natexlab{b}}, Astron.~Nachr., 324, Suppl.Issue~3, 66

\bibitem[{{Wedemeyer} {et~al.}(2004){Wedemeyer}, {Freytag}, {Steffen},
  {Ludwig}, \& {Holweger}}]{2004A&A...414.1121W}
{Wedemeyer}, S., {Freytag}, B., {Steffen}, M., {Ludwig}, H.-G., \& {Holweger},
  H. 2004, \aap, 414, 1121

\bibitem[{{Wedemeyer} \& {Parmer}(2015)}]{ssalmon_alma_imaging15}
{Wedemeyer}, S. \& {Parmer}, A. 2015, ASPCS, Vol. 499, 343

\bibitem[{{Wedemeyer} {et~al.}(2013{\natexlab{a}}){Wedemeyer}, {Scullion},
  {Rouppe van der Voort}, {Bosnjak}, \& {Antolin}}]{2013ApJ...774..123W}
{Wedemeyer}, S., {Scullion}, E., {Rouppe van der Voort}, L., {Bosnjak}, A., \&
  {Antolin}, P. 2013{\natexlab{a}}, \apj, 774, 123

\bibitem[{{Wedemeyer} {et~al.}(2013{\natexlab{b}}){Wedemeyer}, {Scullion},
  {Steiner}, {de la Cruz Rodriguez}, \& {Rouppe van der
  Voort}}]{2013JPhCS.440a2005W}
{Wedemeyer}, S., {Scullion}, E., {Steiner}, O., {de la Cruz Rodriguez}, J., \&
  {Rouppe van der Voort}, L.~H.~M. 2013{\natexlab{b}}, Journal of Physics
  Conference Series, 440, 012005

\bibitem[{{Wedemeyer} \& {Steiner}(2014)}]{2014PASJ...66S..10W}
{Wedemeyer}, S. \& {Steiner}, O. 2014, \pasj, 66, 10

\bibitem[{{Wedemeyer-B{\"o}hm} {et~al.}(2005){Wedemeyer-B{\"o}hm}, {Kamp},
  {Bruls}, \& {Freytag}}]{2005A&A...438.1043W}
{Wedemeyer-B{\"o}hm}, S., {Kamp}, I., {Bruls}, J., \& {Freytag}, B. 2005, \aap,
  438, 1043

\bibitem[{{Wedemeyer-B{\"o}hm} {et~al.}(2006){Wedemeyer-B{\"o}hm}, {Kamp},
  {Freytag}, {Bruls}, \& {Steffen}}]{2006ASPC..354..301W}
{Wedemeyer-B{\"o}hm}, S., {Kamp}, I., {Freytag}, B., {Bruls}, J., \& {Steffen},
  M. 2006, in Astronomical Society of the Pacific Conference Series, Vol. 354,
  Solar MHD Theory and Observations: A High Spatial Resolution Perspective, ed.
  J.~{Leibacher}, R.~F. {Stein}, \& H.~{Uitenbroek}, 301--+

\bibitem[{{Wedemeyer-B{\"o}hm} {et~al.}(2009){Wedemeyer-B{\"o}hm}, {Lagg}, \&
  {Nordlund}}]{2009SSRv..144..317W}
{Wedemeyer-B{\"o}hm}, S., {Lagg}, A., \& {Nordlund}, {\AA}. 2009, Space Science
  Reviews, 144, 317

\bibitem[{{Wedemeyer-B{\"o}hm} {et~al.}(2007){Wedemeyer-B{\"o}hm}, {Ludwig},
  {Steffen}, {Leenaarts}, \& {Freytag}}]{2007A&A...471..977W}
{Wedemeyer-B{\"o}hm}, S., {Ludwig}, H.~G., {Steffen}, M., {Leenaarts}, J., \&
  {Freytag}, B. 2007, \aap, 471, 977

\bibitem[{{Wedemeyer-B{\"o}hm} \& {Rouppe van der
  Voort}(2009)}]{2009A&A...507L...9W}
{Wedemeyer-B{\"o}hm}, S. \& {Rouppe van der Voort}, L. 2009, \aap, 507, L9

\bibitem[{{Wedemeyer-B{\"o}hm} {et~al.}(2012){Wedemeyer-B{\"o}hm}, {Scullion},
  {Steiner}, {van der Voort}, {de La Cruz Rodriguez}, {Fedun}, \&
  {Erd{\'e}lyi}}]{2012Natur.486..505W}
{Wedemeyer-B{\"o}hm}, S., {Scullion}, E., {Steiner}, O., {et~al.} 2012, \nat,
  486, 505

\bibitem[{{Wedemeyer-B{\"o}hm} \& {Steffen}(2007)}]{2007A&A...462L..31W}
{Wedemeyer-B{\"o}hm}, S. \& {Steffen}, M. 2007, \aap, 462, L31

\bibitem[{{Welsh} {et~al.}(2006){Welsh}, {Wheatley}, {Browne}, {Siegmund},
  {Doyle}, {O'Shea}, {Antonova}, {Forster}, {Seibert}, {Morrissey}, \&
  {Taroyan}}]{2006A&A...458..921W}
{Welsh}, B.~Y., {Wheatley}, J., {Browne}, S.~E., {et~al.} 2006, \aap, 458, 921

\bibitem[{{White} {et~al.}(1992){White}, {Kundu}, {Bastian}, {Gary}, {Hurford},
  {Kucera}, \& {Bieging}}]{1992ApJ...384..656W}
{White}, S.~M., {Kundu}, M.~R., {Bastian}, T.~S., {et~al.} 1992, \apj, 384, 656

\bibitem[{{White} {et~al.}(2006){White}, {Loukitcheva}, \&
  {Solanki}}]{2006A&A...456..697W}
{White}, S.~M., {Loukitcheva}, M., \& {Solanki}, S.~K. 2006, \aap, 456, 697

\bibitem[{{Williams} {et~al.}(2002){Williams}, {Mathioudakis}, {Gallagher},
  {Phillips}, {McAteer}, {Keenan}, {Rudawy}, \& {Katsiyannis}}]{williams2002}
{Williams}, D.~R., {Mathioudakis}, M., {Gallagher}, P.~T., {et~al.} 2002,
  \mnras, 336, 747

\bibitem[{{Winebarger} {et~al.}(2014){Winebarger}, {Cirtain}, {Golub},
  {DeLuca}, {Savage}, {Alexander}, \& {Schuler}}]{2014ApJ...787L..10W}
{Winebarger}, A.~R., {Cirtain}, J., {Golub}, L., {et~al.} 2014, \apjl, 787, L10

\bibitem[{{Withbroe} \& {Noyes}(1977)}]{1977ARA&A..15..363W}
{Withbroe}, G.~L. \& {Noyes}, R.~W. 1977, \araa, 15, 363

\bibitem[{{Wittmann}(1969)}]{1969SoPh....7..366W}
{Wittmann}, A. 1969, \solphys, 7, 366

\bibitem[{{Xu} {et~al.}(2012){Xu}, {Cao}, {Jing}, \&
  {Wang}}]{2012ApJ...750L...7X}
{Xu}, Y., {Cao}, W., {Jing}, J., \& {Wang}, H. 2012, \apjl, 750, L7

\bibitem[{{Xu} {et~al.}(2004){Xu}, {Cao}, {Liu}, {Yang}, {Qiu}, {Jing},
  {Denker}, \& {Wang}}]{2004ApJ...607L.131X}
{Xu}, Y., {Cao}, W., {Liu}, C., {et~al.} 2004, \apjl, 607, L131

\bibitem[{{Young} {et~al.}(2014){Young}, {Tian}, \&
  {Jaeggli}}]{2014arXiv1409.8603Y}
{Young}, P., {Tian}, H., \& {Jaeggli}, S. 2014, ArXiv e-prints

\bibitem[{{Yu} {et~al.}(2013){Yu}, {Nakariakov}, {Selzer}, {Tan}, \&
  {Yan}}]{2013ApJ...777..159Y}
{Yu}, S., {Nakariakov}, V.~M., {Selzer}, L.~A., {Tan}, B., \& {Yan}, Y. 2013,
  \apj, 777, 159

\bibitem[{{Yuan} {et~al.}(2014){Yuan}, {Nakariakov}, {Huang}, {Li}, {Su},
  {Yan}, \& {Tan}}]{2014ApJ...792...41Y}
{Yuan}, D., {Nakariakov}, V.~M., {Huang}, Z., {et~al.} 2014, \apj, 792, 41

\bibitem[{{Yurchyshyn} {et~al.}(2015){Yurchyshyn}, {Abramenko}, \&
  {Kilcik}}]{2015ApJ...798..136Y}
{Yurchyshyn}, V., {Abramenko}, V., \& {Kilcik}, A. 2015, \apj, 798, 136

\bibitem[{{Zaitsev} \& {Stepanov}(1982)}]{1982SvAL....8..132Z}
{Zaitsev}, V.~V. \& {Stepanov}, A.~V. 1982, Soviet Astronomy Letters, 8, 132

\bibitem[{{Zaitsev} {et~al.}(2014){Zaitsev}, {Stepanov}, \&
  {Kaufmann}}]{2014SoPh..289.3017Z}
{Zaitsev}, V.~V., {Stepanov}, A.~V., \& {Kaufmann}, P. 2014, \solphys, 289,
  3017

\bibitem[{{Zaitsev} {et~al.}(1998){Zaitsev}, {Stepanov}, {Urpo}, \&
  {Pohjolainen}}]{1998A&A...337..887Z}
{Zaitsev}, V.~V., {Stepanov}, A.~V., {Urpo}, S., \& {Pohjolainen}, S. 1998,
  \aap, 337, 887

\bibitem[{{Zaqarashvili} {et~al.}(2011){Zaqarashvili}, {Khodachenko}, \&
  {Rucker}}]{2011A&A...529A..82Z}
{Zaqarashvili}, T.~V., {Khodachenko}, M.~L., \& {Rucker}, H.~O. 2011, \aap,
  529, A82

\bibitem[{{Zaqarashvili} {et~al.}(2013){Zaqarashvili}, {Khodachenko}, \&
  {Soler}}]{2013A&A...549A.113Z}
{Zaqarashvili}, T.~V., {Khodachenko}, M.~L., \& {Soler}, R. 2013, \aap, 549,
  A113

\bibitem[{{Zheleznyakov}(1962{\natexlab{a}})}]{1962AZh....39....5Z}
{Zheleznyakov}, V.~V. 1962{\natexlab{a}}, Astronomicheskii Zhurnal, 39, 5

\bibitem[{{Zheleznyakov}(1962{\natexlab{b}})}]{1962SvA.....6....3Z}
{Zheleznyakov}, V.~V. 1962{\natexlab{b}}, Soviet Astronomy, 6, 3

\end{thebibliography}

\end{document}